\documentclass[fleqn,usenatbib]{mnras}

\usepackage{newtxtext,newtxmath}

\usepackage[T1]{fontenc}

\DeclareRobustCommand{\VAN}[3]{#2}
\let\VANthebibliography\thebibliography
\def\thebibliography{\DeclareRobustCommand{\VAN}[3]{##3}\VANthebibliography}

\usepackage{graphicx}	
\usepackage{amsmath}	
\graphicspath{{figures}}
\usepackage{url}
\usepackage{subcaption}

\newcommand{\Msun}{\;M$_\odot$}
\newcommand{\gureft}{\textsc{gureft}}
\defcitealias{Rodriguez-Puebla2016}{RP16}
\defcitealias{Sheth2002}{ST}

\title[Ultra-high-$z$ halo demographics \& assembly]{Characterising ultra-high-redshift dark matter halo demographics and assembly histories with the \gureft\ simulations}

\author[L. Y. A. Yung et al.]{L. Y. Aaron\ Yung,$^{1,2}$\thanks{E-mail: aaron.yung@nasa.gov}\thanks{NASA Postdoctoral Fellow}
Rachel S.\ Somerville,$^{3}$
Tri Nguyen,$^{4,5,3}$
Peter Behroozi,$^{6,7}$
Chirag Modi,$^{3,8}$\newauthor
Jonathan P. Gardner,$^{1}$
\\
$^{1}$Astrophysics Science Division, NASA Goddard Space Flight Center, 8800 Greenbelt Rd, Greenbelt, MD 20771, USA\\
$^{2}$Space Telescope Science Institute, 3700 San Martin Drive, Baltimore, MD 21218, USA\\
$^{3}$Center for Computational Astrophysics, Flatiron Institute, 162 5th Ave, New York, NY 10010, USA\\
$^{4}$The NSF AI Institute for Artificial Intelligence and Fundamental Interactions\\
$^{5}$Department of Physics and Kavli Institute for Astrophysics and Space Research, Massachusetts Institute of Technology, \\\;\;\,77 Massachusetts Ave, Cambridge MA 02139, USA\\
$^{6}$ Department of Astronomy, University of Arizona, 933 N Cherry Ave, Tucson, AZ 85721, USA\\
$^{7}$ Division of Science, National Astronomical Observatory of Japan, 2-21-1 Osawa, Mitaka, Tokyo 181-8588, Japan\\
$^{8}$Center for Computational Mathematics, Flatiron Institute, 162 5th Ave, New York, NY 10010, USA
}

\date{Accepted XXX. Received YYY; in original form ZZZ}

\pubyear{\the\year{}}

\begin{document}
\label{firstpage}
\pagerange{\pageref{firstpage}--\pageref{lastpage}}
\maketitle

\begin{abstract}
Dark matter halo demographics and assembly histories are a manifestation of cosmological structure formation and have profound implications for the formation and evolution of galaxies. In particular, merger trees provide fundamental input for several modelling techniques, such as semi-analytic models (SAMs), sub-halo abundance matching (SHAM), and decorated halo occupation distribution models (HODs). Motivated by the new ultra-high-redshift ($z \gtrsim 10$) frontier enabled by \textit{JWST}, we present a new suite of \textit{Gadget at Ultrahigh Redshift with Extra-Fine Timesteps} (\textsc{gureft}) dark matter-only cosmological simulations that are carefully designed to capture halo merger histories and structural properties in the ultra-$z$ universe.
The simulation suite consists of four $1024^3$-particle simulations with box sizes of $5$, $15$, $35$, and $90$ Mpc\,h$^{-1}$, each with 170 snapshots stored between $40 \geq z \geq 6$.
With the unprecedented number of available snapshots and strategically chosen dynamic range covered by these boxes, \gureft\ uncovers the emerging dark matter halo populations and their assembly histories in the earliest epochs of cosmic history. In this work, we present the halo mass functions between $z\sim20$ to 6 down to $\log(M_\text{vir}/\text{M}_\odot)\sim5$, and show that at high redshift, these robust halo mass functions can differ substantially from commonly used analytic approximations or older fitting functions in the literature. 
We also present key physical properties of the ultra-high $z$ halo population, such as concentration and spin, as well as their mass growth and merger rates, and again provide updated fitting functions.
\end{abstract}



\section{Introduction}

Dark matter accounts for the bulk of the matter budget in our Universe and it is the fundamental driver of structure formation across a vast range of scales, from the typical length scales where matter becomes virialized (e.g. a few kilo-parsecs, \citealt{White1991}) to the cosmic web structures (e.g. tens of mega-parsecs, \citealt{Peebles1980}).
Halo structural properties can have strong implications for the properties of the galaxies that form within them. For example, halo spin and concentration can affect galaxy sizes and rotation velocities \citep{Mo1998, Bullock2001a, Somerville2008a}; halo circular velocity sets the depth of the potential well that gas must escape in order to be ejected by feedback processes, and concentration, density profile, and substructure tie closely to the number and distribution of satellite galaxies \citep{Navarro1997}.
Understanding the evolution of the overall demographics of dark matter halos is key to interpreting the evolution of the number density, spatial distribution, and clustering of observed galaxies across cosmic time. 
In addition, the assembly and merger histories of dark matter halos also have strong impacts on the star formation histories of enclosed galaxies \citep[e.g.][]{Conselice2003} and the growth of the supermassive black holes hosted in their nuclei \citep[e.g.][]{Hopkins2007b}.
Overall, dark matter halos provide a fundamental basic framework for modelling and interpreting observed galaxy populations (see \citealt{Wechsler2018} for a review).

Cosmic microwave background observations and galaxy surveys have provided strong evidence that our Universe is composed of mostly dark energy ($\Lambda$) and cold dark matter (CDM), and have provided fairly precise constraints on the fundamental cosmological parameters \citep[e.g.][]{Spergel2003, Planck2014}.
This paradigm is commonly referred to as the $\Lambda$CDM model.
Over the years, state-of-the-art dark matter-only cosmological simulations, such as Millennium \citep{Springel2005}, Millennium-II \citep{Boylan-Kolchin2009}, Bolshoi \citep{Klypin2011}, MultiDark and Bolshoi-Planck \citep{Prada2012, Klypin2016}, Uchuu \citep{Ishiyama2021}, Euclid Flagship \citep{Potter2017}, AEMULUS \citep{DeRose2019}, and \textsc{AbacusSummit} \citep{Garrison2016, Maksimova2021}, have been conducted to gain insights into the formation of large-scale structures and the assembly history of halos under the paradigm of dark matter dominated hierarchical structure formation.
$\Lambda$CDM-based cosmological-scale hydrodynamic and $N$-body simulations incorporating baryonic physics have been shown to be quite successful at reproducing many observed properties of galaxies across cosmic time \citep[for reviews see][]{Somerville2015a,Naab2017}.
Furthermore, simulated halo populations from this standard cosmological framework have also been extensively used to build galaxy models with semi-analytic (e.g. \texttt{galform} \citep{Cole2000}, L-Galaxies \citep{Henriques2015}, \texttt{SAGE} \citep{Croton2016}, \textsc{Galacticus} \citep{Benson2010}, \textit{Delphi} \citep{Dayal2014, Dayal2019}, Santa Cruz SAM \citep{Somerville2008,Somerville2015}) and (semi-)empirical modelling techniques (e.g. \textsc{UniverseMachine} \citep{Behroozi2019,Behroozi2020}, DREaM \citep{Drakos2021}, \textsc{Trinity} \citep{Zhang2023}, \textsc{Emerge} \citep{Moster2018, OLeary2023}).

In particular, dark matter halo merger and assembly histories, also commonly referred to as merger trees, are the backbone for semi-analytic models for galaxy formation and evolution \citep[for reviews see][]{Baugh2006, Benson2010, Somerville2015a}.
These merger trees can either be extracted from $N$-body simulations \citep{Behroozi2013c, Rodriguez-Gomez2015, Elahi2019} or constructed with a probabilistic, Monte Carlo approach \citep{Lacey1994, Somerville1999a, Somerville2008, Parkinson2008, Jiang2014} based on the Extended Press-Schechter (EPS) formalism \citep{Press1974, Bond1991, Lacey1993}. At lower redshifts ($z \lesssim 6$), it has been shown that the predictions of semi-analytic models run within EPS trees differ from those run within $N$-body trees at a level that is within the uncertainties on the parameters of the baryonic physics models; i.e., the differences in the key predictions are relatively small, and can be largely removed by re-calibrating the model parameters \citep[see e.g.][]{Gabrielpillai2022}.
However, \citet{Yung2023a} showed that the SAM results based on EPS trees can diverge significantly from those using $N$-body trees at $z \gtrsim 10$, especially in massive halos.
We also note that it is difficult to directly assess the performance of EPS merger trees at high- to- ultra-high redshift (e.g. $z\gtrsim10$) as the typical number of snapshots available and mass range covered by currently available N-body simulations are not optimised for the ultra-high-redshift universe.

The new frontier in the observable universe enabled by \textit{JWST} presents new challenges and needs for galaxy formation simulations and theory.
Numerical simulations inevitably face limitations from the tension between simulated volume and mass resolution, limited by the number of particles or grid cells that can be held in the memory of the largest available computers.
A large simulated volume is essential to capture the large-scale spatial distribution of structures, as well as providing robust statistical samples for rare objects.
On the other hand, high mass resolution is required to properly resolve both the low-mass halo populations and the progenitors of the massive halos.
Current state-of-the-art cosmological simulations cover volume and mass resolutions that are sub-optimal for tracking the assembly history of halos in the high (e.g. $z \gtrsim 10$) to ultrahigh (e.g. $z \gtrsim 15$) redshift universe.
While some of the latest cosmological simulations are reaching large volumes and dynamic range (e.g. Uchuu \citep{Ishiyama2021}, MilleniumTNG \citep{Pakmor2022}) thanks to the fast-growing capabilities of computing facilities, it is often impractical to overcome this bottleneck with brute force as the tension becomes even more pronounced in the early universe, as the emerging halo populations are expected to be orders of magnitude rarer and lower mass than their low-redshift counterparts. In addition, the typical number of high- and ultra-$z$ snapshots stored by large cosmological simulations are sparse and insufficient for reconstructing halo merger histories, as the cost for storage is also a major limiting factor.

In light of the above considerations, we introduce a new suite of dark matter-only, cosmological simulations designed to collectively address this dynamic range problem. This work provides simulated halo mass functions and halo structural properties across a wide mass range as well as high temporal resolution halo merger trees to aid galaxy formation modelling in this new era.
These simulations are extremely timely for the interpretation of recent \textit{JWST} observations of the ultra-high redshift universe. With the first year of scientific operations, \textit{JWST} has already revealed surprising and exciting results about the early universe. Early Release Observations (ERO) and Early Release Science (ERS) observations such as SMACS0723, GLASS, and CEERS almost immediately yielded galaxy candidates that broke existing redshift records, pushing into the $z\gtrsim 10$ regime \citep[e.g.][]{Finkelstein2022a,Finkelstein2022b,Naidu2022,Castellano2022,Atek2022,Adams2022,Donnan2022,Harikane2022}.
This progress has continued rapidly with additional surveys such as NGDEEP \citep{Bagley2023, Leung2023a} and JADES \citep{Curtis-Lake2022, Robertson2023}, and with spectroscopic confirmation of many of the $z\gtrsim10$ candidates \citep{ArrabalHaro2023a, Fujimoto2023, Harikane2023}.  \textit{JWST} has also revealed that there are a surprising number of obscured active galactic nuclei (AGN) and accreting supermassive black holes (SMBH) in the very early universe \citep[e.g.][]{Kocevski2023, Kocevski2022, Yang2023, Barro2023, Larson2023}.

The results presented in this work serve as a fundamental building block for the next generation of semi-analytic models for galaxies and black holes at ultra-high-redshift. The predictions of the Santa Cruz SAM with the current configuration and calibrations \citep{Gabrielpillai2022} at ultra-high redshifts (e.g. $8 \lesssim z \lesssim 17$) are presented in \citet{Yung2023a}. In a series of planned works, we will explore the impact of a wide variety of physical processes that operate in the ultra-high-$z$ universe, such as black hole seeding and growth, on the formation and evolution of galaxies across cosmic time.
In the same spirit as the \textit{Semi-analytic forecasts for the Universe} series \citep{Yung2019, Yung2019a, Yung2020, Yung2022, Yung2023}, we will also explore the co-evolution of early supermassive black holes and their host galaxies via various feedback channels, and present predictions and forecasts for current and future telescopes \citep{Yung2021, Yung_JWST2021}. Additionally, the \textsc{gureft} suite is being used to develop machine learning based techniques to very rapidly generate accurate halo merger histories over a large dynamic range in mass and redshift \citep{Nguyen2023}.

The structure of this paper is as follows: the design and specifications of the \gureft\ simulation suite are presented in Section \ref{sec:gureft}. We then the halo mass functions and key present key physical properties of the simulated halo populations in Section \ref{sec:results}. We discuss our results in Section \ref{sec:discussion}, and provide a summary and conclusions in Section \ref{sec:summary}.

\begin{table}
    \centering
    \caption{Specifications of the \gureft\ simulation suite and the Bolshoi-P and SMDPL simulations from the MultiDark suite. 
    }
    \label{tab:gureft_specs}
    \begin{tabular}{lcccccc}
        \hline
        & Box size          & $M_\text{DM}$            & N   & $\epsilon$      \\
        & [Mpc\,$h^{-1}$]   & [M$_\odot$\,$h^{-1}$]    &     & [kpc\,$h^{-1}$] \\
        \hline
        \gureft-05          & 5     & $9.92\times10^3$ & $1024^3$ & 0.16  \\
        \gureft-15          & 15    & $2.68\times10^5$ & $1024^3$ & 0.49  \\
        \gureft-35          & 35    & $3.40\times10^6$ & $1024^3$ & 1.14  \\
        \gureft-90          & 90    & $5.78\times10^7$ & $1024^3$ & 2.93  \\
        \hline
        BolshoiP            & 250   & $1.55\times10^8$ & $2048^3$ & 1.0   \\
        SMDPL               & 400   & $9.63\times10^7$ & $3840^3$ & 1.5   \\
        VSMDPL              & 160   & $6.20\times10^6$ & $3840^3$ & 2.0   \\
        \hline
    \end{tabular}
\end{table}

\begin{figure*}
    \includegraphics[width=2.1\columnwidth]{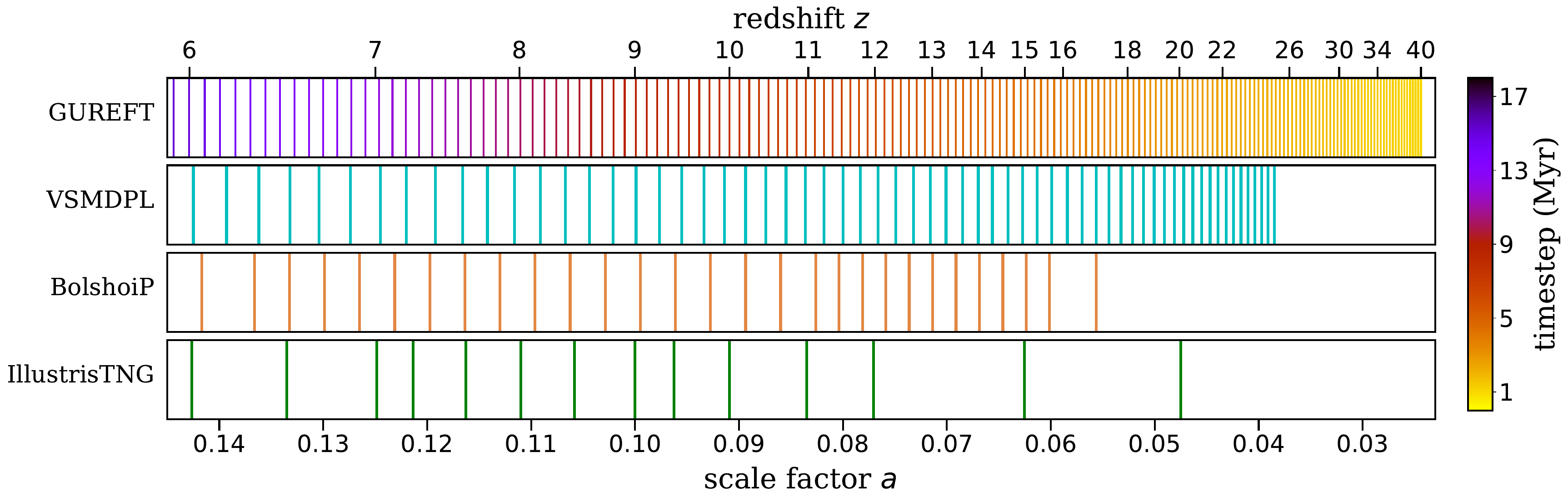}
    \caption{
        This figure shows the redshifts and scale factors at which snapshots are stored for the \gureft\ simulations (\textit{top}), which may be compared to same quantities for the \textsc{vsmdpl} (\textit{second row}), Bolshoi-Planck (\textit{third row}), and IllustrisTNG (\textit{bottom}) simulations over the same redshift range. For the \gureft\ simulations, we colour-coded the lines by the time elapsed between snapshots. Many previous $N$-body simulations have stored only a sparse number of snapshots at high redshifts.
    }
    \label{fig:barcode_plot}
\end{figure*}

\section{The GUREFT simulation suite}
\label{sec:gureft}

The \textsc{gadget} at Ultrahigh Redshift with Extra-Fine Timesteps (\gureft; pronounced \textit{graft}) is a suite of cosmological simulations designed to resolve the merger and assembly history of dark matter halos emerging in the early Universe.
Throughout this work, we adopt cosmological parameters $\Omega_\text{m} = 0.307$, $\Omega_\Lambda = 0.693$, $H_0 = 67.8$\,km\,s$^{-1}$\,Mpc$^{-1}$, $\sigma_8 = 0.829$, and $n_s = 0.960$; which are broadly consistent with the constraints reported by the Planck Collaboration (Planck Collaboration XIII \citeyear{Planck2016}). These are identical to the cosmological parameters adopted in the Bolshoi-Planck \citep[][hereafter BolshoiP]{Klypin2016}, and Small and Very Small Multidark (SMDPL\footnote{\url{https://www.cosmosim.org/metadata/smdpl/}} and VSMDPL\footnote{\url{https://www.cosmosim.org/metadata/vsmdpl/}}; see also \citealp{Klypin2016}) simulations, which we use as a comparison and complement throughout this work.

The \gureft\ suite consists of four dark matter-only $N$-body simulations carried out with the publicly available version of \textsc{gadget-2} code\footnote{\url{https://wwwmpa.mpa-garching.mpg.de/galform/gadget/}} \citep{Springel2005a}.
The nested grid initial conditions for our high-resolution simulations are generated using the Multi-Scale Initial Conditions \citep[MUSIC;][]{Hahn2011} at $z_\text{init} = 200$ using the second-order Lagrangian perturbation theory (2LPT), which is more accurate than the Zel'dovich approximation \citep{Scoccimarro1998, Crocce2006}.
In a set of pilot simulations, we carried out a convergence test with initial redshifts at $z_\text{init} = 100$, 200, and 400, and find no noticeable effects on the output halo population at $z < 40$ (see Appendix \ref{sec:AppD} for details).

Our suite of four simulations each consists of $N = 1024^3$ DM particles, representing matter distributed across simulated volumes with 5, 15, 35, and 90 Mpc $h^{-1}$ on a side. These simulations are labelled \gureft-05, \gureft-15, \gureft-35, \gureft-90 accordingly. The specifications of these simulations are summarised in Table \ref{tab:gureft_specs}, along with the corresponding specifications for BolshoiP, SMDPL, and VSMDPL\footnote{We note that these specifications are provided in units scaled by the Hubble parameter $h$, following the convention commonly used in the literature for cosmological simulations. Throughout the rest of this work, we use physical units (without the $h$ scaling) as we feel that this is more intuitive and the historic convention of scaling with $h$ is no longer well justified.}. We report on the size of the simulated volume, mass of DM particle ($M_\text{DM}$), number of particles, and the gravity softening length ($\epsilon$). The redshifts of the output snapshots and the rationale for the chosen box sizes are further detailed in Sections \ref{sec:timesteps} and \ref{sec:boxsizes}.

Halos are identified with the six-dimensional phase-space halo finder \textsc{rockstar}\footnote{\url{https://bitbucket.org/gfcstanford/rockstar/src/main/}} and a gravitationally consistent merger tree construction algorithm \textsc{consistent trees}\footnote{\url{https://bitbucket.org/pbehroozi/consistent-trees/src/main/}} \citep{Behroozi2013b, Behroozi2013c}.
The \textsc{rockstar} type approach is more robust at high redshift compared to the commonly used alternative "friends-of-friends" based halo finding algorithm, which tends to spuriously group halos in elongated (filamentary) structures together \citep{Klypin2011}.

The halo properties presented in this work are estimated by \textsc{rockstar} with its default configurations, where physical properties are computed only for bound particles.
For halo mass, \textsc{rockstar} computes the spherical over-densities according to a given density threshold relative to the background or critical density. Here, we adopted the \citet{Bryan1998} virial mass definition, which is defined relative to the cosmic critical matter density.
In addition, we require at least 100 DM particles for a halo to be considered resolved.
The maximum circular velocity, $V_\text{max}$, is given by the maximum of the quantity $\sqrt{(GM(r)r^{-1})}$.
The angular momentum of halos, $J$, which is required for computing the spin parameter (see Section \ref{sec:properties} for detail), is calculated using bound particles out to $R_\text{vir}$.
The scale radius, $R_s$ is computed by fitting an NFW \citep{Navarro1997} profile to the density profile of bound particles within the halos.
We refer the reader to \citet{Behroozi2013b} and the \textsc{rockstar} documentation for the full technical details related to the \textsc{rockstar} halo finder.
We include \textit{all} dark matter halos and subhalos found by \textsc{rockstar} in our analysis, which is consistent with the past analysis on the MultiDark suite that we adopt for comparison with our results \citep{Rodriguez-Puebla2016}.

\subsection{Spacing of Output Snapshots}
\label{sec:timesteps}

The temporal resolution of merger trees (which is limited by the spacing between output snapshots from $N$-body simulations) remains a major limitation in reconstructing ultra high-$z$ merger histories from simulation outputs. 
Most previous large volume $N$-body simulations have stored sparsely spaced snapshots at high redshifts (e.g. $z > 8$) and increased the number of snapshots stored towards lower redshifts (e.g. $z < 3$), as demonstrated in Fig.~\ref{fig:barcode_plot}.

For the \gureft\ simulations, to ensure that we properly resolve the formation history of halos out to $z\sim 20$, we chose a spacing between snapshots of one-tenth of the halo dynamical time at the output redshift, and store a total of 171 snapshots between $z \sim 40$ to 6. 
The halo dynamical time is defined as $t_\text{dyn} \equiv R_\text{vir}/V_\text{vir}$, where $R_\text{vir}$ is the virial radius and $V_\text{vir}$ is the halo virial velocity.
The redshifts and scale factors of stored snapshots are shown in Fig.~\ref{fig:barcode_plot}. We also show snapshots available from Bolshoi-Plank and IllustrisTNG in the same redshift range for comparison. The more finely spaced snapshots of \gureft\ allow for the reconstruction of detailed merger histories out to ultra high redshifts.

\begin{figure*}
    \centering
    \begin{subfigure}[b]{0.44\textwidth}
        \centering
        \includegraphics[width=\textwidth]{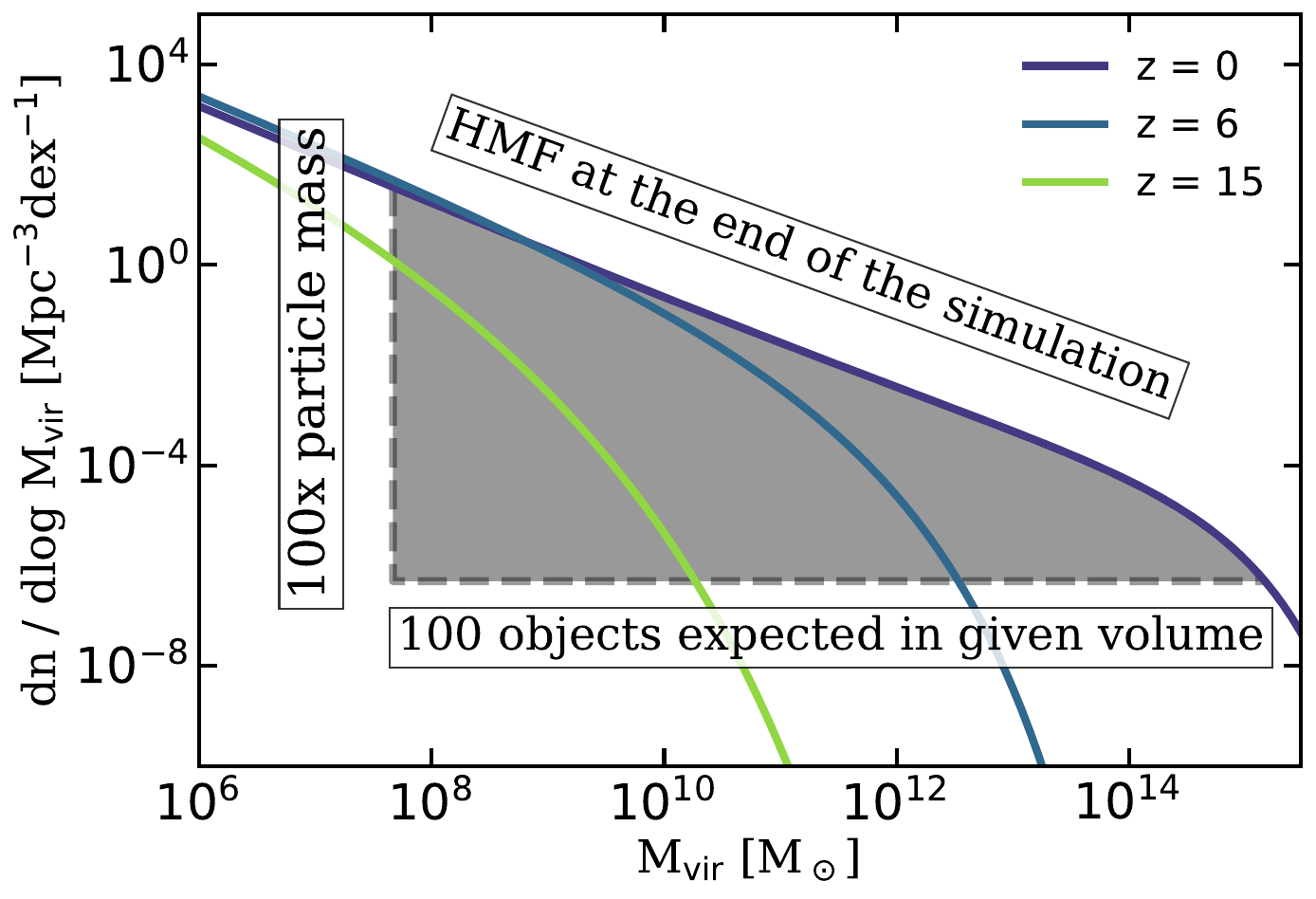}
    \end{subfigure}
    \hspace{0.01in}
    \begin{subfigure}[b]{0.55\textwidth}
        \centering
        \includegraphics[width=\textwidth]{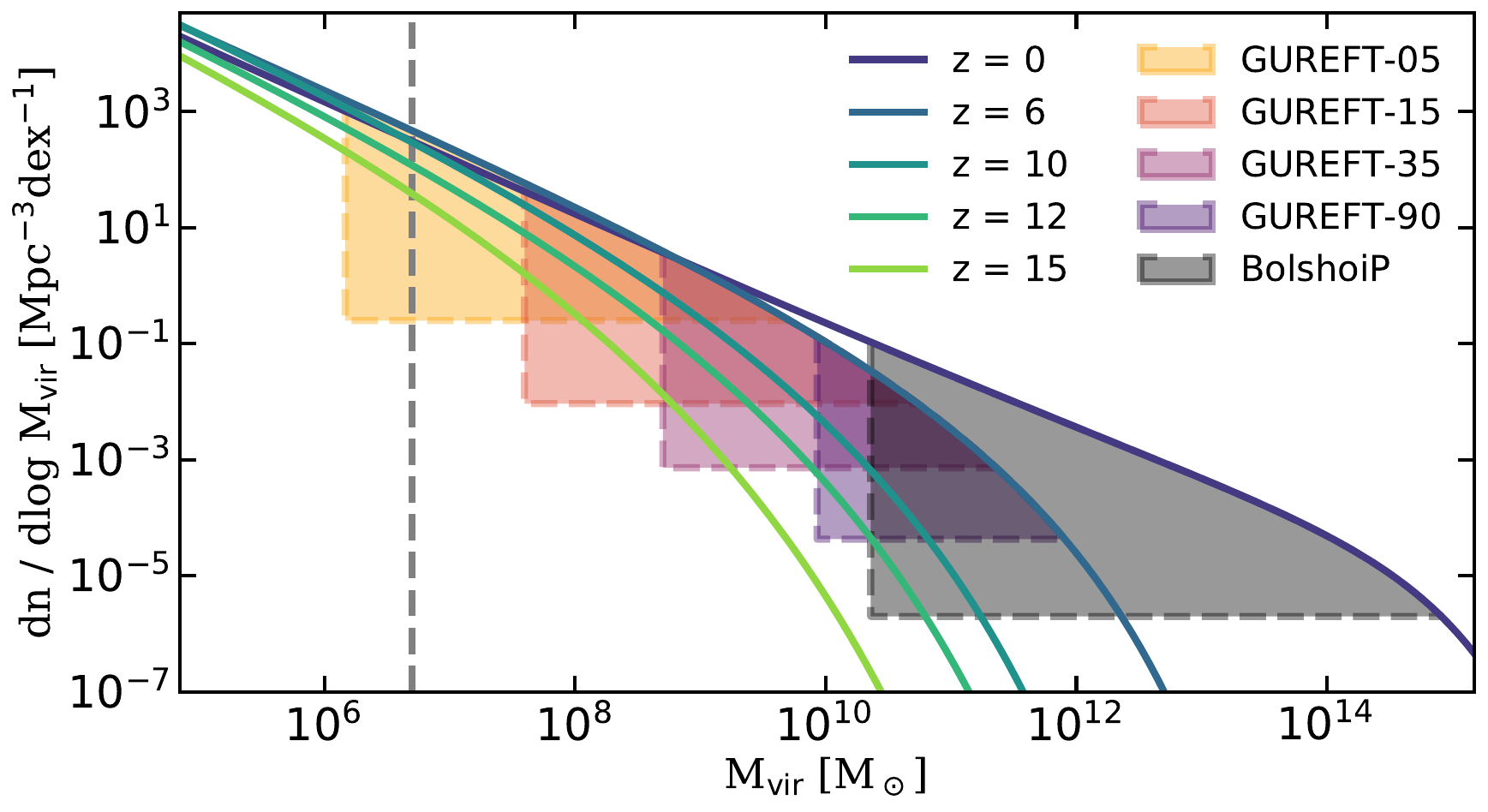}
    \end{subfigure}
    \caption{
        This figure shows `coverage triangles' that illustrate the rationale for choosing the mass resolution and volume of the \gureft\ suite. In the \textit{left panel}, we show an example \textit{`coverage triangle'} in the virial mass-number density space, which highlights the mass range of halos that are expected to be resolved by a cosmological simulation over time. The left boundary of this triangle is the mass where halos are well-resolved for the simulation $M_\text{vir} = 100 M_\text{DM}$ (vertical dashed line). The lower bound is where 100 objects are expected for the simulated volume, and the upper bound is the expected number density of halos at the redshift of the final output of the simulation. Objects to the left of the triangle fall below the mass resolution of the simulation and those below the triangle are too rare to be found in large numbers in the simulated volume. 
        In the \textit{right panel}, we show colour-coded coverage triangles for the suite of \gureft\ simulations (which terminate at $z=6$) and Bolshoi-Planck (which terminates at $z\sim0$). The vertical line marks the atomic cooling limit, below which galaxy formation is expected to be inefficient.
        In both panels, we show HMFs at selected redshifts from past studies to qualitatively represent how the halo population evolve over time.
    }
    \label{fig:coverage}
\end{figure*}

\subsection{Box sizes and resolution}
\label{sec:boxsizes}

The box sizes and resolution of the four \gureft\ simulations are selected to capture the assembly histories of halos across a wide mass range while providing an adequate sample size for statistical studies.
In addition, the mass range covered by these boxes are designed to \textit{overlap} with each other, which allows us to carry out convergence tests and to \textit{graft} together the merger trees extracted from these simulations using machine learning approaches \citep[][T. Nguyen et al. in prep]{Nguyen2023}.

We introduce a set of \textit{`coverage triangles'} in the virial mass-number density space to illustrate the rationale for the choice of survey volume and corresponding mass resolution. In this work, we assume that halos with 100 particles are well-resolved (we test this assumption in subsequent sections), implying a mass threshold of $M_\text{vir, res} = 100$\,$M_\text{DM}$. The leftmost edge of the triangles in Fig.~\ref{fig:coverage} show this limit for each of the simulation volumes. The horizontal line marks the number density which would yield at least 100 halos in the given volume. 
Given a fixed number of particles, the simulation volume is inversely related to its mass resolution.
This restriction is reflected in the area of the triangles.
The number of particles in a simulation is limited by the computational resources available and the efficiency of numerical methods.
The hypotenuse of the coverage triangles marks the HMF at a particular epoch. In Fig.~\ref{fig:coverage}, we show HMFs at selected redshifts to illustrate how the mass distribution of halo populations evolves over cosmic time. We note that the HMFs shown here are based on fits to previous simulations, and are extrapolated in some regions. They are meant as a qualitative illustrations and are not necessarily highly accurate. For $z \leq 10$, we show the HMF fits from the MultiDark simulations \citep{Rodriguez-Puebla2016, Rodriguez-Puebla2017} and at $z > 10$ we show the HMFs fits presented in \citet{Yung2020a}, which are obtained by combining BolshoiP \citep{Klypin2016} with very high resolution, small volume boxes from \citet{Visbal2018}. 

In the right panel of Fig.~\ref{fig:coverage}, we show the series of coverage triangles for the \gureft\ simulations and BolshoiP.
As illustrated, the suite of four simulations are designed to progressively increase in size as the mass resolution becomes coarser, with adequate overlap between boxes in the mass range where halos are both well-resolved and are present in sufficient numbers to comprise statistically robust samples.
The goal of this configuration is to maximise coverage of the halo population that is most important for modelling galaxy and BH formation physics over this redshift interval. 
For example, \gureft-05 is designed to resolve the low-mass halo populations (e.g. $6 \lesssim \log(M_\text{vir}/\text{M}_\odot) \lesssim 9$) between $18 \gtrsim z \gtrsim 10$, which may host Population III stars and black hole seeds. More massive halos are covered with the adjacent box \gureft-15, resolving halo masses $7.5 \lesssim \log(M_\text{vir}/\text{M}_\odot) \lesssim 11$ between $12 \gtrsim z \gtrsim 6$, which reaches down to approximately the halo mass range where metal-free atomic cooling is efficient ($T_{\rm vir} \simeq 10^4$ K. 
Similarly, the \gureft-35 and \gureft-90 boxes contain the more massive halos that are likely to host galaxies that are observable with \textit{JWST} \citep{Yung2023}.

\begin{figure*}
    \centering
    \begin{subfigure}[b]{0.49\textwidth}
        \centering
        \includegraphics[width=\textwidth]{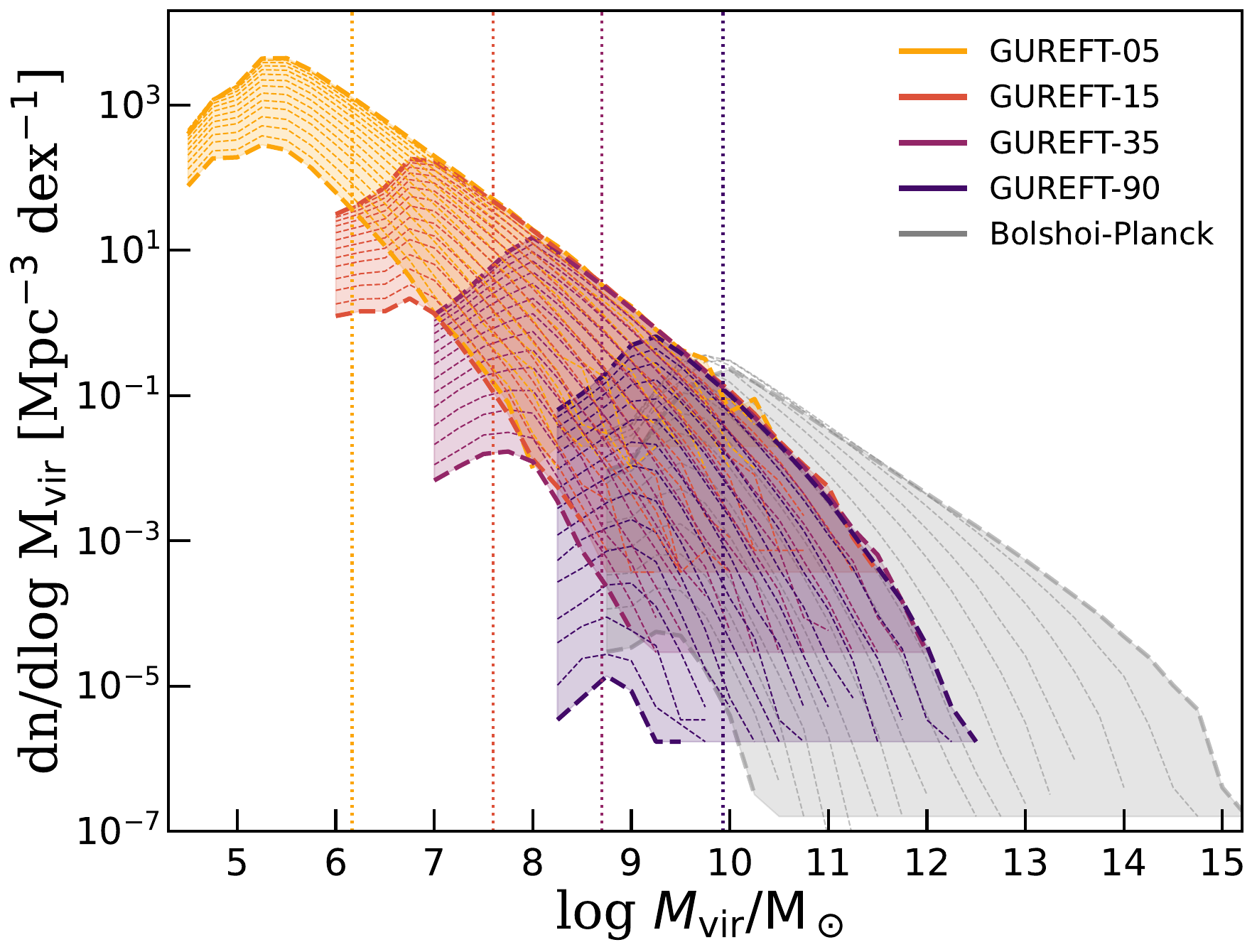}
    \end{subfigure}
    \hspace{0.01in}
    \begin{subfigure}[b]{0.488\textwidth}
        \centering
        \includegraphics[width=\textwidth]{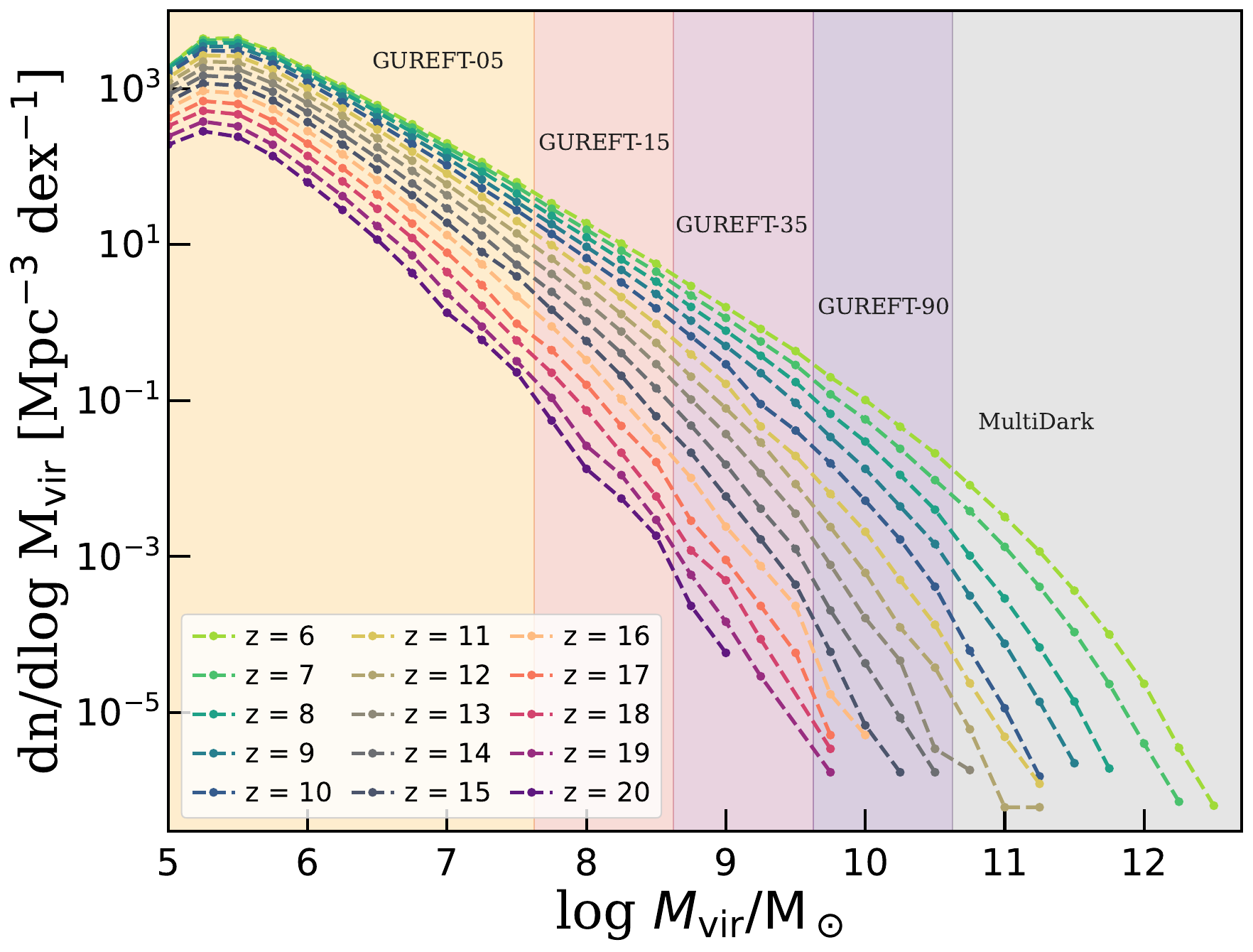}
    \end{subfigure}
    \caption{
        In this figure, we show the HMFs between $z = 6$ to 20 predicted from the full suite of \gureft\ simulations. In the \textit{left panel}, we show the halo mass functions colour-coded by the \gureft\ boxes. The shaded region shows the range between $z = 6$ to 20, with individual HMFs shown by colour-matched dashed lines, with thicker lines at the boundaries ($z = 6$ and 20). 
        The vertical dotted lines mark the mass thresholds corresponding to 100 DM particles for the \gureft\ boxes in matching colours.
        In addition, we show halo mass functions from the Bolshoi-Planck simulation between $z = 0$ to 15 (shown in grey with matching style). In the \textit{right panel}, we show the HMFs over a wide mass range obtained by combining well-resolved halo populations sourced from the four \gureft\ boxes. In addition, the massive end of the HMFs are supplemented with the MultiDark boxes (Bolshoi-Planck for $z \lesssim 10$ and VSMDPL for $z > 10$). We show shaded regions with matching colour in the right panel to indicate which simulation the halo population was drawn from. The combined \gureft\ and Multidark simulation suites allow us to probe halo properties over an extremely broad dynamic range.
    } 
    \label{fig:hmf}
\end{figure*}

\begin{figure}
    \includegraphics[width=\columnwidth]{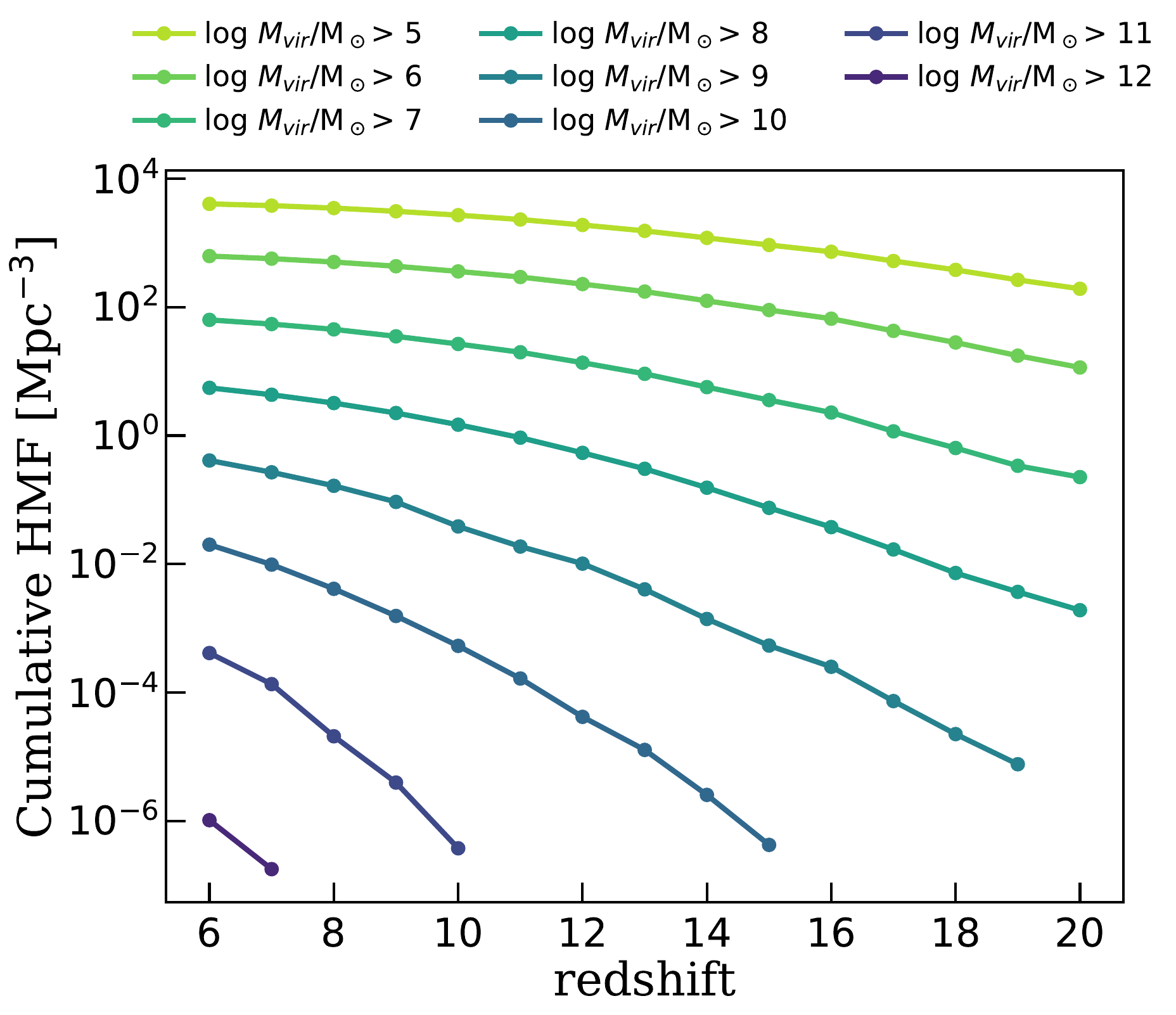}
    \caption{
       Cumulative number density of halos with mass above a given $M_\text{h}$ as a function of redshift, $n(>M_\text{vir},z)$, calculated based on the \gureft-MultiDark combined HMFs as shown in the right panel of Fig.~\ref{fig:hmf}. This figure provides a benchmark for the expected abundances of the dark matter halos that host galaxies and SMBH in the early Universe.
    }
    \label{fig:cumulative_hmf}
\end{figure}

\begin{figure*}
    \includegraphics[width=1.8\columnwidth]{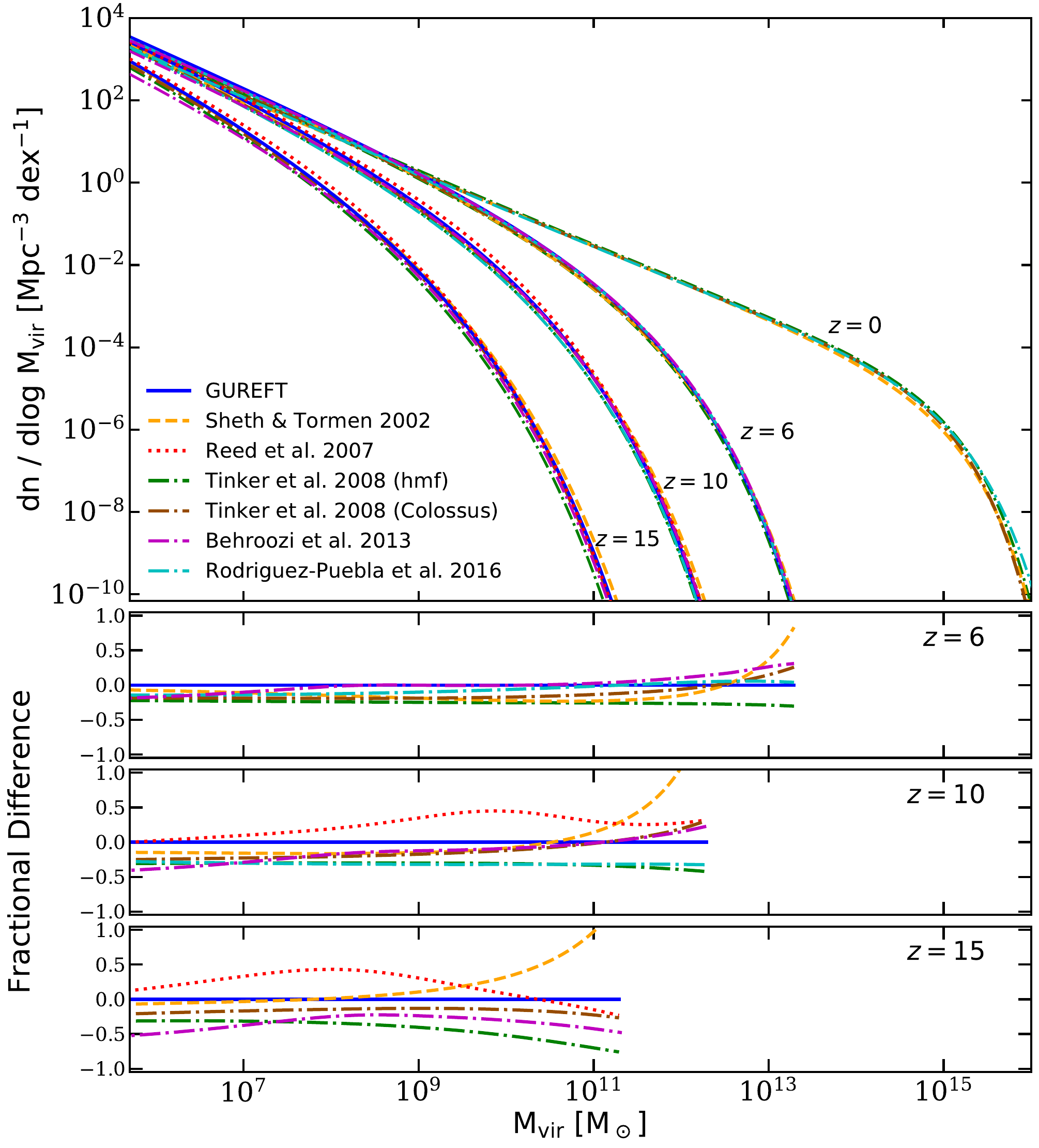}
    \caption{
        This figure compares the results from the \gureft\ simulation suite (blue solid lines) at $z = 6$, 10, and 15 with the HMFs commonly adopted by other high redshift-focused studies. 
        The fits to simulations in the MultiDark suite \citep[BolshoiP, SMDPL, and MDPL;][]{Klypin2016}, which adopted the \citeauthor{Tinker2008} HMF parameterization and cosmological parameters matching the ones adopted by the \gureft\ simulation suite, are shown for $z = 6$ and 10 \citep[][cyan dot-dashed lines]{Rodriguez-Puebla2016}. The \citet[][magenta dot-dashed]{Behroozi2013a} HMFs also adopted the \citeauthor{Tinker2008} parameterization, with corrections applied for 2LPT initial conditions.
        In additional, we also show the \texttt{hmf} and \textsc{Colossus} python packages implementations of the {Tinker2008} HMFs with cosmological parameters updated to match the ones adopted in this work with the green and brown dot-dashed lines, respectively.
        We also show HMFs from the analytic model of \citet[][yellow dashed lines]{Sheth2002}, updated to cosmological parameters and matter power spectrum reported by the \citet[XIII][]{Planck2016}. Results from the high-resolution $N$-body simulations of \citet[][red dotted]{Reed2007}, which adopted earlier WMAP cosmological parameters, are shown at $z = 10$ and 15.
        In the top panel, we show a direct comparison of these HMFs across a wide mass range. In addition, we show a comparison among MultiDark, Tinker, and \citetalias{Sheth2002} at $z = 0$.
        The bottom panels show the fractional difference of the number density of halos from these HMFs relative to the \gureft\ outputs $(\phi_{X}-\phi_\text{\gureft})/\phi_\text{\gureft}$ at $z=6$, 10, and 15. These commonly used analytic models and fitting functions can differ from our results by up to an order of magnitude at high redshift. 
    }
    \label{fig:hmf_compare}
\end{figure*}

\section{Results}
\label{sec:results}
In this section, we present results from the suite of \gureft\ simulations. In particular, we highlight the unique results that are enabled by these new suite of simulations, and make comparisons and connections with previous work. In section \ref{sec:hmf}--\ref{sec:properties}, we present the halo mass functions, halo mass vs. $V_{\rm max}$ relations, concentration vs. mass relations, and the distributions of halo spin parameters at a series of output times. In \ref{sec:mahs}, we present the halo mass assembly histories.

\subsection{Halo mass functions and number densities}
\label{sec:hmf}
One-point distribution functions of halo virial mass, also known as halo mass functions (HMFs), are an important, quick diagnostic tool quantifying halo populations. 
In Fig.~\ref{fig:hmf}, we show the HMFs from the full suite of \gureft\ simulations between $z = 6$ to 20. In the left panel, we show individual HMFs from the four \gureft\ boxes. We show the full mass range of detected halos in these boxes, including where the measurements become incomplete due to the mass resolution limit, manifested by the turnover in the HMFs. In addition, we also show HMFs from the BolshoiP simulations.

In the right panel of Fig.~\ref{fig:hmf}, we show HMFs that are constructed by combining results across the \gureft\ suite.
Here, halos with $\log(M_\text{h}/\text{M}_\odot) \leq 7.5$ are sourced from \gureft-05, $7.5 < \log(M_\text{h}/\text{M}_\odot) \leq 8.5$ from \gureft-15,  $8.5 < \log(M_\text{h}/\text{M}_\odot) \leq 9.5$ from \gureft-35, and $\log(M_\text{h}/\text{M}_\odot) > 9.5$ are sourced from \gureft-90.
Our goal is to source halos from each box in a mass range that are both well resolved and are statistically robust, except for the low-mass end of \gureft-05, for which we show the turnover to mark where the simulations are limited by resolution, and the massive end of \gureft-90, where the most massive bins may contain only a handful of halos. 
Fig.~\ref{fig:cumulative_hmf} shows the cumulative number density of halos above a given mass as a function of redshift, $n(>M_\text{h},z)$, calculated based on the combined HMFs (see right panel in Fig.~\ref{fig:hmf}).

In Fig.~\ref{fig:hmf_compare}, we compare the \gureft\ HMFs at $z = 6$, 10, and 15 to a number of fitting functions that are widely used in high- to ultrahigh-redshift focused studies.
We show HMFs from the analytic model of \citet{Sheth2002}, which is based on the excursion set approach \citep[e.g.][]{Bond1991, Sheth1998} that allows ellipsoidal collapse. 
In this comparison, we incorporated the cosmological parameters and matter power spectrum reported by the \citet[XIII][]{Planck2016}, matching the ones adopted by \gureft. 
\citeauthor{Tinker2008} provided a HMF parameterization and HMFs that were fitted to a wide range of cosmological simulations adopting cosmological parameters reported by early phase of the WMAP mission.
\citet{Rodriguez-Puebla2016} adopted the \citeauthor{Tinker2008} parameterization to fit to simulations in the MultiDark suite \citep[BolshoiP, SMDPL, and MDPL;][]{Klypin2016} up to $z\sim 10$, which adopts cosmological parameters matching the ones adopted by the \gureft\ simulation suite.  
The \gureft\ results match BolshoiP by design given the matching cosmological parameters and the same halo finding methods and definitions. We note that in the \citet{Rodriguez-Puebla2016} parameterization, the parameter $b$ becomes negative and the $f(\sigma)$ function becomes undefined at $z>12.63$, and therefore is not valid for redshifts above this value.

In addition, we also show HMFs from \citet{Behroozi2013a} that adopted the same parameterization and had corrections applied for 2LPT initial conditions and the implementation of \citet{Tinker2008} HMFs from the \texttt{hmf}\footnote{\url{https://hmf.readthedocs.io/}} package, with cosmological parameters updated to match the ones adopted by this work using a package-provided function (shown as green dot-dashed lines). We note that the \citeauthor{Tinker2008} HMFs shown in this comparison are specific to the implementation adopted by the \texttt{hmf} package. Other implementations, such as \citet{Klypin2016}, have been shown to yield better agreement with $N$-body simulations that adopt Planck-compatible cosmological parameters up to $z \sim 5.5$ with updated fitting parameters for $f(\sigma)$.

Results from the high-resolution $N$-body simulations of \citet{Reed2007}, which were carried out over $z\sim 30$ to 10 with earlier WMAP3 cosmological parameters are shown at $z = 10$ and 15 for comparison.

The comparison in Fig.~\ref{fig:hmf_compare} shows that all modelled and fitted HMFs are in superb agreement with each other at $z = 0$ and are in fairly good agreement with \gureft\ and each other at $z\sim6$ (e.g. within $\pm 50\%$ from \gureft). 
However, these HMFs can evolve very differently towards the higher redshifts as they are extrapolated to regimes that were previously unconstrained. Thus, the discrepancy among them continues to widen towards ultra-high redshifts, and the differences can be up to a factor of two in both directions relative to the \gureft\ results.
We also add that despite the updated cosmological parameters in the \citet{Tinker2008} lines in the \texttt{hmf} and \textsc{Colossus} implementations, the fitting function $f(\sigma)$ is not universal to changes in cosmology and would need to be refitted \citep[e.g.][and this work]{Behroozi2013a, Rodriguez-Puebla2016}.

We further note that a number of recent studies that attempt to interpret observed galaxy populations at ultra-high redshift \citep[e.g.][]{Boylan-Kolchin2022, Mason2022, Ferrara2022, Dekel2023, Harikane2023, Munoz2023, Padmanabhan2023} adopt these fits in their work. Furthermore, these results have important implications for the use of abundance matching approaches to link observed galaxy populations to dark matter halos. 
A fitting function and best-fit parameters for the \gureft\ HMFs are provided in Appendix \ref{sec:AppA}.

\begin{figure}
    \includegraphics[width=\columnwidth]{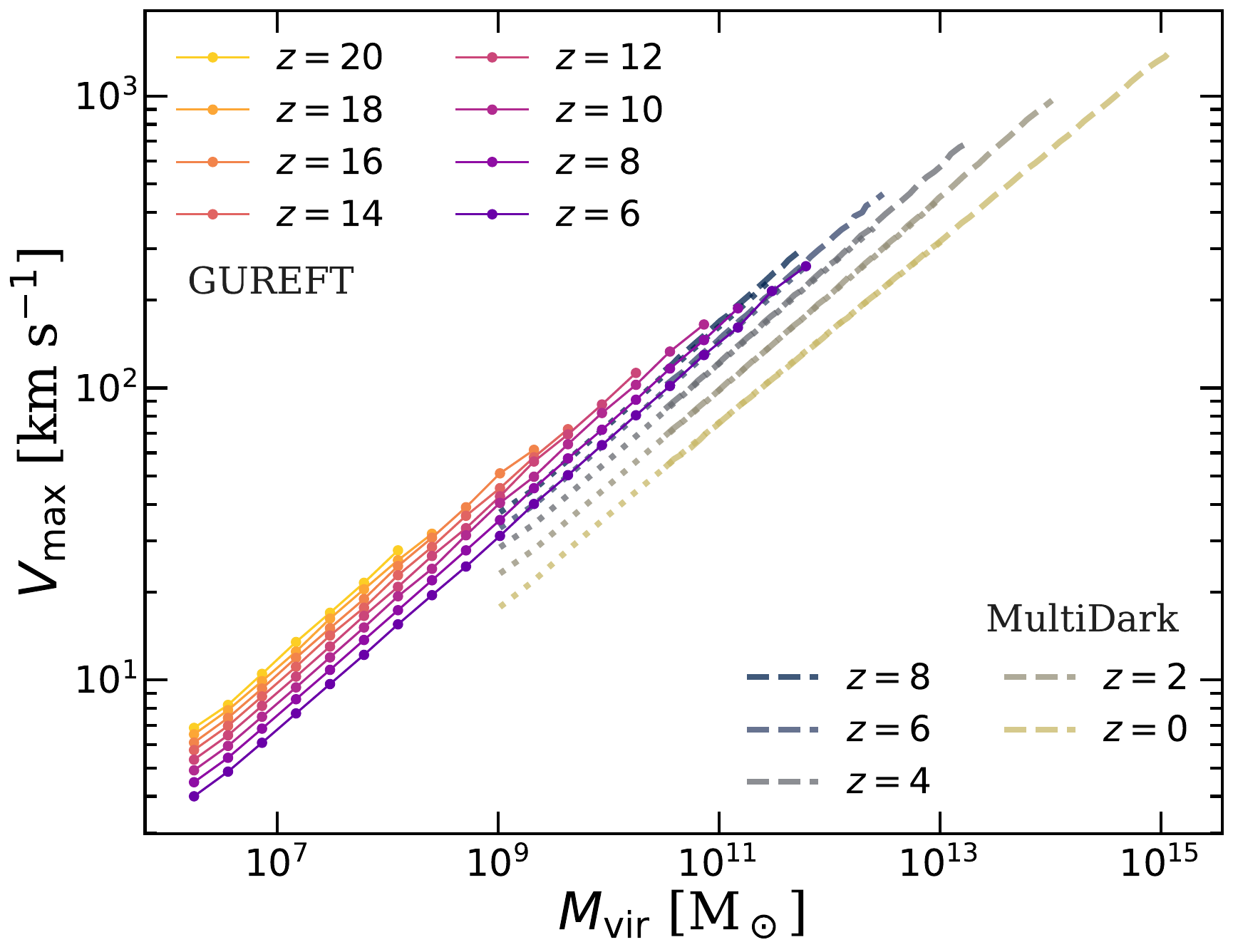}
    \caption{
        The median of maximum halo circular velocity, $V_\text{max}$, as a function of $M_\text{vir}$ for halos in the \gureft\ simulation suite between $z = 6$ to 20 (dotted lines). These results are compared to the SMDPL (dashed lines) and VSMDPL (dotted lines, \citetalias{Rodriguez-Puebla2016}) simulations from the MultiDark suite from $z = 0$ to 6. \gureft\ shows a continuation of the trend of increasing $V_\text{max}$ at fixed halo mass towards earlier cosmic times that was seen at lower redshift in previous simulations. 
    }
    \label{fig:gureft_Vmax_Mvir}
\end{figure}

\begin{figure}
    \includegraphics[width=\columnwidth]{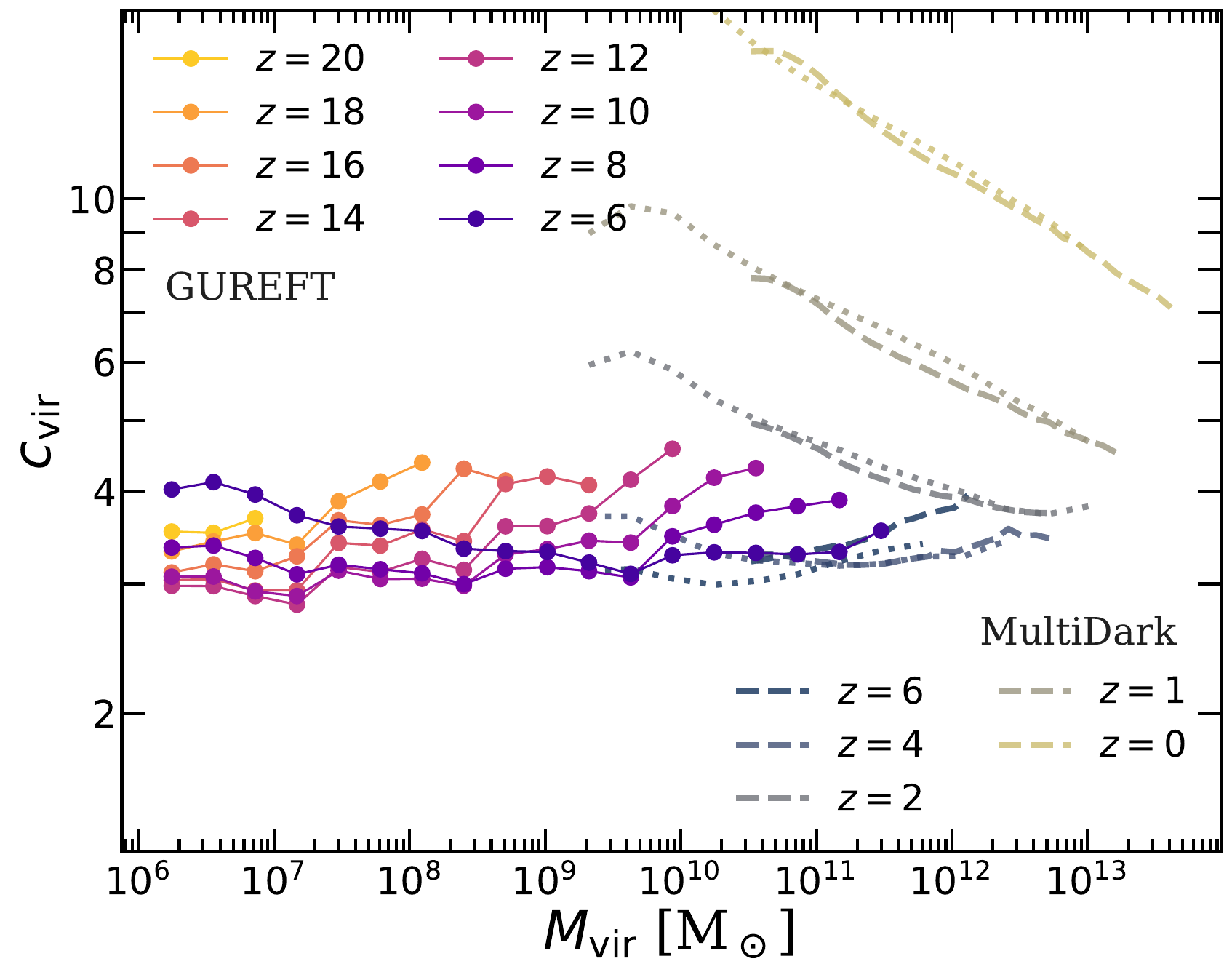}
    \caption{
        The median of halo concentration as a function of $M_\text{vir}$ between $z = 6$ to 20 from the \gureft\ suite, compared to results from the SMDPL (dashed lines, \citetalias{Rodriguez-Puebla2016}) and VSMDPL (dotted lines) simulations from the MultiDark suite from $z = 0$ to 6.
        Previous simulations showed that the concentration-mass relation decreased in normalization and flattened with increasing redshift to $z\sim 6$. \gureft\ reveals a reversal in this trend at $z \gtrsim 6$, with the concentration at fixed mass increasing slightly towards earlier times. 
    }
    \label{fig:gureft_cvir_Mvir}
\end{figure}

\subsection{Structural properties of ultra-high-redshift halos}
\label{sec:properties}
In this subsection, we show the evolution of scaling relations among key structural properties of halos across high- to ultra-high redshifts. Similar to that of \citetalias{Rodriguez-Puebla2016} or other studies, we adopt a 100-particle threshold for halos in the \gureft\ simulation suite to be considered well-resolved.

$V_\text{max}$ is an important quantity that is often utilised in sub-halo abundance matching (SHAM) and (semi-)empirical techniques \citep{Behroozi2019, Behroozi2020, Zhang2023}, and is also used by some semi-analytic models. It is defined as the rotation velocity of the halo at the maximum value of the rotation curve. In Fig.~\ref{fig:gureft_Vmax_Mvir}, we show $V_\text{max}$ as a function of $M_\text{vir}$ between $z = 6$ and 20. The \gureft\ results are combined in a similar manner to the approach used to construct the composite HMFs as demonstrated in the right panel of Fig.~\ref{fig:hmf}.
These results are compared to the ones from the SMDPL simulation between $z = 0$ to 8 as shown in \citetalias{Rodriguez-Puebla2016} and from the VSMDPL simulation accessed through the MultiDark Database \citep{Riebe2013}. 
We show that \gureft\ and MultiDark are in superb agreement in mass ranges and redshifts where the two overlap.

We follow the same fitting functions as presented in \citetalias{Rodriguez-Puebla2016}, where $E(z)$ is the expansion rate for a flat universe: 
\begin{equation}
E(z) = H/H_0 = \sqrt{\Omega_{\Lambda,0}+\Omega_{m,0}(1+z)^3} \text{.}
\end{equation}
With this parameterization, the best-fit $V_\text{max}$--$M_\text{vir}$ relation based on \gureft\ for halos between $6 \lesssim z \lesssim 14$ is expressed as:
\begin{equation}
\label{eqn:Vmax_Mvir}
\begin{split}
    V_\text{max}(M_\text{vir}, z) &= \beta(z) (M_\text{vir,12} \; E(z) ) ^{\alpha(z)} \text{, with}\\
    \alpha(z)     &= 0.339 - 0.053a + 0.079a^2\\
    \log \beta(z) &= 2.171 - 0.114a - 0.085a^2\text{,}
\end{split}
\end{equation}
where $M_\text{vir,12} \equiv M_\text{vir}/10^{12}$ and $a = 1/(1+z)$. We note that the fitting function presented in \citetalias{Rodriguez-Puebla2016} and other works are for mass units of M$_\odot h^{-1}$. The fitting parameters in equation \ref{eqn:Vmax_Mvir} assumes $M_\text{vir}$ in physical units M$_\odot$, the same as those plotted in Fig.~\ref{fig:gureft_Vmax_Mvir}.

Concentration and spin are two key physical properties for characterising the internal structure of dark matter halos \citep[e.g.][]{Bullock2001b, Bullock2001a}. Fig.~\ref{fig:gureft_cvir_Mvir} shows concentration assuming a \citet*[NFW;][]{Navarro1997} halo profile, defined as $c_\text{vir} \equiv R_\text{vir}/R_s$, where $R_\text{vir}$ is the virial radius and $R_\text{s}$ is the scale radius. 
Given the large scatter in this relation and limited dynamic range covered by these boxes, combining results from halos across multiple boxes can be tricky. For instance, the low-mass end of the \textit{`effective mass range'} covered by these boxes is limited by the number of particles required to resolve a halo and reliably measure its properties. On the other hand, the massive end is limited by the number of halos available in these boxes. While well-resolved halos identified with tens of thousands of particles can have properties measured more robustly, these halos are also very rare and become insufficient for robust statistical studies. 
We conducted a controlled experiment with the simulated $M_\text{vir}$--$c_\text{vir}$ from the \gureft\ boxes to inform the set of criteria for combining these boxes and present the detailed findings in Appendix \ref{sec:AppB}.

In order to utilise predictions based on the full suite of \gureft\ simulations, we combined together the scaling relations predicted by these boxes following the following criteria:
\begin{enumerate}
    \item Only well-resolved halos identified with at least 100 DM particles are considered;
    \item When computing scaling relations, only mass bins containing over 100 halos are included
    \item In the mass range where two boxes overlap, the larger box containing more halos is preferred over the smaller box with fewer halos.
\end{enumerate}
Appendix \ref{sec:AppB} shows that the concentration-mass relation measured in the overlapping regions of different boxes agree well when these criteria are applied.

We also note that in the halo profile fitting carried out by \textsc{rockstar}, $R_\text{s}$ is capped at $R_\text{vir}$. Halos with profiles that are not well-fit by an NFW profile (e.g. recent mergers, or halos with multiple density peaks) can yield unphysical measurements of $R_\text{s} \approx R_\text{vir}$. We remove these halos from the sample by requiring $(R_\text{vir} - R_\text{s})/R_\text{vir} > 0.1$. Given that there are only very few halos with naturally occurring $c_\text{vir} < 1.1$, this cutoff removes mostly these halos with spuriously low concentration estimates. 
We add that while this selection criterion cuts off the most unrelaxed halos, some remaining unrelaxed halos may affect the distributions of concentration and spin parameter \citep{Neto2007}.

Fig.~\ref{fig:gureft_cvir_Mvir} shows the $M_\text{vir}$--$c_\text{vir}$ between $z = 6$ to 20 measured from halos across the \gureft\ boxes. 
This is compared with halos from the SMDPL simulations as shown by \citetalias{Rodriguez-Puebla2016} and calculated from VSMDPL.
Once again, we find that \gureft\ and MultiDark are in superb agreement where they overlap. The \gureft\ predictions also agree with the evolution described in \citet{Diemer2015}. These previous studies showed that the $M_\text{vir}$--$c_\text{vir}$ relation flattens back to $z\sim 4$, and halo concentration decreases back in time at fixed halo mass back to $z\sim 4$. There was already a hint of a reversal in this evolution from $z \sim 4$ to 6, with halos at $z\sim 6$ having slightly higher concentrations than halos of the same mass at $z\sim 4$, but the effect was subtle. Our results more definitively show that halo concentrations at fixed mass continue to \emph{increase} towards earlier cosmic times back to $z\sim 20$. 
We discuss the physical interpretation of this behaviour in Section~\ref{sec:discussion}.

\begin{figure}
    \centering
    \begin{subfigure}[b]{0.45\textwidth}
        \centering
        \includegraphics[width=\textwidth]{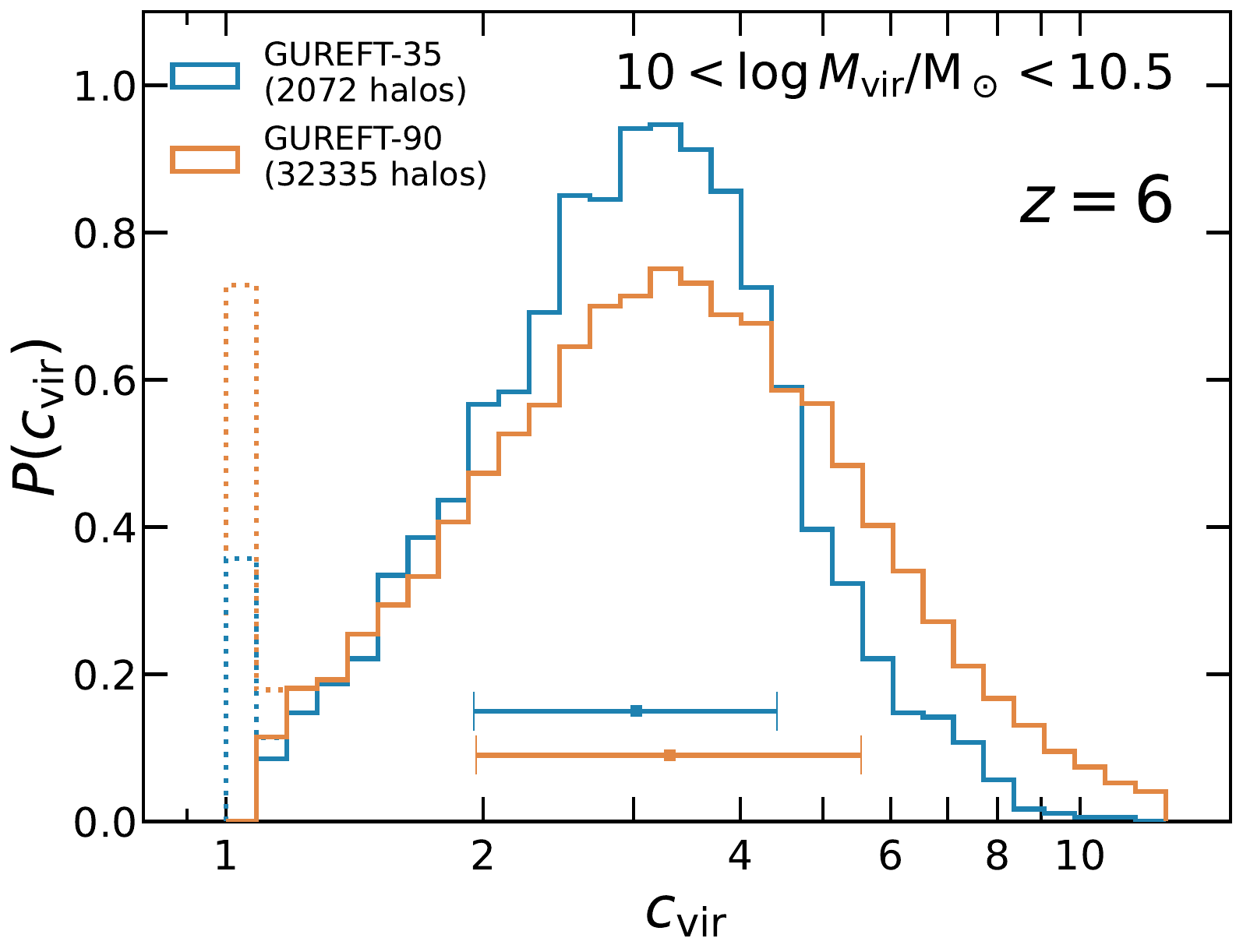}
    \end{subfigure}
    \begin{subfigure}[b]{0.45\textwidth}
        \centering
        \includegraphics[width=\textwidth]{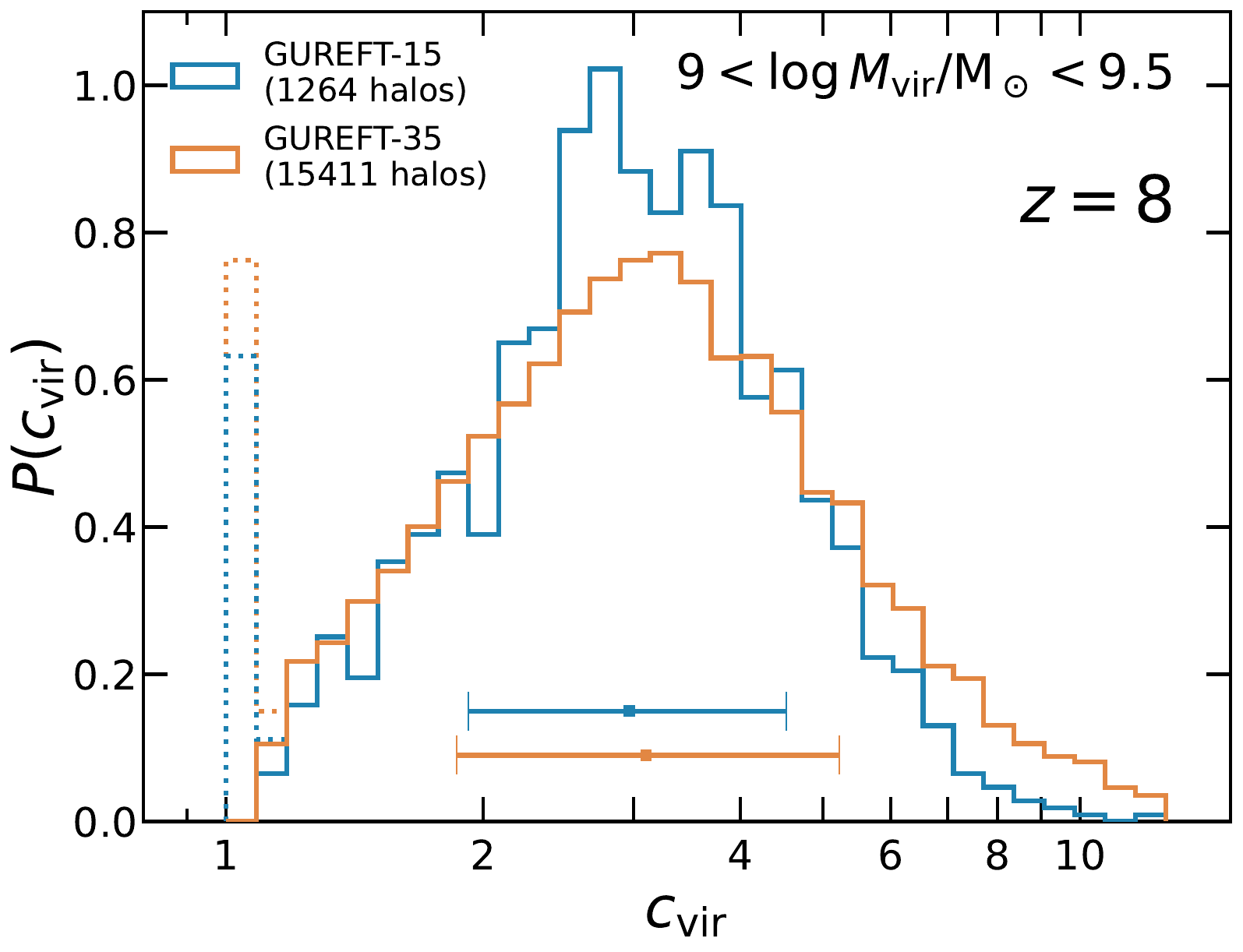}
    \end{subfigure}
    \begin{subfigure}[b]{0.45\textwidth}
        \centering
        \includegraphics[width=\textwidth]{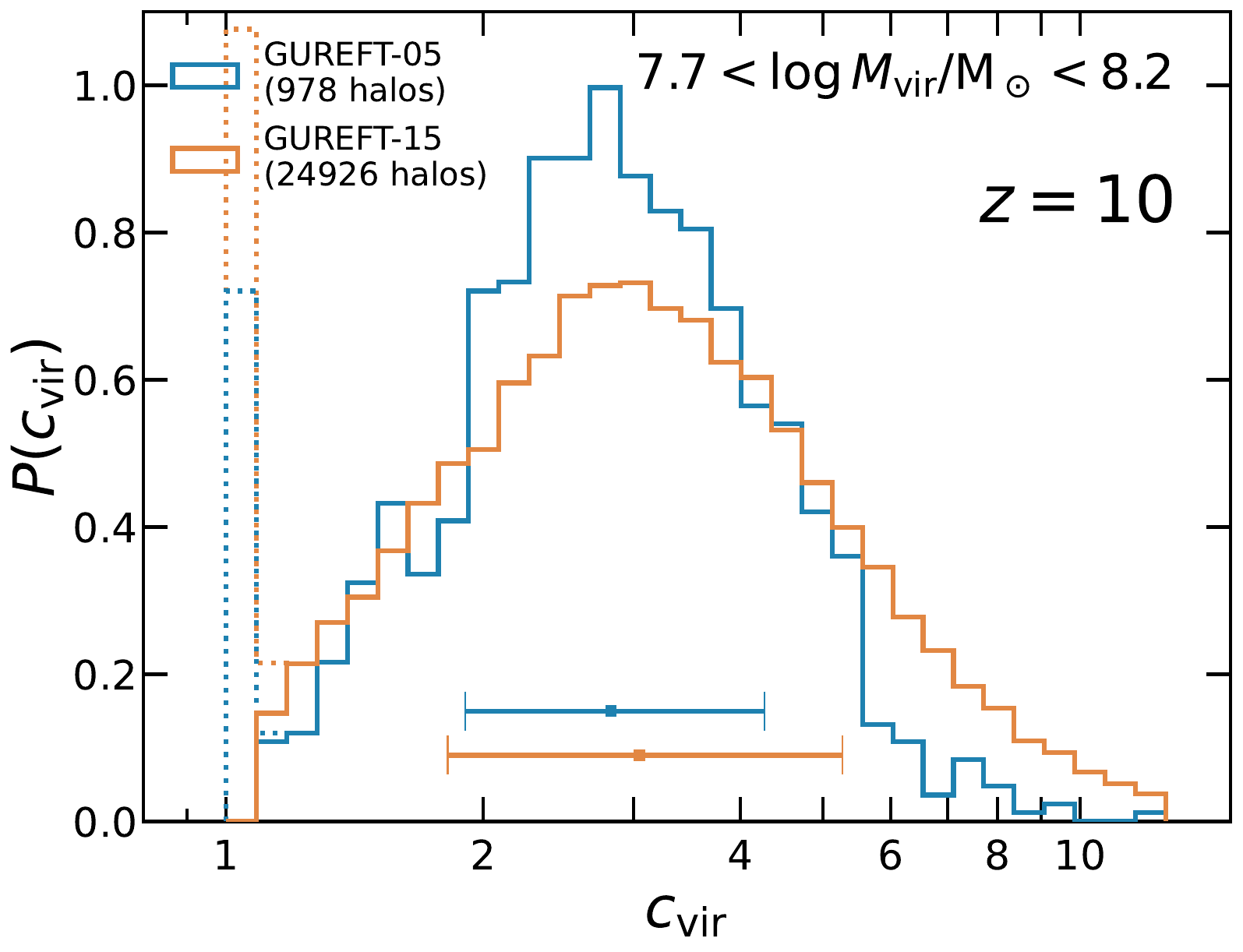}
    \end{subfigure}
    \caption{
        This figure shows the distribution of $c_\text{vir}$ for halos in the mass ranges where two adjacent \gureft\ boxes, including $10 < \log M_\text{h}/\text{M}_\odot < 10.5$ for \gureft-35 and \gureft-90 at $z = 6$, $9 < \log M_\text{h}/\text{M}_\odot < 9.5$ for \gureft-15 and \gureft-35 at $z = 8$, and $7.7 < \log M_\text{h}/\text{M}_\odot < 8.2$ for \gureft-05 and \gureft-15 at $z = 10$. The larger/lower (smaller/higher) resolution box in the pair is shown in orange (blue). The $c_\text{vir}$ bins that do not satisfy the selection criteria $(R_\text{vir} - R_\text{s})/R_\text{vir} > 0.1$ detailed in the text are shown with dotted lines, and are excluded from the rest of our analysis.
        The number of halos available for comparison is indicated in the upper-left corner of each plot. The data points and error bars in matching colour mark the median and 16th and 84th percentiles of the distributions. 
        We find that the mass resolution of a cosmological simulation can have a mild impact on the resultant distribution of $c_\text{vir}$, where higher mass resolution yields lower concentration. This mostly affects halos in the high concentration tail, and does not have a large effect on the median concentration.
    }
    \label{fig:cvir_hist}
\end{figure}

In Fig.~\ref{fig:cvir_hist}, we show the distribution of $c_\text{vir}$ for halos in the mass ranges where two adjacent \gureft\ boxes overlap, including $10 < \log M_\text{h}/\text{M}_\odot < 10.5$ for \gureft-35 and \gureft-90 at $z = 6$, $9 < \log M_\text{h}/\text{M}_\odot < 9.5$ for \gureft-15 and \gureft-35 at $z = 8$, and $7.7 < \log M_\text{h}/\text{M}_\odot < 8.2$ for \gureft-05 and \gureft-15 at $z = 10$. This demonstrates how the distribution of $c_\text{vir}$ can be affected by the mass resolution and the number of halos, which is limited by the simulated volume.
Fig.~\ref{fig:cvir_hist} also presents a resolution study, where we can compare the halo populations in similar mass ranges across pairs of boxes with different mass resolutions. We find that the median value of concentration is quite insensitive to resolution. However, the high concentration end of the distribution is noticeably impacted by resolution, with simulations with higher mass resolution (smaller volume) producing a narrower concentration distribution with a less pronounced tail towards high concentrations.

The angular momentum profile of halos is characterised by the dimensionless spin parameter, originally defined by \citet{Peebles1969} as 
\begin{equation}
    \lambda_\text{P} \equiv \frac{J|E|^{1/2}}{GM^{5/2}} \text{,}
\end{equation} 
where $J$, $E$, and $M$ are the total angular momentum, energy, and mass of the system, respectively; and $G$ is the universal gravitational constant.
\citet{Bullock2001a} presented an alternative definition, which is also widely adopted:
\begin{equation}
    \lambda_{B} \equiv \frac{J}{\sqrt{2}MVR} \text{,}
\end{equation} 
where $J$ is the angular momentum inside a sphere of radius $R$ containing mass $M$; and $V$ is the halo circular velocity at radius $R$.

\begin{figure}
    \centering
    \begin{subfigure}[b]{0.47\textwidth}
        \centering
        \includegraphics[width=\textwidth]{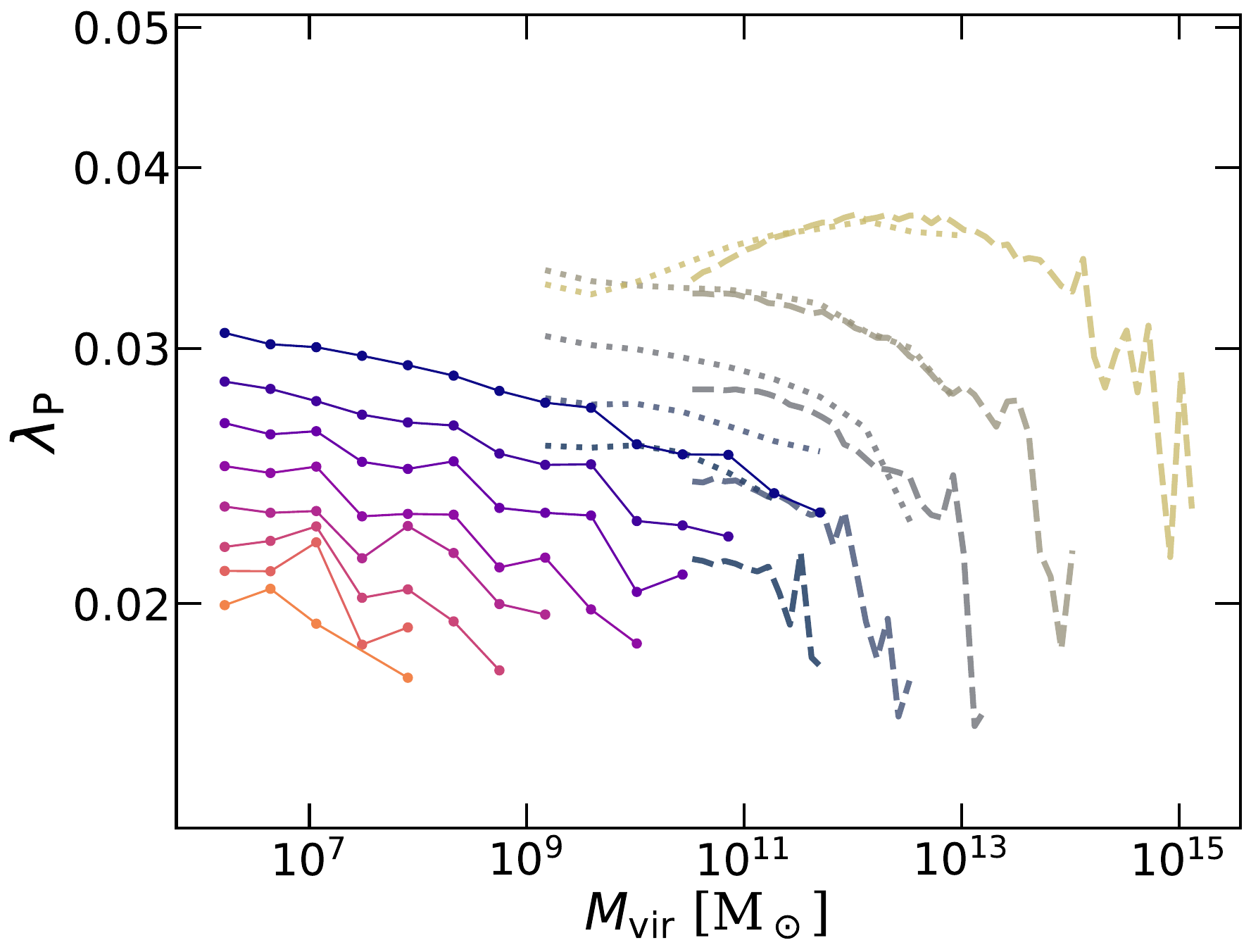}
    \end{subfigure}
    \begin{subfigure}[b]{0.47\textwidth}
        \centering
        \includegraphics[width=\textwidth]{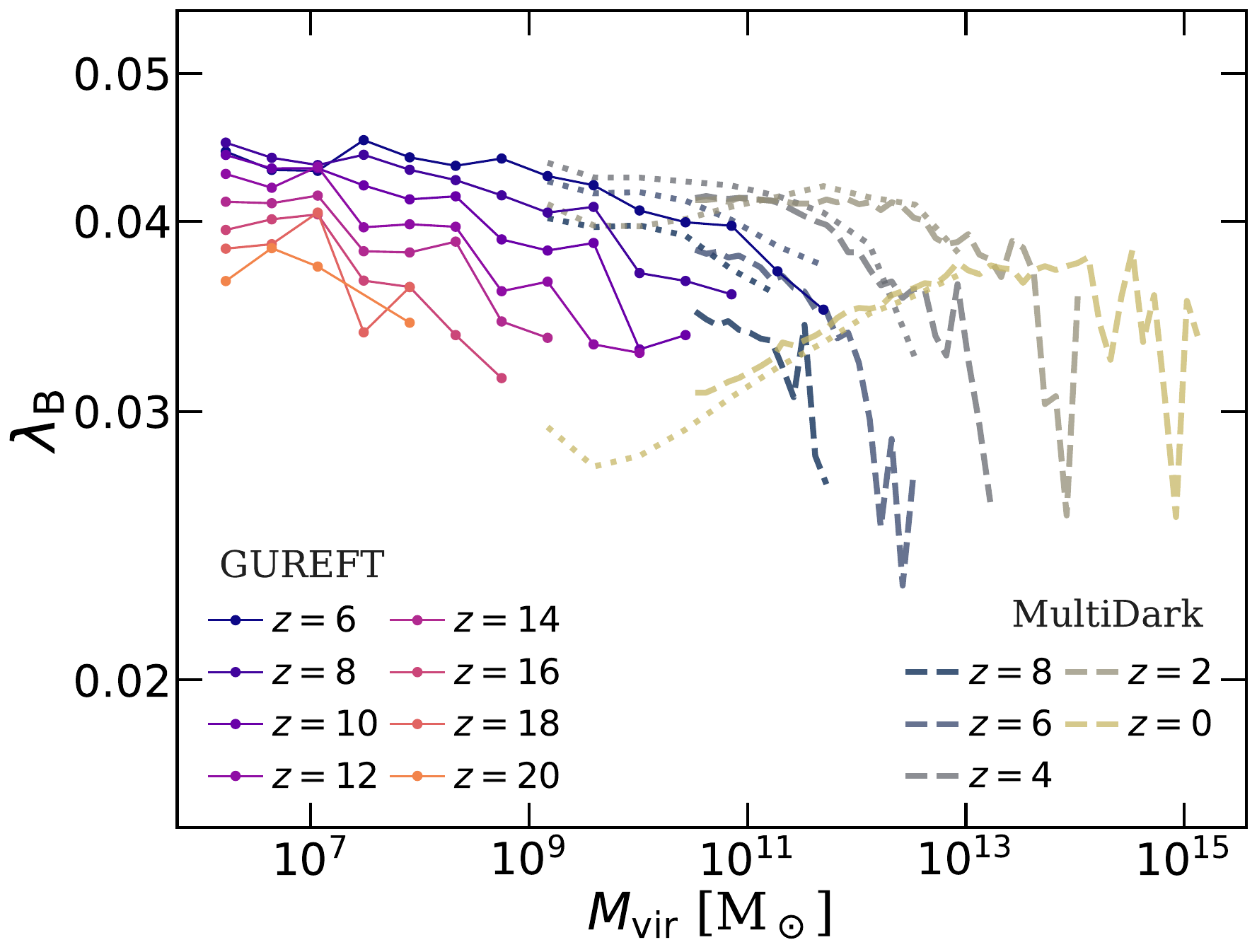}
    \end{subfigure}
    \caption{
        The median of spin parameter $\lambda$ calculated based on the definitions from \citep[][\textit{top panel}]{Peebles1969} and \citep[][\textit{bottom panel}]{Bullock2001a} as a function of $M_\text{vir}$ for halos in the \gureft\ simulation suite. These results are compared to the ones from the SMDPL (dashed lines, \citetalias{Rodriguez-Puebla2016}) and VSMDPL (dotted lines) simulations from the MultiDark suite between $z=0$ to 8. 
        \gureft\ shows a continuation of the trend of decreasing spin across all halo masses towards earlier cosmic times that was seen at lower redshift in previous simulations.
    }
    \label{fig:gureft_Spin_Mvir}
\end{figure}

In Fig.~\ref{fig:gureft_Spin_Mvir}, we show the spin parameter calculated based on both definitions, \citeauthor{Peebles1969} ($\lambda_\text{P}$) and \citeauthor{Bullock2001a} ($\lambda_\text{B}$), as a function of $M_\text{vir}$, between $z = 6$ to 20. 
The overall higher values of $\lambda_\text{B}$ compared with $\lambda_\text{P}$ is consistent with findings from the MultiDark simulations presented by \citetalias{Rodriguez-Puebla2016}.
We also show results at $z = 0$ to 8 based on halos from the SMDPL simulations as shown by \citetalias{Rodriguez-Puebla2016} and from VSMDPL. We add that the noisy behavior on the massive end of the SMDPL results is likely due to small samples of halos in this mass range. We note that the scale we use to plot $\lambda$ has a much smaller range than the one used in \citetalias{Rodriguez-Puebla2016}, which accounts for the apparently stronger mass dependence of $\lambda$. However our results are, again, fully consistent with those of \citetalias{Rodriguez-Puebla2016} where they overlap in mass and redshift.

\begin{figure}
    \centering
    \begin{subfigure}[b]{0.45\textwidth}
        \centering
        \includegraphics[width=\textwidth]{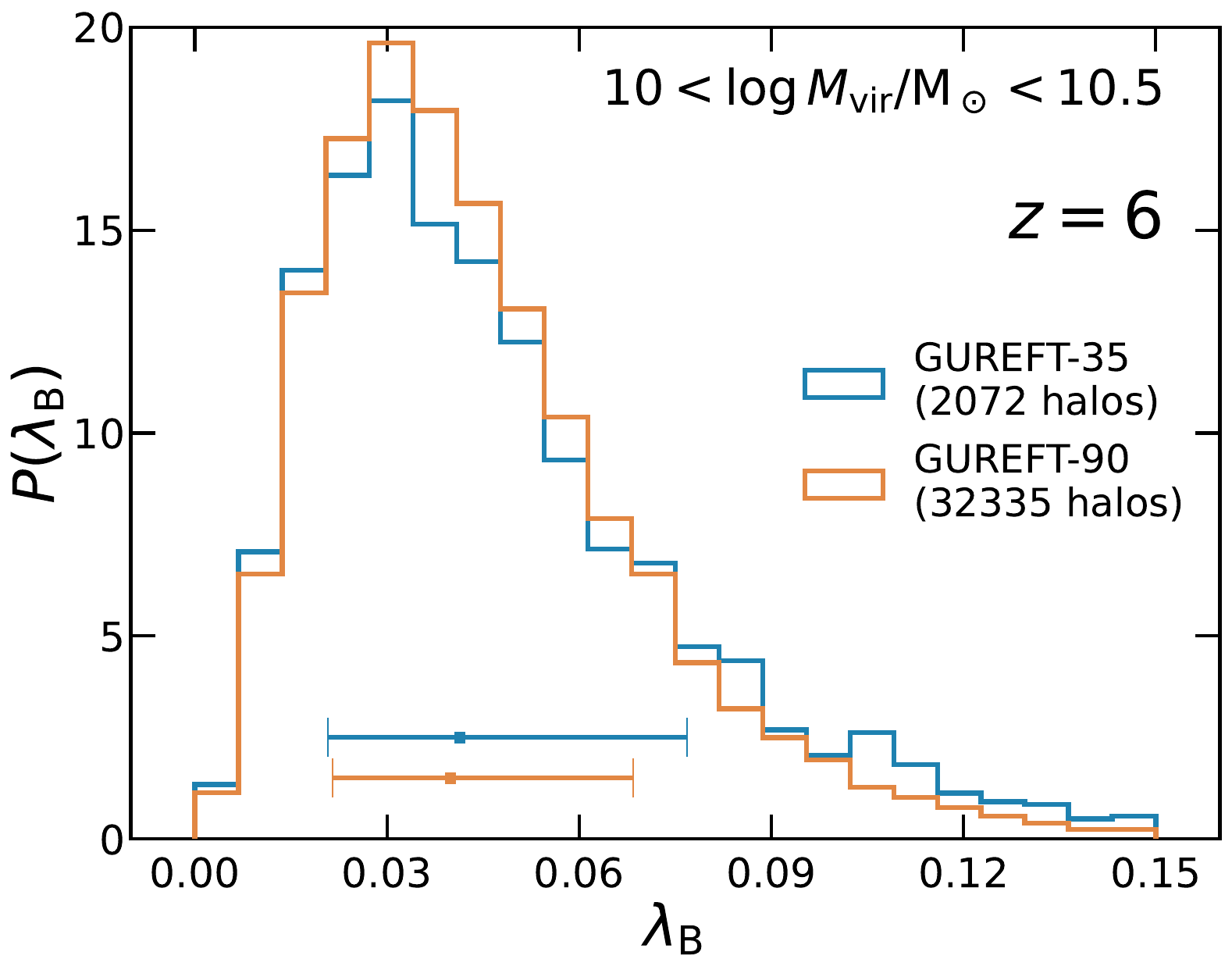}
    \end{subfigure}
    \begin{subfigure}[b]{0.45\textwidth}
        \centering
        \includegraphics[width=\textwidth]{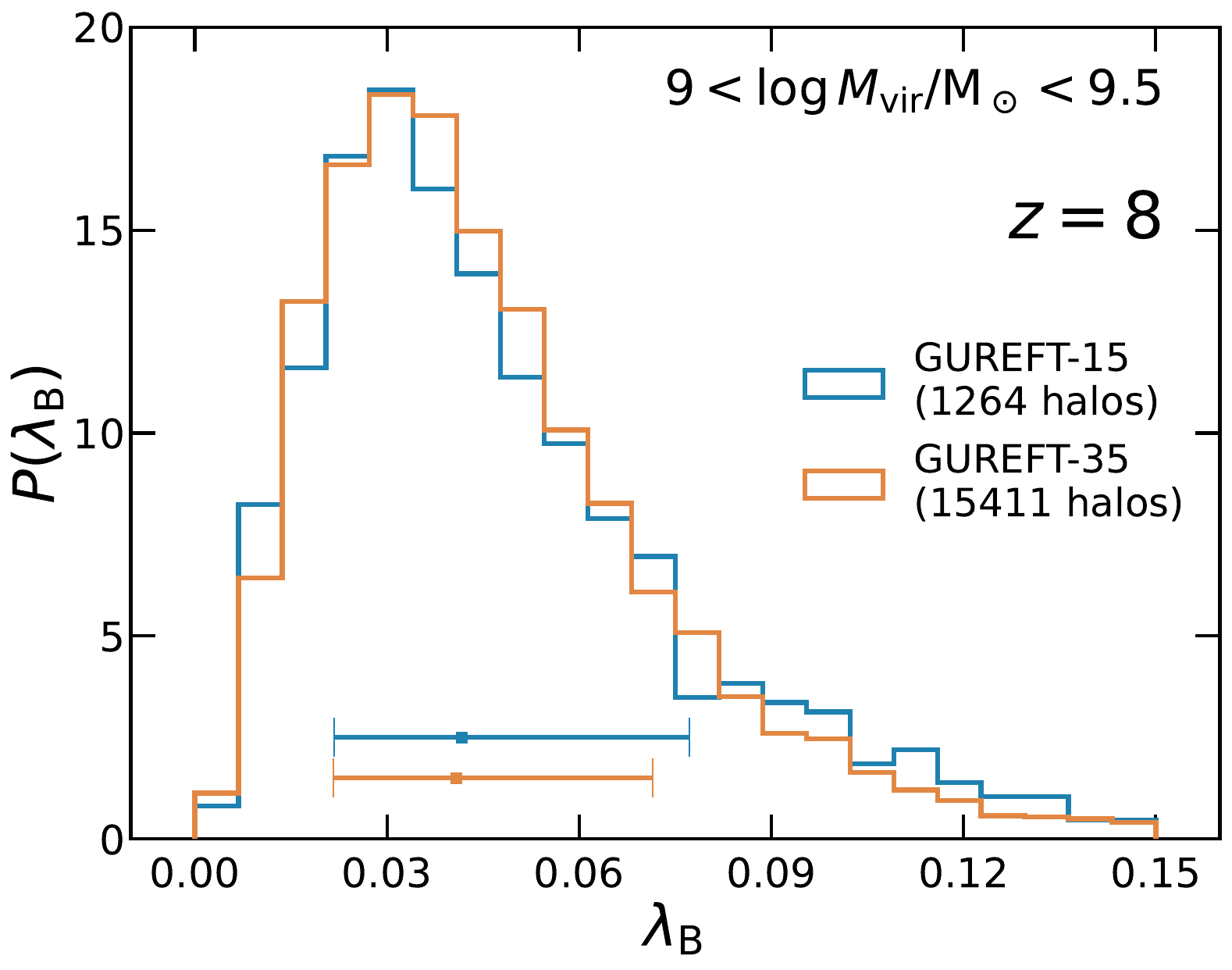}
    \end{subfigure}
    \begin{subfigure}[b]{0.45\textwidth}
        \centering
        \includegraphics[width=\textwidth]{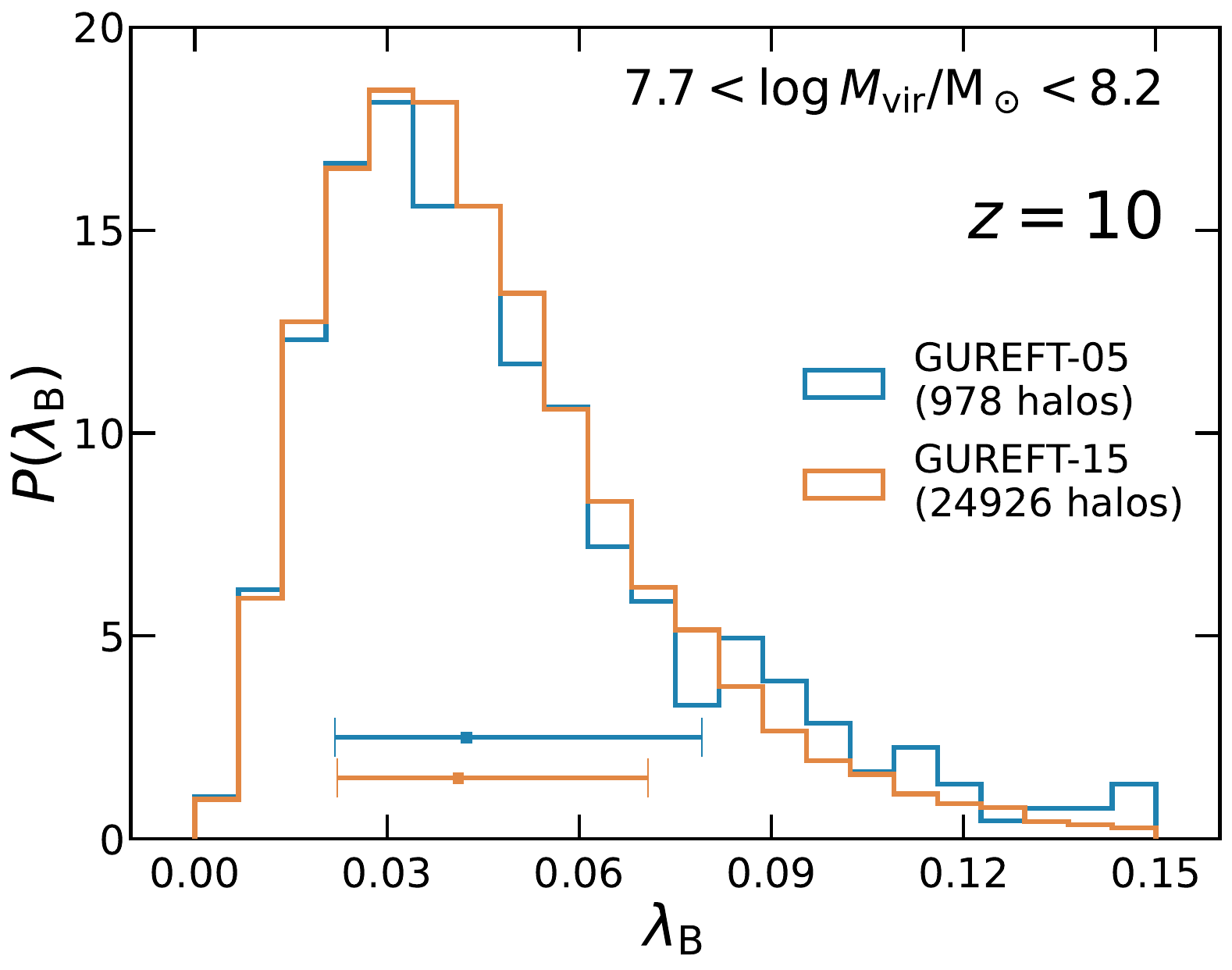}
    \end{subfigure}
    \caption{
        This figure shows the distribution of $\lambda_\text{B}$ for halos in the mass ranges where two adjacent \gureft\ boxes, including $10 < \log M_\text{h}/\text{M}_\odot < 10.5$ for \gureft-35 and \gureft-90 at $z = 6$, $9 < \log M_\text{h}/\text{M}_\odot < 9.5$ for \gureft-15 and \gureft-35 at $z = 8$, and $7.7 < \log M_\text{h}/\text{M}_\odot < 8.2$ for \gureft-05 and \gureft-15 at $z = 10$. The larger/lower resolution (smaller/higher resolution) box in the pair is shown in orange (blue). 
        The number of halos available for comparison is indicated in the centre-right of each plot. The data points and error bars in matching colour mark the median and 16th and 84th percentiles of the distributions. 
        We find that the mass resolution of a cosmological simulation can have a mild impact on the resultant distribution of $\lambda_\text{B}$, where higher mass resolution yields larger $\lambda_\text{B}$. This mostly affects halos in the high-$\lambda$ tail of the distribution, and does not have a large impact on the median spin.
    }
    \label{fig:spin_hist}
\end{figure}

Similar to that for $c_\text{vir}$ above, we present a resolution study for $\lambda_\text{B}$ in Fig.~\ref{fig:spin_hist}. We find that the mass resolution of the simulation has less impact on the distribution of $\lambda_\text{B}$ than was the case for $c_\text{vir}$, with the overall distribution showing only a weak dependence on the mass resolution. We note that the comparisons presented in Figs.~\ref{fig:cvir_hist} and \ref{fig:spin_hist} show the combined effect of differences in box sizes and mass resolution. While some studies have shown that finite box size could have an impact on the statistical properties of simulated dark matter halos \citep[e.g.][]{Power2006}, we restricted the halo samples to mass ranges where the two compared \gureft\ boxes overlap, where halos are still fairly abundant in the larger of the pairs of boxes and are well-resolved in the smaller of the pairs. Thus, the impact on simulated halo properties due to different box sizes should be minor compared to the impact of mass resolution.

\begin{figure*}
    \centering
    \begin{subfigure}[b]{0.47\textwidth}
        \centering
        \includegraphics[width=\textwidth]{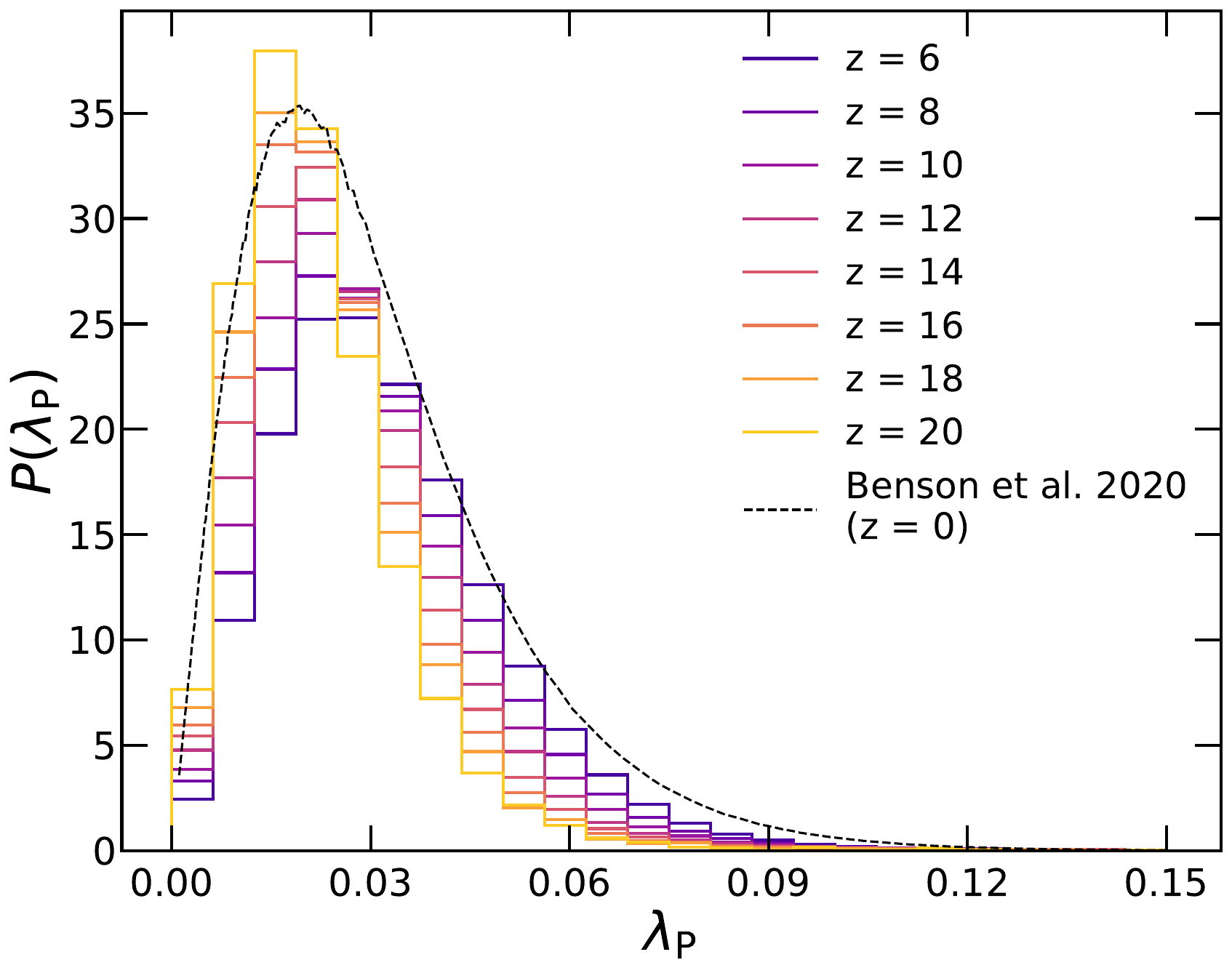}
    \end{subfigure}
    \hspace{0.15in}
    \begin{subfigure}[b]{0.47\textwidth}
        \centering
        \includegraphics[width=\textwidth]{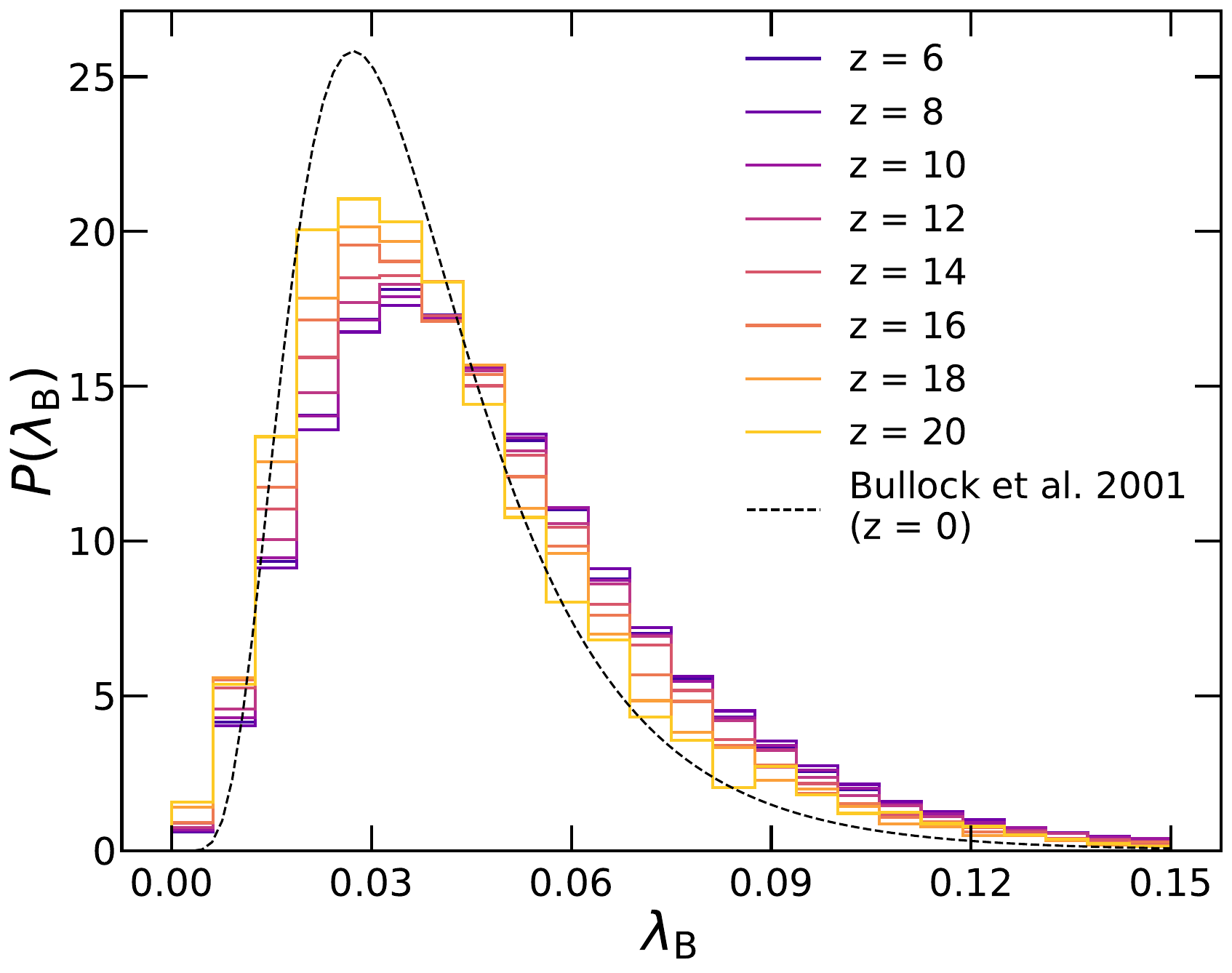}
    \end{subfigure}
    \caption{
        Normalised distributions of the spin parameter $\lambda$, calculated based on the \citet{Peebles1969} definition (\textit{left panel}) and the \citet{Bullock2001a} definition (\textit{right panel}), between $z = 6$ to 20. The best-fit log-normal distribution fitted to $z\sim0$ halos for \citet[][$\lambda_\text{P}$]{Benson2020} and \citet[][$\lambda_\text{B}$]{Bullock2001a} is also shown. 
        \gureft\ reveals a mild evolution in $\lambda_\text{B}$, especially at $z \lesssim 16$, and a stronger evolution in $\lambda_\text{P}$ across the full redshift range explored in this work.
    }
    \label{fig:Spin_distribution}
\end{figure*}

In Fig.~\ref{fig:Spin_distribution}, we show the evolution across redshift of the spin parameter distributions obtained by combining the four \gureft\ boxes, as before, calculated based on both the \citeauthor{Peebles1969} and \citeauthor{Bullock2001a} definitions. For both definitions, we see moderate evolution in this distribution, with a shift towards lower median values of the spin parameter at earlier cosmic times. This is a continuation of the trend in $\lambda$ with redshift seen in \citetalias{Rodriguez-Puebla2016} and \citet{Somerville2018}.
We note that neither a log-normal distribution \citep[see][]{Bullock2001a} nor a double-Schechter function fit (see \citetalias{Rodriguez-Puebla2016}) provides an accurate fit to the distribution of $\lambda$ at $z > 6$. A new fitting function and best-fit parameters are provided in Appendix \ref{sec:AppC}.

\subsection{Halo growth, merger, and assembly histories}
\label{sec:mahs}
In this section, we investigate the predicted growth rates for halos at high- to ultra-high redshifts from the \gureft\ simulations. The growth of dark matter halos from the suite of \gureft\ simulations is further investigated in a companion work \citep{Nguyen2023}.

\begin{figure*}
    \centering
    \begin{subfigure}[b]{0.485\textwidth}
        \centering
        \includegraphics[width=\textwidth]{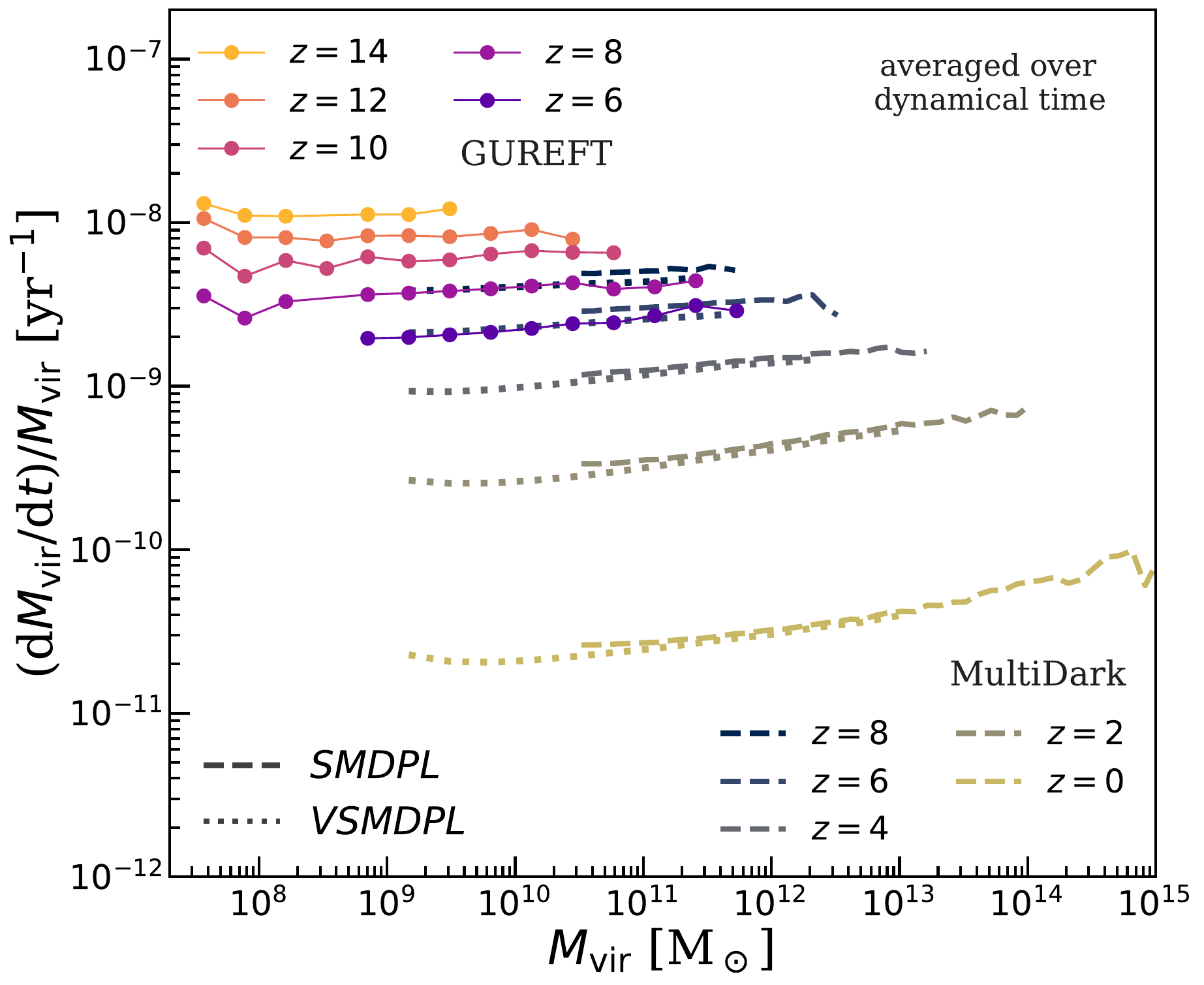}
    \end{subfigure}
    \hspace{0.15in}
    \begin{subfigure}[b]{0.485\textwidth}
        \centering
        \includegraphics[width=\textwidth]{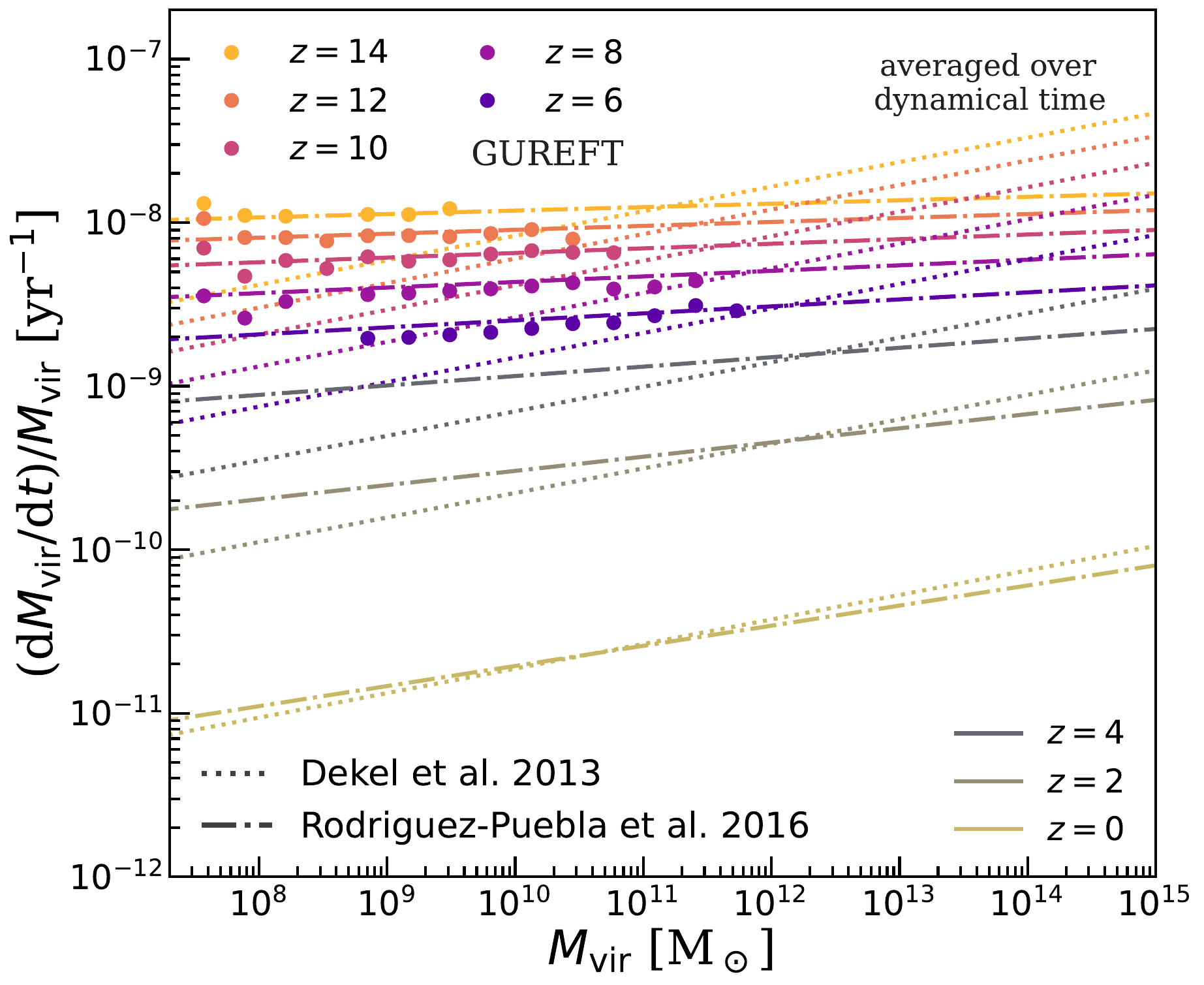}
    \end{subfigure}
    \caption{
        The \textit{left panel} shows halo mass accretion rate normalised to halo mass, ($dM_\text{vir}/dt$)/$M_\text{vir}$, averaged over a dynamical time, as a function of $M_\text{vir}$ between $z = 6$ and $14$, compared to results from the SMDPL (dashed lines, \citetalias{Rodriguez-Puebla2016}) and VSMDPL (dotted lines) simulations from the MultiDark suite between $z = 0$ to 8. In the \textit{right panel}, we show the analytic fitting function from \citet{Dekel2013} and \citet{Rodriguez-Puebla2016} computed for $z = 0$, 2, 4, 6, 8, 10, 12, and 14, shown by lines in colours matching those of the $N$-body simulations (MultiDark for $z<5$ and \gureft\ for $z>5$). 
        \gureft\ reveals a continuation of the trend of increasing $dM_\text{vir}/dt$ at fixed halo mass towards earlier cosmic times that was seen at lower redshift in previous simulations. The comparison with fitting functions from the literature shows that fitting functions based on lower redshifts and larger halo masses are not necessarily accurate in this regime.
    }
    \label{fig:gureft_dMvirdt_Mvir}
\end{figure*}

Fig.~\ref{fig:gureft_dMvirdt_Mvir} shows the specific halo mass accretion rate averaged over a dynamical time as a function of $M_\text{vir}$. 
These results combine halos from across the \gureft\ boxes following the same procedures for producing the $M_\text{vir}$--$V_\text{max}$ relation presented in Fig.~\ref{fig:gureft_Vmax_Mvir}. 
To put these high-redshift results in the context of the evolution across cosmic time, we also show the evolution between $z = 0$ to 8 from SMDPL (\citetalias{Rodriguez-Puebla2016}) and VSMDPL. 
We also show the analytic fitting function from \citet{Dekel2013} for comparison, and find that the analytic function that reproduces the lower redshift results from the MultiDark simulations gradually starts to deviate from the simulation results towards higher redshift and lower halo masses. We find that the slope of the $M_\text{vir}$--$dM_\text{vir}/dt$ relation from \citet{Dekel2013} is slightly steeper than the results from the \gureft\ simulations. We also show the fitting function from \citet{Rodriguez-Puebla2016} that was fitted to the MultiDark simulations from $z=0$ to 8. Based on the same parameterization and best-fit parameters, we also show extrapolations up to $z=14$. The comparison of these fitting functions show that while they may agree well at low redshift, they can behave very differently when extrapolated to ultra-high redshift.
We provide an updated fitting function below.

With a parameterization similar to that of the $V_\text{max}$--$M_\text{vir}$ relation presented in Equation \ref{eqn:Vmax_Mvir}, the best-fit equation for the $dM_\text{vir}/dt$--$M_\text{vir}$ relation for halos between $6 \lesssim z \lesssim 14$ is
\begin{equation}
\label{eqn:dMdt_Mvir}
\begin{split}
    dM_\text{vir}/dt(M_\text{vir}, z) &= \beta(z) (M_\text{vir,12} \; E(z) ) ^{\alpha(z)} \text{, with}\\
    \alpha(z)     &= 0.858 + 1.554a - 1.176a^2\\
    \log \beta(z) &= 2.578 - 0.989a - 1.545a^2\text{,}
\end{split}
\end{equation}
where $M_\text{vir,12} \equiv M_\text{vir}/(10^{12}$ \Msun) and $a = 1/(1+z)$. Once again the fitting function assumes $M_\text{vir}$ is in physical units, M$_\odot$, instead of conventional simulation mass units, M$_\odot h^{-1}$. The fitting function is based only on the four \gureft\ boxes and should only be used in the corresponding halo mass range.

\begin{figure*}
    \includegraphics[width=2\columnwidth]{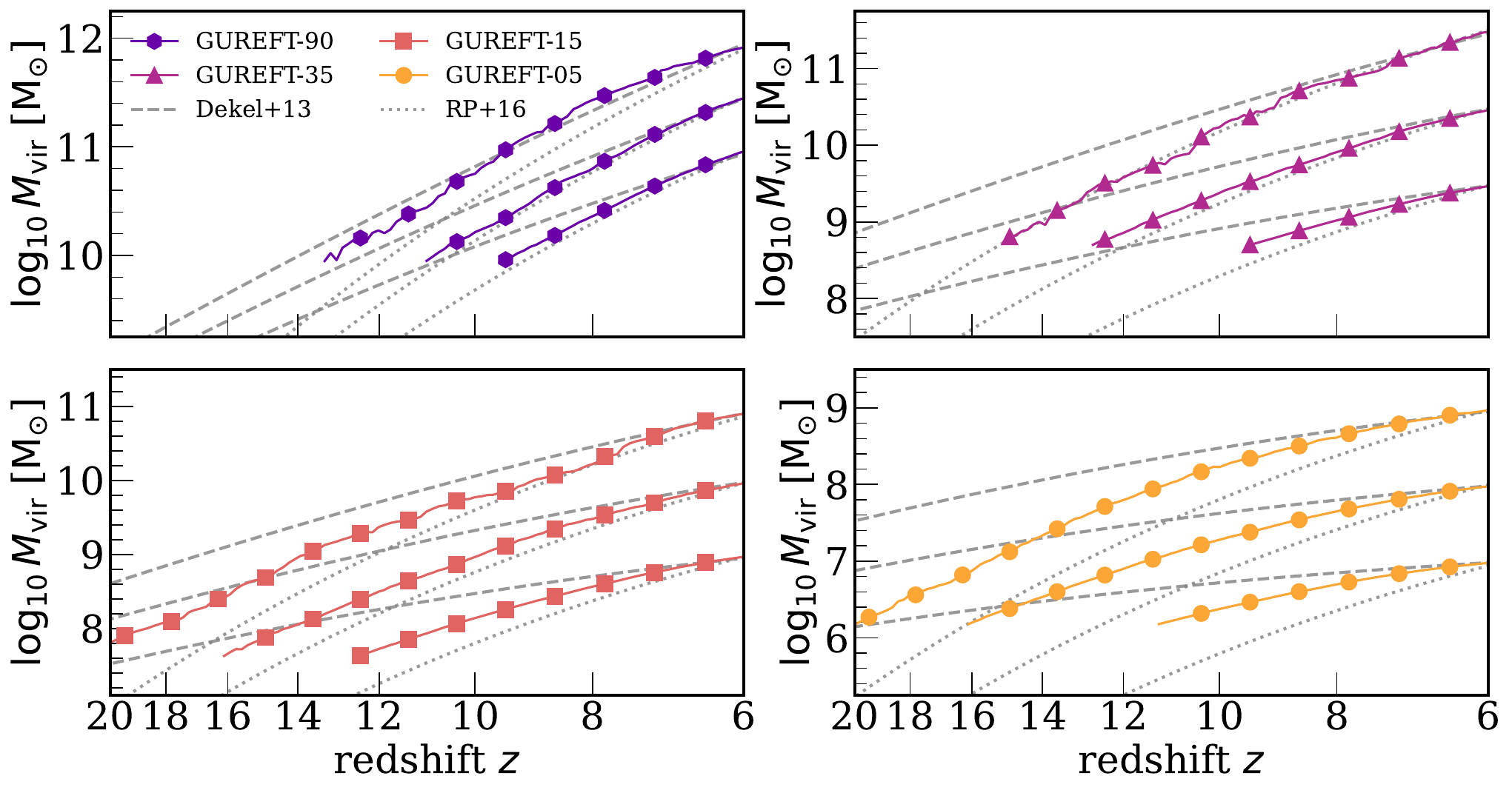}
    \caption{
        The median of the largest progenitor halo mass traced back in time for halos arriving at a final $M_\text{vir}$ at $z = 6$ for halos from the \gureft\ simulations. These results are compared to the fitting function presented by \citet[][grey dashed lines]{Dekel2013} and \citet[][grey dotted lines]{Rodriguez-Puebla2016}.
        \gureft\ reveals the mass accretion histories of halos in mass and redshift ranges that are first probed by our simulations. Once again, fitting functions from the literature do not necessarily perform well when extrapolated to mass and redshift ranges beyond those on which they were calibrated. The extrapolation of the \citet{Dekel2013} fitting function in general predicts earlier halo formation than our results indicate, while the \citet{Rodriguez-Puebla2016} fits perform fairly well for the higher mass halos in our GUREFT-35 and GUREFT-90 boxes, but predict later formation of the lower mass halos in GUREFT-15 and GUREFT-5.
    }
    \label{fig:Mvir_z}
\end{figure*}

Fig.~\ref{fig:Mvir_z} shows the median of the halo mass growth history for halos arriving at a final $M_\text{vir}$ at $z = 6$. This is calculated by tracking the largest progenitor halo back in time using the \textsc{consistent trees} merger trees.
For comparison, we show the analytic model from \citet{Dekel2013}, where the growth of halo mass is given by $M = M_0\;e^{-\alpha(z-z_0)}$, where $\alpha = (3/2) s\;t_1 \simeq 0.79$, $s = 0.030$  Gyr$^{-1}$ is the inverse of the accretion time-scale and $t_1$ is the Hubble time. 
Similar to the previous figure, we find that halos at $z \gtrsim 6$ in \gureft\ have higher growth rates than the \citeauthor{Dekel2013} fitting functions imply.
We attempted to fit the halo growth histories from \gureft\ with the same functional as \citet{Dekel2013} and found that our results can't be well fit by simple exponential growth.

\section{Discussion}
\label{sec:discussion}

In this section, we discuss the results from the new suite of \gureft\ simulations and the wide range of astrophysical modelling that they enable.
We also discuss the implications of these new results in the context of the new era of ultra-high redshift exploration with \textit{JWST} and the upcoming \textit{Roman Space Telescope}. We also discuss the caveats and uncertainties in the results presented in this work.

\subsection{Implications for the interpretation of high redshift populations}

In the new era of ultra-high-redshift exploration enabled by \textit{JWST} \citep{Gardner2006, Gardner2023}, it is necessary to lay the ground work to support modelling and interpreting galaxies and their embedded supermassive black holes in the ultra high-$z$ universe. Shortly after the first images from \textit{JWST} were released, \citet{Labbe2022} presented an analysis that claimed the discovery of several galaxies at $z\sim 8$--10 with stellar mass estimates of $\gtrsim 10^{10} M_{\rm \odot}$, with a few galaxies having estimated stellar masses of $\sim 10^{11} M_{\rm \odot}$. The question that immediately arose was whether the existence of such massive galaxies so early in the Universe would violate fundamental constraints from $\Lambda$CDM. For example, \citet{Boylan-Kolchin2022} claimed that these galaxies were several orders of magnitude more abundant than would be expected within a standard $\Lambda$CDM paradigm. Based on an Extreme Value Statistics analysis, \citet{Lovell2022} found that these objects were in tension with $\Lambda$CDM at the $3 \sigma$ level. However, \citet{Boylan-Kolchin2022} used the predictions of the \citet{Sheth1999} model, which are similar to the \citet{Sheth2002} model. \citet{Lovell2022} use an extrapolated version of the \citet{Tinker2008} halo mass function fits, which were calibrated at much lower redshifts. As shown in Fig.~\ref{fig:hmf_compare}, these analytic models and extrapolated fits can differ from our measured halo mass functions by up to an order of magnitude. Although the stellar mass estimates of the objects analyzed by \citet{Labbe2022} have now been significantly revised downwards, such that there is no longer any tension with $\Lambda$CDM (in fact many of these objects are now thought to be obscured AGN; see \citealt{Barro2023}), the point remains that accurate HMF estimates should be used for interpreting future observations. The fits that we provide in this work are intended to facilitate this.  Moreover, the ultra high-$z$ matter assembly history captured by the suite of \gureft\ simulations provide results critical to the development of physical models that can explain the physics that drives the early formation of these galaxies and black holes \citep[e.g.][]{Yung2023a}.

A new generation of exascale cosmological simulations containing trillions of particles enabled by the fast-growing high performance computing capabilities, such as the new suite of Uchuu simulations \citep{Ishiyama2021}, has the capability of simulating multi-hundred-Mpc cosmological volumes with unprecedented mass resolution. However, it can cost upwards of tens of millions of CPU-hours and generate multiple petabytes of data, which will cost a significant additional amount of computing resources to analyse.
In addition, many existing halo finders and merger tree construction codes are not optimised to handle data of such volume. Therefore, solving this problem by brute force is not practical.
The tiered box approach explored in this work provides a computationally efficient and elegant alternative.

\begin{figure}
    \includegraphics[width=1\columnwidth]{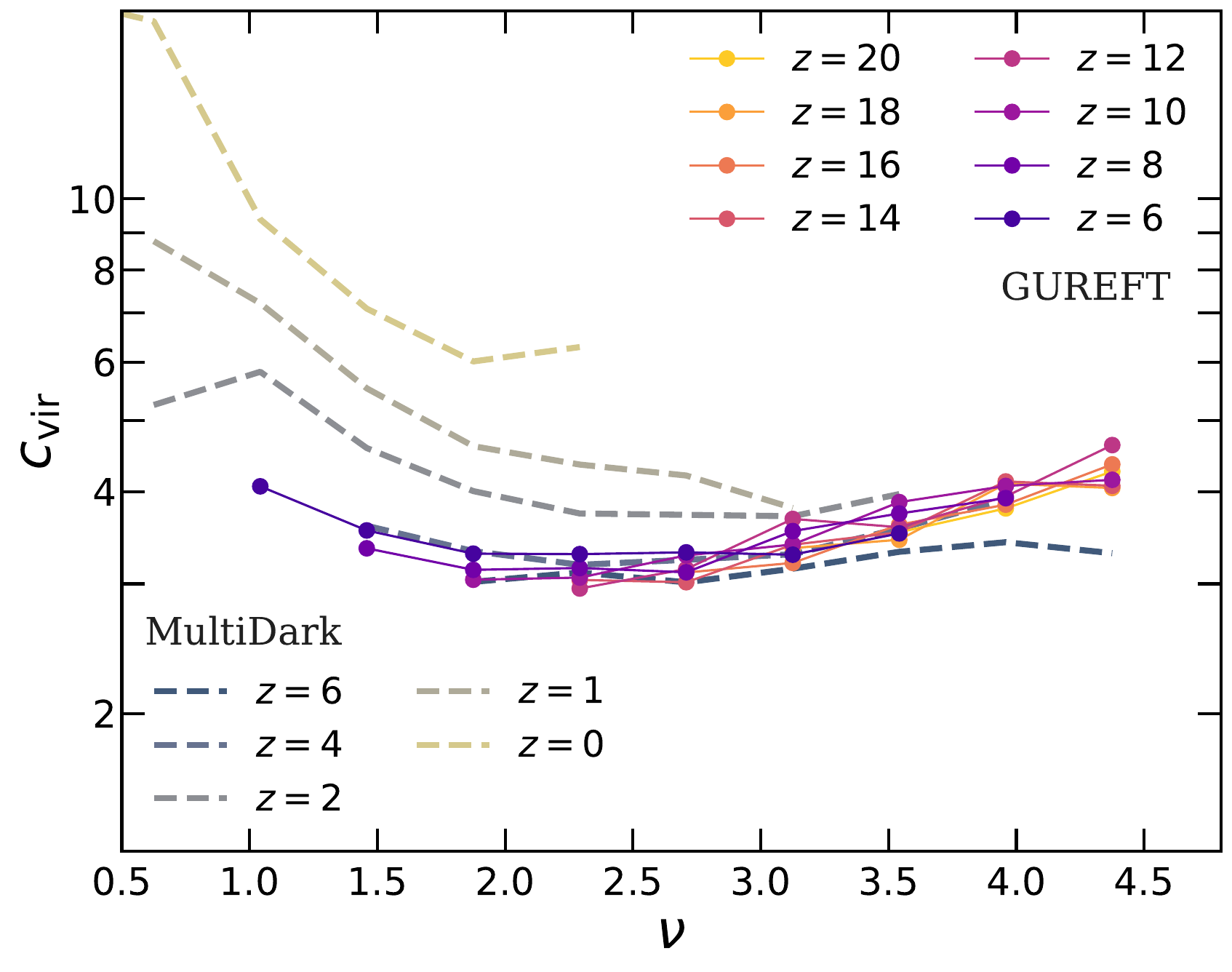}
    \caption{
        The median of halo concentration as a function of peak height  $\nu$, between $z = 6$ to 20 for halos in the \gureft\ simulation suite. 
        We also computed this relation for halos in the VSMDPL simulations (dashed lines) from the MultiDark suite between $z = 0$ to 6 for comparison.
        This illustrates that halo concentration is self-similar and remains invariant at ultra-high $z$ and across a wide redshift range.
        }
    \label{fig:gureft_nu_Mvir}
\end{figure}

\subsection{Interpretation of the trends in halo properties}

Given the limitations in mass resolution and stored snapshots from past cosmological simulations, the key structural properties and assembly histories of these halos, are often inferred by extrapolating fitting functions that were previously fitted to low redshift to untested territories, both into the earlier universe and mass range below the resolution limits of these simulations. In this work, we directly simulate these halos and provide a first look into the distribution of ultra-high $z$ redshift halos by mass, as well as their key structural properties and mass assembly histories. We have shown that these commonly used extrapolations often do not accurately represent the ultra-high redshift halo populations.

The suite of \gureft\ simulations, for the first time, demonstrates the evolution of various halo properties out to ultra-high redshifts and demonstrated that some relations seem to continue the trends seen at lower redshift (e.g. $M_\text{vir}$--$V_\text{max}$, $M_\text{vir}$--$dM_\text{vir}/dt$, and $M_\text{vir}$--$dM_\text{vir}/dt$). 
However, \gureft\ results also reveal a reversal in the $M_\text{vir}$--$c_\text{vir}$ relation at $z\gtrsim6$, where the median concentration for intermediate-mass halos (e.g. $9 \lesssim \log(M_\text{h}/\text{\Msun}) \lesssim 12$) seems to increase between $6 \lesssim z \lesssim 14$.

Numerous past studies have explored the origin of halo density profiles and halo concentration-mass relations \citep[e.g.][]{Navarro1996a, Bullock2001b}, finding strong ties between these properties and halo mass accretion history \citep{Wechsler2002,Zhao2003}. Halo mass accretion history in turn is known to be highly correlated with the height of the density peak within which the halo forms. Assuming that matter is distributed following a linear Gaussian overdensity field, the peak height quantifies how statistically rare a halo of a given mass at a given cosmic epoch is, and is defined as $\nu \equiv \delta_\text{c}/\sigma(M_\text{h},z)$, where $\delta_\text{c}$ is the critical over-density required for spherical collapse at $z$ and $\sigma(M_\text{h},z)$ is the root-mean-square fluctuation of the smoothed density field. Several numerical studies have shown that expressing halo mass in terms of peak height removes much of the evolution of concentration as a function of redshift \citep[e.g.][]{Dalal2008,Zhao2009, Prada2012, Ludlow2014, Dutton2014}. 
Building on this work, \citet{Diemer2015} presented a model in which halo concentration is a function of peak height and the local slope of the power spectrum, where the latter is needed to explain the redshift evolution in the $c$--$\nu$ relation seen between $z \sim 0$ to 6.

To further investigate the underlying drivers for the redshift evolution in halo structural properties presented in Section \ref{sec:properties}, we show in Fig.~\ref{fig:gureft_nu_Mvir} halo concentration as a function of peak height corresponding to a given halo mass.
In our analysis, this quantity is computed using the \texttt{peakHeight} function from the  \textsc{Colossus}\footnote{\url{https://bdiemer.bitbucket.io/colossus/index.html}} python package \citep{Diemer2018}. 
This calculation adopts a value for uniform-density spherical collapse, $\delta_\text{c}\approx 1.686$, derived from the spherical top-hat collapse model in an Einstein-de Sitter universe \citep{Gunn1972}.
We note that the median $c_\text{vir}$--$\nu$ relation presented in Fig.~\ref{fig:gureft_nu_Mvir} is binned separately after peak height is computed for individual halos, rather than translating the binned $c_\text{vir}$--$M_\text{vir}$ relation.
We also calculate this quantity for halos in VSMDPL for comparison.

As demonstrated in Fig.~\ref{fig:gureft_nu_Mvir}, the $c_\text{vir}$--$\nu$ relation exhibits negligible evolution over the redshift range $6 \lesssim z \lesssim 20$.
This result is highly consistent with the prediction of the \citet{Diemer2015} model, as the other main dependency in their universal model, the local slope of the power spectrum, defined as $n(k)\equiv d\ln P(k) / d\ln k$, evolves very little across this redshift range (i.e., the power spectrum is close to self-similar in this regime).

\subsection{Caveats and limitations}

As part of this work, we conducted a few controlled experiments to better characterise the effective mass range covered by individual simulations and explored the impact on the predicted halo populations due to the volumes and resolution of the simulations. 
In addition to the previously mentioned dynamic range limitations due to simulated volume and resolution, we also find that the mass resolution may have an impact on the predicted physical properties of halos. We make use of the overlapping mass range between adjacent \gureft\ boxes to study the potential impact on the simulated halo populations due to differences in mass resolution. The simulated results overall converge extremely well across boxes, including halo number density (see Fig.~\ref{fig:hmf} for HMF) and the median of scaling relations (see Fig.~\ref{fig:gureft_stitch} for $M_\text{vir}$--$c_\text{vir}$ relation).

As shown in Figs.~\ref{fig:cvir_hist} and \ref{fig:spin_hist}, we show the normalised distribution functions for $c_\text{vir}$ and $\lambda_\text{B}$ for halos in overlapping mass range between two adjacent \gureft\ boxes.
We find that halos in the lower resolution box (plotted in orange) tend to have slightly overall higher $c_\text{vir}$ and lower $\lambda_\text{B}$ than halos in the simulated volume with higher mass resolution. 
Halos in simulations with lower mass resolution tend to have their scale radius under-estimated.
While the difference in the resultant distribution is rather subtle, we need to exercise some caution when combining result across different boxes.
We also note that in the context of \gureft\ (or other cosmological simulations when considering the trade off between simulated volume and mass resolution), while halos are better-resolved and better-characterised with a higher number of particles in boxes with higher mass resolutions, for a fixed number of particles the number of halos available in the higher-resolution (hence smaller) boxes would be significantly fewer than in the larger box, which could lead to an incomplete sampling of the true dispersion in the distributions and scaling relations (see Fig.~\ref{fig:gureft_stitch}).
When combining scaling relations across \gureft\ boxes, we prioritise results from larger boxes over smaller ones in order to obtain better statistics.

We also note the mass ranges covered by the \gureft\ suite, especially \gureft-05 and \gureft-15, are sensitive to the small-scale behaviour of dark matter. As shown by \citep{Kulkarni2022}, fuzzy dark matter (FDM; e.g. \citealt{Khlopov1985, Hui2017}) can cause deviations in the HMF at $z = 0$ for halos with $M_\text{h} < 10^{11}$ \Msun\ and a difference in halo number density up to $\sim 2$ dex for $M_\text{h} \sim 10^7$ \Msun. \citeauthor{Kulkarni2022} also showed that it can substantially impact the low-mass end slope ($\log M_\text{h}/\text{M}_\odot \lesssim 11$) of the HMF at $z = 0$. At $z\sim 4$, virtually no halos with $M_\text{h} < 10^{8}$ \Msun\ are able to form \citep{Schive2016}. Similarly, in a warm dark matter (WDM; e.g. \citealt{Dodelson1994}) scenario, the halo number density at $M_\text{h} < 10^9$ \Msun\ can significantly deviate from that in the CDM model \citep{Gilman2020}.
Deep-field galaxy surveys with \textit{JWST} may provide insights that can help indirectly constrain dark matter models.
Like all state-of-the-art dark matter-only cosmological simulations, \gureft\ does not capture the potential back-reaction on dark matter due to baryonic processes, which can modify the halo mass functions, the maximum circular velocity distributions among halos, and their concentration \citep[e.g.][]{Duffy2010, Schaller2015, Sawala2016a, Schaye2023}.

\section{Summary and Conclusions}
\label{sec:summary}

This paper presents the new \gureft\ suite of four dark matter-only, cosmological $N$-body simulations, which were carefully designed to capture the emerging halo populations and their merger histories at high to ultra-high redshifts ($6 \lesssim z \lesssim 20$). The suite of \gureft\ simulations fill in the gap where the earliest episode of halo assembly histories are not covered by existing cosmological-scale dark matter-only simulations.
We provide a comprehensive overview of the predicted distribution functions for key physical properties, including virial mass $M_\text{vir}$, concentration $c_\text{vir}$, and spin $\lambda$, and scaling relations between $M_\text{vir}$ and maximum rotation velocity $V_\text{max}$, $c_\text{vir}$, $\lambda$, and halo growth rate $dM_\text{vir}/dt$. In addition, we present accurate updated fitting functions over the halo mass range and redshift range that is relevant at high redshift. 

We summarise our main conclusions below.

\begin{enumerate}
 
    \item By combining simulated halo populations from all four \gureft\ volumes, we presented halo mass functions (HMF) between $z\sim6$ to 20 for halos across a very wide mass range $\log M_\text{h}/\text{M}_\odot \sim 5$ to 12. In addition, we provide an updated fitting function that accurately describes the HMF over this mass and redshift range. 
    
    \item Fitting functions for the HMF from the literature, as well as analytic models based on the Extended Press Schechter formalism \citep[e.g.][]{Sheth2001} can disagree with the results from \gureft\ by several tenths of a dex up to 1 dex at ultra-high redshifts ($z\gtrsim 10$).

    \item The normalisation of the halo $M_\text{vir}$ vs. $V_\text{max}$ relation continues to increase monotonically from $z\sim 6$--20, continuing the trend previously seen from $z\sim0$--6.
    We find that the relation increases by $\sim 0.25$ dex from $z\sim0$ to 20.

    \item \gureft\ reveals a continuation of the trend seen in previous studies of decreasing spin across all halo masses towards higher redshifts. The probability distribution of $\lambda_\text{B}$ evolves only mildly between $6 \lesssim z \lesssim 20$, whilst the distribution of $\lambda_\text{P}$ evolves more significantly. We provide fitting functions for these distributions.

    \item We find that the $M_\text{vir}$--$c_\text{vir}$ relation, which was known to flatten at $z \sim 4$--6, remains rather flat towards higher redshifts $6$--20. In addition, the normalization of the relation increases mildly, such that halos at earlier cosmic epochs have higher concentration at fixed mass. 

    \item Specific halo mass accretion rates $\dot{M}/M$ at fixed halo mass continue to increase monotonically with increasing redshift from $6$--20, again continuing the trend at lower redshift, but the slope of the $\dot{M}/M$ relation flattens slightly at earlier times.

    \item Fitting functions for $M_\text{vir}$--$dM_\text{vir}/dt$ and the median of the largest progenitor mass over cosmic time
    derived from simulations of larger mass halos at later times are not accurate when extrapolated to the smaller halo masses and higher redshifts that are probed by \gureft. We provide updated fits for this regime.
    
    \item We used halos in the overlapping mass and redshift ranges for two adjacent \gureft\ boxes to perform a limited convergence study. We found that the differences in mass resolution has a small but systematic impact on the predicted distribution of concentration and spin, mainly affecting the highest concentration and highest spin halos. These extreme halos have lower estimated $c_\text{vir}$ and higher $\lambda_\text{B}$ in the simulations with higher mass resolution.
\end{enumerate}

These new dark matter-only simulations provide a framework for modelling and interpreting the first galaxies and black holes in the early universe that can be observed with \textit{JWST}.

\section*{Acknowledgements}
The analysis in this work was carried out with \texttt{astropy} \citep{Robitaille2013, Price-Whelan2018}, \texttt{pandas} \citep{Reback2022}, \texttt{numpy} \citep{VanderWalt2011}, and \texttt{scipy} \citep{Virtanen2020}.

The authors of this paper would like to thank Andrey Kravtsov, Aldo Rodr\'{i}guez-Puebla, Laura Sommovigo, Eli Visbal, and David Spergel for useful discussions.
The \gureft\ simulation suite was run on computing cluster \textit{rusty} managed by the Scientific Computing Core (SCC) of the Flatiron Institute.
AY is supported by an appointment to the NASA Postdoctoral Program (NPP) at NASA Goddard Space Flight Center, administered by Oak Ridge Associated Universities under contract with NASA.
Support from program numbers ERS-01345 and AR-02108 was provided through a grant from the Space Telescope Science Institute under NASA contract NAS5-03127.
RSS acknowledges support from the Simons Foundation.
AY and RSS also thank the Aspen Center for Physics, which is supported by National Science Foundation grant PHY-2210452, for their hospitality during a portion of the creation of this work.

\section*{Data Availability}
The simulated data underlying this paper are stored in a private repository and will be made available upon request. The derived data products will be released through a web portal.



\bibliographystyle{mnras}
\bibliography{ultra-z-halos} 

\begin{thebibliography}{}
\makeatletter
\relax
\def\mn@urlcharsother{\let\do\@makeother \do\$\do\&\do\#\do\^\do\_\do\%\do\~}
\def\mn@doi{\begingroup\mn@urlcharsother \@ifnextchar [ {\mn@doi@}
  {\mn@doi@[]}}
\def\mn@doi@[#1]#2{\def\@tempa{#1}\ifx\@tempa\@empty \href
  {http://dx.doi.org/#2} {doi:#2}\else \href {http://dx.doi.org/#2} {#1}\fi
  \endgroup}
\def\mn@eprint#1#2{\mn@eprint@#1:#2::\@nil}
\def\mn@eprint@arXiv#1{\href {http://arxiv.org/abs/#1} {{\tt arXiv:#1}}}
\def\mn@eprint@dblp#1{\href {http://dblp.uni-trier.de/rec/bibtex/#1.xml}
  {dblp:#1}}
\def\mn@eprint@#1:#2:#3:#4\@nil{\def\@tempa {#1}\def\@tempb {#2}\def\@tempc
  {#3}\ifx \@tempc \@empty \let \@tempc \@tempb \let \@tempb \@tempa \fi \ifx
  \@tempb \@empty \def\@tempb {arXiv}\fi \@ifundefined
  {mn@eprint@\@tempb}{\@tempb:\@tempc}{\expandafter \expandafter \csname
  mn@eprint@\@tempb\endcsname \expandafter{\@tempc}}}

\bibitem[\protect\citeauthoryear{Adams et~al.,}{Adams et~al.}{2022}]{Adams2022}
Adams N.~J.,  et~al., 2022, \mn@doi [MNRAS] {10.1093/mnras/stac3347}, 518, 4755

\bibitem[\protect\citeauthoryear{{Arrabal Haro} et~al.,}{{Arrabal Haro}
  et~al.}{2023}]{ArrabalHaro2023a}
{Arrabal Haro} P.,  et~al., 2023, \mn@doi [ApJL] {10.3847/2041-8213/acdd54},
  951, L22

\bibitem[\protect\citeauthoryear{Atek et~al.,}{Atek et~al.}{2022}]{Atek2022}
Atek H.,  et~al., 2022, \mn@doi [MNRAS] {10.1093/mnras/stac3144}, 519, 1201

\bibitem[\protect\citeauthoryear{Bagley et~al.,}{Bagley
  et~al.}{2024}]{Bagley2023}
Bagley M.~B.,  et~al., 2024, \mn@doi [ApJL] {10.3847/2041-8213/ad2f31}, 965, L6

\bibitem[\protect\citeauthoryear{Barro et~al.,}{Barro et~al.}{2024}]{Barro2023}
Barro G.,  et~al., 2024, \mn@doi [ApJ] {10.3847/1538-4357/ad167e}, 963, 128

\bibitem[\protect\citeauthoryear{Baugh, Benson, Cole, Frenk  \& Lacey}{Baugh
  et~al.}{2006}]{Baugh2006}
Baugh C.~M.,  Benson A.~J.,  Cole S.,  Frenk C.~S.,   Lacey C.,  2006, \mn@doi
  [Mass Galaxies Low High Redshift] {10.1007/10899892_22}, pp 91--96

\bibitem[\protect\citeauthoryear{Behroozi, Wechsler  \& Wu}{Behroozi
  et~al.}{2013a}]{Behroozi2013b}
Behroozi P.~S.,  Wechsler R.~H.,   Wu H.-Y.,  2013a, \mn@doi [ApJ]
  {10.1088/0004-637X/762/2/109}, 762, 109

\bibitem[\protect\citeauthoryear{Behroozi, Wechsler, Wu, Busha, Klypin  \&
  Primack}{Behroozi et~al.}{2013b}]{Behroozi2013c}
Behroozi P.~S.,  Wechsler R.~H.,  Wu H.-Y.,  Busha M.~T.,  Klypin A.~A.,
  Primack J.~R.,  2013b, \mn@doi [ApJ] {10.1088/0004-637X/763/1/18}, 763, 18

\bibitem[\protect\citeauthoryear{Behroozi, Wechsler  \& Conroy}{Behroozi
  et~al.}{2013c}]{Behroozi2013a}
Behroozi P.~S.,  Wechsler R.~H.,   Conroy C.,  2013c, \mn@doi [ApJ]
  {10.1088/0004-637X/770/1/57}, 770, 57

\bibitem[\protect\citeauthoryear{Behroozi, Wechsler, Hearin  \&
  Conroy}{Behroozi et~al.}{2019}]{Behroozi2019}
Behroozi P.,  Wechsler R.~H.,  Hearin A.~P.,   Conroy C.,  2019, \mn@doi
  [MNRAS] {10.1093/mnras/stz1182}, 488, 3143

\bibitem[\protect\citeauthoryear{Behroozi et~al.,}{Behroozi
  et~al.}{2020}]{Behroozi2020}
Behroozi P.,  et~al., 2020, \mn@doi [MNRAS] {10.1093/mnras/staa3164}, 499, 5702

\bibitem[\protect\citeauthoryear{Benson}{Benson}{2010}]{Benson2010}
Benson A.~J.,  2010, \mn@doi [New Astron.] {10.1016/j.newast.2011.07.004}, 17,
  175

\bibitem[\protect\citeauthoryear{Benson, Behrens  \& Lu}{Benson
  et~al.}{2020}]{Benson2020}
Benson A.,  Behrens C.,   Lu Y.,  2020, \mn@doi [MNRAS]
  {10.1093/MNRAS/STAA1777}, 496, 3371

\bibitem[\protect\citeauthoryear{Bond, Cole, Efstathiou  \& Kaiser}{Bond
  et~al.}{1991}]{Bond1991}
Bond J.~R.,  Cole S.,  Efstathiou G.,   Kaiser N.,  1991, \mn@doi [ApJ]
  {10.1086/170520}, 379, 440

\bibitem[\protect\citeauthoryear{Boylan-Kolchin}{Boylan-Kolchin}{2023}]{Boylan-Kolchin2022}
Boylan-Kolchin M.,  2023, \mn@doi [Nat. Astron.] {10.1038/s41550-023-01937-7},
  7, 731

\bibitem[\protect\citeauthoryear{Boylan-Kolchin, Springel, White, Jenkins  \&
  Lemson}{Boylan-Kolchin et~al.}{2009}]{Boylan-Kolchin2009}
Boylan-Kolchin M.,  Springel V.,  White S. D.~M.,  Jenkins A.,   Lemson G.,
  2009, \mn@doi [MNRAS] {10.1111/j.1365-2966.2009.15191.x}, 398, 1150

\bibitem[\protect\citeauthoryear{Bryan \& Norman}{Bryan \&
  Norman}{1998}]{Bryan1998}
Bryan G.~L.,  Norman M.~L.,  1998, \mn@doi [ApJ] {10.1086/305262}, 495, 80

\bibitem[\protect\citeauthoryear{Bullock, Kolatt, Sigad, Somerville, Kravtsov,
  Klypin, Primack  \& Dekel}{Bullock et~al.}{2001a}]{Bullock2001b}
Bullock J.~S.,  Kolatt T.~S.,  Sigad Y.,  Somerville R.~S.,  Kravtsov A.~V.,
  Klypin A.~A.,  Primack J.~R.,   Dekel A.,  2001a, \mn@doi [MNRAS]
  {10.1046/j.1365-8711.2001.04068.x}, 321, 559

\bibitem[\protect\citeauthoryear{Bullock, Dekel, Kolatt, Kravtsov, Klypin,
  Porciani  \& Primack}{Bullock et~al.}{2001b}]{Bullock2001a}
Bullock J.~S.,  Dekel A.,  Kolatt T.~S.,  Kravtsov A.~V.,  Klypin A.~A.,
  Porciani C.,   Primack J.~R.,  2001b, \mn@doi [ApJ] {10.1086/321477}, 555,
  240

\bibitem[\protect\citeauthoryear{Castellano et~al.,}{Castellano
  et~al.}{2022}]{Castellano2022}
Castellano M.,  et~al., 2022, \mn@doi [ApJL] {10.3847/2041-8213/ac94d0}, 938,
  L15

\bibitem[\protect\citeauthoryear{Cole, Lacey, Baugh  \& Frenk}{Cole
  et~al.}{2000}]{Cole2000}
Cole S.,  Lacey C.,  Baugh C.,   Frenk C.,  2000, \mn@doi [MNRAS]
  {10.1046/j.1365-8711.2000.03879.x}, 319, 168

\bibitem[\protect\citeauthoryear{Conselice, Chapman  \& Windhorst}{Conselice
  et~al.}{2003}]{Conselice2003}
Conselice C.~J.,  Chapman S.~C.,   Windhorst R.~a.,  2003, Astrophys. J. ApJ,
  596, 2001

\bibitem[\protect\citeauthoryear{Crocce, Pueblas  \& Scoccimarro}{Crocce
  et~al.}{2006}]{Crocce2006}
Crocce M.,  Pueblas S.,   Scoccimarro R.,  2006, \mn@doi [MNRAS]
  {10.1111/j.1365-2966.2006.11040.x}, 373, 369

\bibitem[\protect\citeauthoryear{Croton et~al.,}{Croton
  et~al.}{2016}]{Croton2016}
Croton D.~J.,  et~al., 2016, \mn@doi [ApJS] {10.3847/0067-0049/222/2/22}, 222,
  22

\bibitem[\protect\citeauthoryear{Curtis-Lake et~al.,}{Curtis-Lake
  et~al.}{2022}]{Curtis-Lake2022}
Curtis-Lake E.,  et~al., 2022, arXiv:2212.04568

\bibitem[\protect\citeauthoryear{Dalal, White, Bond  \& Shirokov}{Dalal
  et~al.}{2008}]{Dalal2008}
Dalal N.,  White M.,  Bond J.~R.,   Shirokov A.,  2008, \mn@doi [ApJ]
  {10.1086/591512}, 687, 12

\bibitem[\protect\citeauthoryear{Dayal, Ferrara, Dunlop  \& Pacucci}{Dayal
  et~al.}{2014}]{Dayal2014}
Dayal P.,  Ferrara A.,  Dunlop J.~S.,   Pacucci F.,  2014, \mn@doi [MNRAS]
  {10.1093/mnras/stu1848}, 445, 2545

\bibitem[\protect\citeauthoryear{Dayal, Rossi, Shiralilou, Piana, Choudhury  \&
  Volonteri}{Dayal et~al.}{2019}]{Dayal2019}
Dayal P.,  Rossi E.~M.,  Shiralilou B.,  Piana O.,  Choudhury T.~R.,
  Volonteri M.,  2019, \mn@doi [MNRAS] {10.1093/mnras/stz897}, 486, 2336

\bibitem[\protect\citeauthoryear{DeRose et~al.,}{DeRose
  et~al.}{2019}]{DeRose2019}
DeRose J.,  et~al., 2019, \mn@doi [ApJ] {10.3847/1538-4357/ab1085}, 875, 69

\bibitem[\protect\citeauthoryear{Dekel, Zolotov, Tweed, Cacciato, Ceverino  \&
  Primack}{Dekel et~al.}{2013}]{Dekel2013}
Dekel A.,  Zolotov A.,  Tweed D.,  Cacciato M.,  Ceverino D.,   Primack J.~R.,
  2013, \mn@doi [MNRAS] {10.1093/mnras/stt1338}, 435, 999

\bibitem[\protect\citeauthoryear{Dekel, Sarkar, Birnboim, Mandelker  \&
  Li}{Dekel et~al.}{2023}]{Dekel2023}
Dekel A.,  Sarkar K.~C.,  Birnboim Y.,  Mandelker N.,   Li Z.,  2023, \mn@doi
  [MNRAS] {10.1093/mnras/stad1557}, 523, 3201

\bibitem[\protect\citeauthoryear{Diemer}{Diemer}{2018}]{Diemer2018}
Diemer B.,  2018, \mn@doi [ApJS] {10.3847/1538-4365/aaee8c}, 239, 35

\bibitem[\protect\citeauthoryear{Diemer \& Kravtsov}{Diemer \&
  Kravtsov}{2015}]{Diemer2015}
Diemer B.,  Kravtsov A.~V.,  2015, \mn@doi [ApJ] {10.1088/0004-637X/799/1/108},
  799, 108

\bibitem[\protect\citeauthoryear{Dodelson \& Widrow}{Dodelson \&
  Widrow}{1994}]{Dodelson1994}
Dodelson S.,  Widrow L.~M.,  1994, \mn@doi [Phys. Rev. Lett.]
  {10.1103/PhysRevLett.72.17}, 72, 17

\bibitem[\protect\citeauthoryear{Donnan et~al.,}{Donnan
  et~al.}{2022}]{Donnan2022}
Donnan C.~T.,  et~al., 2022, \mn@doi [MNRAS] {10.1093/mnras/stac3472}, 518,
  6011

\bibitem[\protect\citeauthoryear{Drakos et~al.,}{Drakos
  et~al.}{2022}]{Drakos2021}
Drakos N.~E.,  et~al., 2022, \mn@doi [ApJ] {10.3847/1538-4357/ac46fb}, 926, 194

\bibitem[\protect\citeauthoryear{Duffy, Schaye, Kay, Vecchia, Battye  \&
  Booth}{Duffy et~al.}{2010}]{Duffy2010}
Duffy A.~R.,  Schaye J.,  Kay S.~T.,  Vecchia C.~D.,  Battye R.~A.,   Booth
  C.~M.,  2010, \mn@doi [MNRAS] {10.1111/j.1365-2966.2010.16613.x}, 405, 2161

\bibitem[\protect\citeauthoryear{Dutton \& Macci{\`{o}}}{Dutton \&
  Macci{\`{o}}}{2014}]{Dutton2014}
Dutton A.~A.,  Macci{\`{o}} A.~V.,  2014, \mn@doi [MNRAS]
  {10.1093/mnras/stu742}, 441, 3359

\bibitem[\protect\citeauthoryear{Elahi, Poulton, Tobar, Ca{\~{n}}as, Lagos,
  Power  \& Robotham}{Elahi et~al.}{2019}]{Elahi2019}
Elahi P.~J.,  Poulton R. J.~J.,  Tobar R.~J.,  Ca{\~{n}}as R.,  Lagos C. d.~P.,
   Power C.,   Robotham A. S.~G.,  2019, \mn@doi [PASA] {10.1017/pasa.2019.18},
  36, e028

\bibitem[\protect\citeauthoryear{Ferrara, Pallottini  \& Dayal}{Ferrara
  et~al.}{2023}]{Ferrara2022}
Ferrara A.,  Pallottini A.,   Dayal P.,  2023, \mn@doi [MNRAS]
  {10.1093/mnras/stad1095}, 522, 3986

\bibitem[\protect\citeauthoryear{Finkelstein et~al.,}{Finkelstein
  et~al.}{2022}]{Finkelstein2022a}
Finkelstein S.~L.,  et~al., 2022, \mn@doi [ApJL] {10.3847/2041-8213/ac966e},
  940, L55

\bibitem[\protect\citeauthoryear{Finkelstein et~al.,}{Finkelstein
  et~al.}{2023}]{Finkelstein2022b}
Finkelstein S.~L.,  et~al., 2023, \mn@doi [ApJL] {10.3847/2041-8213/acade4},
  946, L13

\bibitem[\protect\citeauthoryear{Fujimoto et~al.,}{Fujimoto
  et~al.}{2023}]{Fujimoto2023}
Fujimoto S.,  et~al., 2023, \mn@doi [ApJL] {10.3847/2041-8213/acd2d9}, 949, L25

\bibitem[\protect\citeauthoryear{Gabrielpillai, Somerville, Genel,
  Rodriguez-Gomez, Pandya, Yung  \& Hernquist}{Gabrielpillai
  et~al.}{2022}]{Gabrielpillai2022}
Gabrielpillai A.,  Somerville R.~S.,  Genel S.,  Rodriguez-Gomez V.,  Pandya
  V.,  Yung L. Y.~A.,   Hernquist L.,  2022, \mn@doi [MNRAS]
  {10.1093/mnras/stac2297}, 517, 6091

\bibitem[\protect\citeauthoryear{Gardner et~al.,}{Gardner
  et~al.}{2006}]{Gardner2006}
Gardner J.~P.,  et~al., 2006, \mn@doi [SSR] {10.1007/s11214-006-8315-7}, 123,
  485

\bibitem[\protect\citeauthoryear{Gardner et~al.,}{Gardner
  et~al.}{2023}]{Gardner2023}
Gardner J.~P.,  et~al., 2023, \mn@doi [PASP] {10.1088/1538-3873/acd1b5}, 135,
  068001

\bibitem[\protect\citeauthoryear{Garrison, Eisenstein, Ferrer, Metchnik  \&
  Pinto}{Garrison et~al.}{2016}]{Garrison2016}
Garrison L.~H.,  Eisenstein D.~J.,  Ferrer D.,  Metchnik M.~V.,   Pinto P.~A.,
  2016, \mn@doi [MNRAS] {10.1093/mnras/stw1594}, 461, 4125

\bibitem[\protect\citeauthoryear{Gilman, Birrer, Nierenberg, Treu, Du  \&
  Benson}{Gilman et~al.}{2020}]{Gilman2020}
Gilman D.,  Birrer S.,  Nierenberg A.,  Treu T.,  Du X.,   Benson A.,  2020,
  \mn@doi [MNRAS] {10.1093/mnras/stz3480}, 491, 6077

\bibitem[\protect\citeauthoryear{Gunn \& Gott}{Gunn \& Gott}{1972}]{Gunn1972}
Gunn J.~E.,  Gott J.~R.,  1972, \mn@doi [ApJ] {10.1086/151605}, 176, 1

\bibitem[\protect\citeauthoryear{Hahn \& Abel}{Hahn \& Abel}{2011}]{Hahn2011}
Hahn O.,  Abel T.,  2011, \mn@doi [MNRAS] {10.1111/j.1365-2966.2011.18820.x},
  415, 2101

\bibitem[\protect\citeauthoryear{Harikane, Nakajima, Ouchi, Umeda, Isobe, Ono,
  Xu  \& Zhang}{Harikane et~al.}{2023a}]{Harikane2023}
Harikane Y.,  Nakajima K.,  Ouchi M.,  Umeda H.,  Isobe Y.,  Ono Y.,  Xu Y.,
  Zhang Y.,  2023a, arXiv:2304.06658

\bibitem[\protect\citeauthoryear{Harikane et~al.,}{Harikane
  et~al.}{2023b}]{Harikane2022}
Harikane Y.,  et~al., 2023b, \mn@doi [ApJS] {10.3847/1538-4365/acaaa9}, 265, 5

\bibitem[\protect\citeauthoryear{Henriques, White, Thomas, Angulo, Guo, Lemson,
  Springel  \& Overzier}{Henriques et~al.}{2015}]{Henriques2015}
Henriques B. M.~B.,  White S. D.~M.,  Thomas P.~A.,  Angulo R.,  Guo Q.,
  Lemson G.,  Springel V.,   Overzier R.,  2015, \mn@doi [MNRAS]
  {10.1093/mnras/stv705}, 451, 2663

\bibitem[\protect\citeauthoryear{Hopkins, Hernquist, Cox, Robertson  \&
  Krause}{Hopkins et~al.}{2007}]{Hopkins2007b}
Hopkins P.~F.,  Hernquist L.,  Cox T.~J.,  Robertson B.,   Krause E.,  2007,
  \mn@doi [ApJ] {10.1086/521601}, 669, 67

\bibitem[\protect\citeauthoryear{Hui, Ostriker, Tremaine  \& Witten}{Hui
  et~al.}{2017}]{Hui2017}
Hui L.,  Ostriker J.~P.,  Tremaine S.,   Witten E.,  2017, \mn@doi [Phys. Rev.
  D] {10.1103/PhysRevD.95.043541}, 95

\bibitem[\protect\citeauthoryear{Ishiyama et~al.,}{Ishiyama
  et~al.}{2021}]{Ishiyama2021}
Ishiyama T.,  et~al., 2021, \mn@doi [MNRAS] {10.1093/mnras/stab1755}, 506, 4210

\bibitem[\protect\citeauthoryear{Jiang \& van~den Bosch}{Jiang \& van~den
  Bosch}{2014}]{Jiang2014}
Jiang F.,  van~den Bosch F.~C.,  2014, \mn@doi [MNRAS] {10.1093/mnras/stu280},
  440, 193

\bibitem[\protect\citeauthoryear{Khlopov, Malomed  \& Zeldovich}{Khlopov
  et~al.}{1985}]{Khlopov1985}
Khlopov M.~Y.,  Malomed B.~A.,   Zeldovich Y.~B.,  1985, \mn@doi [MNRAS]
  {10.1093/mnras/215.4.575}, 215, 575

\bibitem[\protect\citeauthoryear{Klypin, Trujillo-Gomez  \& Primack}{Klypin
  et~al.}{2011}]{Klypin2011}
Klypin A.~A.,  Trujillo-Gomez S.,   Primack J.,  2011, \mn@doi [ApJ]
  {10.1088/0004-637X/740/2/102}, 740, 102

\bibitem[\protect\citeauthoryear{Klypin, Yepes, Gottl{\"{o}}ber, Prada  \&
  He{\ss}}{Klypin et~al.}{2016}]{Klypin2016}
Klypin A.,  Yepes G.,  Gottl{\"{o}}ber S.,  Prada F.,   He{\ss} S.,  2016,
  \mn@doi [MNRAS] {10.1093/mnras/stw248}, 457, 4340

\bibitem[\protect\citeauthoryear{Kocevski et~al.,}{Kocevski
  et~al.}{2023a}]{Kocevski2022}
Kocevski D.~D.,  et~al., 2023a, \mn@doi [ApJL] {10.3847/2041-8213/acad00}, 946,
  L14

\bibitem[\protect\citeauthoryear{Kocevski et~al.,}{Kocevski
  et~al.}{2023b}]{Kocevski2023}
Kocevski D.~D.,  et~al., 2023b, \mn@doi [ApJL] {10.3847/2041-8213/ace5a0}, 954,
  L4

\bibitem[\protect\citeauthoryear{Kulkarni \& Ostriker}{Kulkarni \&
  Ostriker}{2022}]{Kulkarni2022}
Kulkarni M.,  Ostriker J.~P.,  2022, \mn@doi [MNRAS] {10.1093/mnras/stab3520},
  510, 1425

\bibitem[\protect\citeauthoryear{Labb{\'{e}} et~al.,}{Labb{\'{e}}
  et~al.}{2023}]{Labbe2022}
Labb{\'{e}} I.,  et~al., 2023, \mn@doi [Nature] {10.1038/s41586-023-05786-2},
  616, 266

\bibitem[\protect\citeauthoryear{Lacey \& Cole}{Lacey \&
  Cole}{1993}]{Lacey1993}
Lacey C.,  Cole S.,  1993, \mn@doi [MNRAS] {10.1093/mnras/262.3.627}, 262, 627

\bibitem[\protect\citeauthoryear{Lacey \& Cole}{Lacey \&
  Cole}{1994}]{Lacey1994}
Lacey C.,  Cole S.,  1994, \mn@doi [MNRAS] {10.1093/mnras/271.3.676}, 271, 676

\bibitem[\protect\citeauthoryear{Larson et~al.,}{Larson
  et~al.}{2023}]{Larson2023}
Larson R.~L.,  et~al., 2023, \mn@doi [ApJL] {10.3847/2041-8213/ace619}, 953,
  L29

\bibitem[\protect\citeauthoryear{Leung et~al.,}{Leung
  et~al.}{2023}]{Leung2023a}
Leung G. C.~K.,  et~al., 2023, \mn@doi [ApJL] {10.3847/2041-8213/acf365}, 954,
  L46

\bibitem[\protect\citeauthoryear{Lovell, Harrison, Harikane, Tacchella  \&
  Wilkins}{Lovell et~al.}{2022}]{Lovell2022}
Lovell C.~C.,  Harrison I.,  Harikane Y.,  Tacchella S.,   Wilkins S.~M.,
  2022, \mn@doi [MNRAS] {10.1093/mnras/stac3224}, 518, 2511

\bibitem[\protect\citeauthoryear{Ludlow, Navarro, Angulo, Boylan-Kolchin,
  Springel, Frenk  \& White}{Ludlow et~al.}{2014}]{Ludlow2014}
Ludlow A.~D.,  Navarro J.~F.,  Angulo R.~E.,  Boylan-Kolchin M.,  Springel V.,
  Frenk C.,   White S. D.~M.,  2014, \mn@doi [MNRAS] {10.1093/mnras/stu483},
  441, 378

\bibitem[\protect\citeauthoryear{Maksimova, Garrison, Eisenstein, Hadzhiyska,
  Bose  \& Satterthwaite}{Maksimova et~al.}{2021}]{Maksimova2021}
Maksimova N.~A.,  Garrison L.~H.,  Eisenstein D.~J.,  Hadzhiyska B.,  Bose S.,
   Satterthwaite T.~P.,  2021, \mn@doi [MNRAS] {10.1093/mnras/stab2484}, 508,
  4017

\bibitem[\protect\citeauthoryear{Mason, Trenti  \& Treu}{Mason
  et~al.}{2023}]{Mason2022}
Mason C.~A.,  Trenti M.,   Treu T.,  2023, \mn@doi [MNRAS]
  {10.1093/mnras/stad035}, 521, 497

\bibitem[\protect\citeauthoryear{Mo, Mao  \& White}{Mo et~al.}{1998}]{Mo1998}
Mo H.~J.,  Mao S.,   White S. D.~M.,  1998, \mn@doi [MNRAS]
  {10.1046/j.1365-8711.1998.01227.x}, 295, 319

\bibitem[\protect\citeauthoryear{Moster, Naab  \& White}{Moster
  et~al.}{2018}]{Moster2018}
Moster B.~P.,  Naab T.,   White S. D.~M.,  2018, \mn@doi [MNRAS]
  {10.1093/mnras/sty655}, 477, 1822

\bibitem[\protect\citeauthoryear{Mu{\~{n}}oz, Mirocha, Furlanetto  \&
  Sabti}{Mu{\~{n}}oz et~al.}{2023}]{Munoz2023}
Mu{\~{n}}oz J.~B.,  Mirocha J.,  Furlanetto S.,   Sabti N.,  2023,
  arXiv:2306.09403

\bibitem[\protect\citeauthoryear{Naab \& Ostriker}{Naab \&
  Ostriker}{2017}]{Naab2017}
Naab T.,  Ostriker J.~P.,  2017, \mn@doi [ARA&A]
  {10.1146/annurev-astro-081913-040019}, 55, 59

\bibitem[\protect\citeauthoryear{Naidu et~al.,}{Naidu et~al.}{2022}]{Naidu2022}
Naidu R.~P.,  et~al., 2022, \mn@doi [ApJL] {10.3847/2041-8213/ac9b22}, 940, L14

\bibitem[\protect\citeauthoryear{Navarro, Frenk  \& White}{Navarro
  et~al.}{1996}]{Navarro1996a}
Navarro J.~F.,  Frenk C.~S.,   White S. D.~M.,  1996, \mn@doi [ApJ]
  {10.1086/177173}, 462, 563

\bibitem[\protect\citeauthoryear{Navarro, Frenk  \& White}{Navarro
  et~al.}{1997}]{Navarro1997}
Navarro J.~F.,  Frenk C.~S.,   White S. D.~M.,  1997, \mn@doi [ApJ]
  {10.1086/304888}, 490, 493

\bibitem[\protect\citeauthoryear{Neto et~al.,}{Neto et~al.}{2007}]{Neto2007}
Neto A.~F.,  et~al., 2007, \mn@doi [MNRAS] {10.1111/j.1365-2966.2007.12381.x},
  381, 1450

\bibitem[\protect\citeauthoryear{Nguyen, Modi, Yung  \& Somerville}{Nguyen
  et~al.}{2023}]{Nguyen2023}
Nguyen T.,  Modi C.,  Yung L. Y.~A.,   Somerville R.~S.,  2023,
  arXiv:2308.05145

\bibitem[\protect\citeauthoryear{O'Leary, Steinwandel, Moster, Martin  \&
  Naab}{O'Leary et~al.}{2023}]{OLeary2023}
O'Leary J.~A.,  Steinwandel U.~P.,  Moster B.~P.,  Martin N.,   Naab T.,  2023,
  \mn@doi [MNRAS] {10.1093/mnras/stad166}, 520, 897

\bibitem[\protect\citeauthoryear{Padmanabhan \& Loeb}{Padmanabhan \&
  Loeb}{2023}]{Padmanabhan2023}
Padmanabhan H.,  Loeb A.,  2023, arXiv:2306.04684

\bibitem[\protect\citeauthoryear{Pakmor et~al.,}{Pakmor
  et~al.}{2022}]{Pakmor2022}
Pakmor R.,  et~al., 2022, arXiv:2210.10060

\bibitem[\protect\citeauthoryear{Parkinson, Cole  \& Helly}{Parkinson
  et~al.}{2008}]{Parkinson2008}
Parkinson H.,  Cole S.,   Helly J.,  2008, \mn@doi [MNRAS]
  {10.1111/j.1365-2966.2007.12517.x}, 383, 557

\bibitem[\protect\citeauthoryear{Peebles}{Peebles}{1969}]{Peebles1969}
Peebles P. J.~E.,  1969, \mn@doi [ApJ] {10.1086/149876}, 155, 393

\bibitem[\protect\citeauthoryear{Peebles}{Peebles}{1980}]{Peebles1980}
Peebles P. J.~E.,  1980, in , Large-Scale Struct. Universe.
Princeton University Press, Princeton, \url
  {https://ui.adsabs.harvard.edu/abs/1980lssu.book.....P}

\bibitem[\protect\citeauthoryear{{Planck Collaboration}}{{Planck
  Collaboration}}{2014}]{Planck2014}
{Planck Collaboration} 2014, \mn@doi [A&A] {10.1051/0004-6361/201321591}, 571,
  A16

\bibitem[\protect\citeauthoryear{{Planck Collaboration}}{{Planck
  Collaboration}}{2016}]{Planck2016}
{Planck Collaboration} 2016, \mn@doi [A&A] {10.1051/0004-6361/201525830}, 594,
  A13

\bibitem[\protect\citeauthoryear{Potter, Stadel  \& Teyssier}{Potter
  et~al.}{2017}]{Potter2017}
Potter D.,  Stadel J.,   Teyssier R.,  2017, \mn@doi [Comput. Astrophys.
  Cosmol.] {10.1186/s40668-017-0021-1}, 4

\bibitem[\protect\citeauthoryear{Power \& Knebe}{Power \&
  Knebe}{2006}]{Power2006}
Power C.,  Knebe A.,  2006, \mn@doi [MNRAS] {10.1111/j.1365-2966.2006.10562.x},
  370, 691

\bibitem[\protect\citeauthoryear{Prada, Klypin, Cuesta, Betancort-Rijo  \&
  Primack}{Prada et~al.}{2012}]{Prada2012}
Prada F.,  Klypin A.~A.,  Cuesta A.~J.,  Betancort-Rijo J.~E.,   Primack J.,
  2012, \mn@doi [MNRAS] {10.1111/j.1365-2966.2012.21007.x}, 423, 3018

\bibitem[\protect\citeauthoryear{Press \& Schechter}{Press \&
  Schechter}{1974}]{Press1974}
Press W.~H.,  Schechter P.,  1974, \mn@doi [ApJ] {10.1086/152650}, 187, 425

\bibitem[\protect\citeauthoryear{Price-Whelan et~al.,}{Price-Whelan
  et~al.}{2018}]{Price-Whelan2018}
Price-Whelan A.~M.,  et~al., 2018, \mn@doi [AJ] {10.3847/1538-3881/aabc4f},
  156, 123

\bibitem[\protect\citeauthoryear{Reback et~al.,}{Reback
  et~al.}{2022}]{Reback2022}
Reback J.,  et~al., 2022, {pandas-dev/pandas: Pandas},
  \mn@doi{10.5281/zenodo.6408044}, \url
  {https://doi.org/10.5281/zenodo.6408044}

\bibitem[\protect\citeauthoryear{Reed, Bower, Frenk, Jenkins  \& Theuns}{Reed
  et~al.}{2007}]{Reed2007}
Reed D.~S.,  Bower R.,  Frenk C.~S.,  Jenkins A.,   Theuns T.,  2007, \mn@doi
  [MNRAS] {10.1111/j.1365-2966.2006.11204.x}, 374, 2

\bibitem[\protect\citeauthoryear{Riebe et~al.,}{Riebe et~al.}{2013}]{Riebe2013}
Riebe K.,  et~al., 2013, \mn@doi [Astron. Nachrichten]
  {10.1002/asna.201211900}, 334, 691

\bibitem[\protect\citeauthoryear{Robertson et~al.,}{Robertson
  et~al.}{2023}]{Robertson2023}
Robertson B.~E.,  et~al., 2023, \mn@doi [Nat. Astron.]
  {10.1038/s41550-023-01921-1}, 7, 611

\bibitem[\protect\citeauthoryear{Robitaille et~al.,}{Robitaille
  et~al.}{2013}]{Robitaille2013}
Robitaille T.~P.,  et~al., 2013, \mn@doi [A&A] {10.1051/0004-6361/201322068},
  558, A33

\bibitem[\protect\citeauthoryear{Rodriguez-Gomez et~al.,}{Rodriguez-Gomez
  et~al.}{2015}]{Rodriguez-Gomez2015}
Rodriguez-Gomez V.,  et~al., 2015, \mn@doi [MNRAS] {10.1093/mnras/stv264}, 449,
  49

\bibitem[\protect\citeauthoryear{Rodr{\'{i}}guez-Puebla, Behroozi, Primack,
  Klypin, Lee  \& Hellinger}{Rodr{\'{i}}guez-Puebla
  et~al.}{2016}]{Rodriguez-Puebla2016}
Rodr{\'{i}}guez-Puebla A.,  Behroozi P.,  Primack J.,  Klypin A.,  Lee C.,
  Hellinger D.,  2016, \mn@doi [MNRAS] {10.1093/mnras/stw1705}, 462, 893

\bibitem[\protect\citeauthoryear{Rodr{\'{i}}guez-Puebla, Primack, Avila-Reese
  \& Faber}{Rodr{\'{i}}guez-Puebla et~al.}{2017}]{Rodriguez-Puebla2017}
Rodr{\'{i}}guez-Puebla A.,  Primack J.~R.,  Avila-Reese V.,   Faber S.~M.,
  2017, \mn@doi [MNRAS] {10.1093/mnras/stx1172}, 470, 651

\bibitem[\protect\citeauthoryear{Sawala et~al.,}{Sawala
  et~al.}{2016}]{Sawala2016a}
Sawala T.,  et~al., 2016, \mn@doi [MNRAS] {10.1093/mnras/stw145}, 457, 1931

\bibitem[\protect\citeauthoryear{Schaller et~al.,}{Schaller
  et~al.}{2015}]{Schaller2015}
Schaller M.,  et~al., 2015, \mn@doi [MNRAS] {10.1093/mnras/stv1067}, 451, 1247

\bibitem[\protect\citeauthoryear{Schaye et~al.,}{Schaye
  et~al.}{2023}]{Schaye2023}
Schaye J.,  et~al., 2023, \mn@doi [MNRAS] {10.1093/mnras/stad2419}, 526, 4978

\bibitem[\protect\citeauthoryear{Schive, Chiueh, Broadhurst  \& Huang}{Schive
  et~al.}{2016}]{Schive2016}
Schive H.-Y.,  Chiueh T.,  Broadhurst T.,   Huang K.-W.,  2016, \mn@doi [ApJ]
  {10.3847/0004-637x/818/1/89}, 818, 89

\bibitem[\protect\citeauthoryear{Scoccimarro}{Scoccimarro}{1998}]{Scoccimarro1998}
Scoccimarro R.,  1998, \mn@doi [MNRAS] {10.1046/j.1365-8711.1998.01845.x}, 299,
  1097

\bibitem[\protect\citeauthoryear{Sheth}{Sheth}{1998}]{Sheth1998}
Sheth R.~K.,  1998, \mn@doi [MNRAS] {10.1046/j.1365-8711.1998.01976.x}, 300,
  1057

\bibitem[\protect\citeauthoryear{Sheth \& Tormen}{Sheth \&
  Tormen}{1999}]{Sheth1999}
Sheth R.~K.,  Tormen G.,  1999, \mn@doi [MNRAS]
  {10.1046/j.1365-8711.1999.02692.x}, 308, 119

\bibitem[\protect\citeauthoryear{Sheth \& Tormen}{Sheth \&
  Tormen}{2002}]{Sheth2002}
Sheth R.~K.,  Tormen G.,  2002, \mn@doi [MNRAS]
  {10.1046/j.1365-8711.2002.04950.x}, 329, 61

\bibitem[\protect\citeauthoryear{Sheth, Mo  \& Tormen}{Sheth
  et~al.}{2001}]{Sheth2001}
Sheth R.~K.,  Mo H.~J.,   Tormen G.,  2001, \mn@doi [MNRAS]
  {10.1046/j.1365-8711.2001.04006.x}, 323, 1

\bibitem[\protect\citeauthoryear{Somerville \& Dav{\'{e}}}{Somerville \&
  Dav{\'{e}}}{2015}]{Somerville2015a}
Somerville R.~S.,  Dav{\'{e}} R.,  2015, \mn@doi [ARA&A]
  {10.1146/annurev-astro-082812-140951}, 53, 31

\bibitem[\protect\citeauthoryear{Somerville \& Kolatt}{Somerville \&
  Kolatt}{1999}]{Somerville1999a}
Somerville R.~S.,  Kolatt T.~S.,  1999, \mn@doi [MNRAS]
  {10.1046/j.1365-8711.1999.02154.x}, 305, 1

\bibitem[\protect\citeauthoryear{Somerville, Hopkins, Cox, Robertson  \&
  Hernquist}{Somerville et~al.}{2008a}]{Somerville2008}
Somerville R.~S.,  Hopkins P.~F.,  Cox T.~J.,  Robertson B.~E.,   Hernquist L.,
   2008a, \mn@doi [MNRAS] {10.1111/j.1365-2966.2008.13805.x}, 391, 481

\bibitem[\protect\citeauthoryear{Somerville et~al.,}{Somerville
  et~al.}{2008b}]{Somerville2008a}
Somerville R.~S.,  et~al., 2008b, \mn@doi [ApJ] {10.1086/523661}, 672, 776

\bibitem[\protect\citeauthoryear{Somerville, Popping  \& Trager}{Somerville
  et~al.}{2015}]{Somerville2015}
Somerville R.~S.,  Popping G.,   Trager S.~C.,  2015, \mn@doi [MNRAS]
  {10.1093/mnras/stv1877}, 453, 4338

\bibitem[\protect\citeauthoryear{Somerville et~al.,}{Somerville
  et~al.}{2018}]{Somerville2018}
Somerville R.~S.,  et~al., 2018, \mn@doi [MNRAS] {10.1093/mnras/stx2040}, 473,
  2714

\bibitem[\protect\citeauthoryear{Spergel et~al.,}{Spergel
  et~al.}{2003}]{Spergel2003}
Spergel D.~N.,  et~al., 2003, \mn@doi [ApJS] {10.1086/377226}, 148, 175

\bibitem[\protect\citeauthoryear{Springel}{Springel}{2005}]{Springel2005a}
Springel V.,  2005, \mn@doi [MNRAS] {10.1111/j.1365-2966.2005.09655.x}, 364,
  1105

\bibitem[\protect\citeauthoryear{Springel}{Springel}{2015}]{Springel2015}
Springel V.,  2015, Astrophys. Source Code Libr. Rec. ascl1502.003

\bibitem[\protect\citeauthoryear{Springel et~al.,}{Springel
  et~al.}{2005}]{Springel2005}
Springel V.,  et~al., 2005, \mn@doi [Nature] {10.1038/nature03597}, 435, 629

\bibitem[\protect\citeauthoryear{Tinker, Kravtsov, Klypin, Abazajian, Warren,
  Yepes, Gottl{\"{o}}ber  \& Holz}{Tinker et~al.}{2008}]{Tinker2008}
Tinker J.,  Kravtsov A.~V.,  Klypin A.,  Abazajian K.,  Warren M.,  Yepes G.,
  Gottl{\"{o}}ber S.,   Holz D.~E.,  2008, \mn@doi [ApJ] {10.1086/591439}, 688,
  709

\bibitem[\protect\citeauthoryear{Virtanen et~al.,}{Virtanen
  et~al.}{2020}]{Virtanen2020}
Virtanen P.,  et~al., 2020, \mn@doi [Nat. Methods] {10.1038/s41592-019-0686-2},
  17, 261

\bibitem[\protect\citeauthoryear{Visbal, Haiman  \& Bryan}{Visbal
  et~al.}{2018}]{Visbal2018}
Visbal E.,  Haiman Z.,   Bryan G.~L.,  2018, \mn@doi [MNRAS]
  {10.1093/mnras/sty142}, 475, 5246

\bibitem[\protect\citeauthoryear{Wechsler \& Tinker}{Wechsler \&
  Tinker}{2018}]{Wechsler2018}
Wechsler R.~H.,  Tinker J.~L.,  2018, \mn@doi [ARA&A]
  {10.1146/annurev-astro-081817-051756}, 56, 435

\bibitem[\protect\citeauthoryear{Wechsler, Bullock, Primack, Kravtsov  \&
  Dekel}{Wechsler et~al.}{2002}]{Wechsler2002}
Wechsler R.~H.,  Bullock J.~S.,  Primack J.~R.,  Kravtsov A.~V.,   Dekel A.,
  2002, \mn@doi [ApJ] {10.1086/338765}, 568, 52

\bibitem[\protect\citeauthoryear{White \& Frenk}{White \&
  Frenk}{1991}]{White1991}
White S. D.~M.,  Frenk C.~S.,  1991, \mn@doi [ApJ] {10.1086/170483}, 379, 52

\bibitem[\protect\citeauthoryear{Yang et~al.,}{Yang et~al.}{2023}]{Yang2023}
Yang G.,  et~al., 2023, \mn@doi [ApJL] {10.3847/2041-8213/acd639}, 950, L5

\bibitem[\protect\citeauthoryear{Yung, Somerville, Finkelstein, Popping  \&
  Dav{\'{e}}}{Yung et~al.}{2019a}]{Yung2019}
Yung L. Y.~A.,  Somerville R.~S.,  Finkelstein S.~L.,  Popping G.,   Dav{\'{e}}
  R.,  2019a, \mn@doi [MNRAS] {10.1093/mnras/sty3241}, 483, 2983

\bibitem[\protect\citeauthoryear{Yung, Somerville, Popping, Finkelstein,
  Ferguson  \& Dav{\'{e}}}{Yung et~al.}{2019b}]{Yung2019a}
Yung L. Y.~A.,  Somerville R.~S.,  Popping G.,  Finkelstein S.~L.,  Ferguson
  H.~C.,   Dav{\'{e}} R.,  2019b, \mn@doi [MNRAS] {10.1093/mnras/stz2755}, 490,
  2855

\bibitem[\protect\citeauthoryear{Yung, Somerville, Popping  \&
  Finkelstein}{Yung et~al.}{2020a}]{Yung2020}
Yung L. Y.~A.,  Somerville R.~S.,  Popping G.,   Finkelstein S.~L.,  2020a,
  \mn@doi [MNRAS] {10.1093/mnras/staa714}, 494, 1002

\bibitem[\protect\citeauthoryear{Yung, Somerville, Finkelstein, Popping,
  Dav{\'{e}}, Venkatesan, Behroozi  \& Ferguson}{Yung
  et~al.}{2020b}]{Yung2020a}
Yung L. Y.~A.,  Somerville R.~S.,  Finkelstein S.~L.,  Popping G.,  Dav{\'{e}}
  R.,  Venkatesan A.,  Behroozi P.,   Ferguson H.~C.,  2020b, \mn@doi [MNRAS]
  {10.1093/mnras/staa1800}, 496, 4574

\bibitem[\protect\citeauthoryear{Yung et~al.,}{Yung
  et~al.}{2021a}]{Yung_JWST2021}
Yung L. Y.~A.,  et~al., 2021a, JWST Propos. ID 2108 Cycle 1 AR/Theory

\bibitem[\protect\citeauthoryear{Yung, Somerville, Finkelstein, Hirschmann,
  Dav{\'{e}}, Popping, Gardner  \& Venkatesan}{Yung et~al.}{2021b}]{Yung2021}
Yung L. Y.~A.,  Somerville R.~S.,  Finkelstein S.~L.,  Hirschmann M.,
  Dav{\'{e}} R.,  Popping G.,  Gardner J.~P.,   Venkatesan A.,  2021b, \mn@doi
  [MNRAS] {10.1093/mnras/stab2761}, 508, 2706

\bibitem[\protect\citeauthoryear{Yung et~al.,}{Yung et~al.}{2022}]{Yung2022}
Yung L. Y.~A.,  et~al., 2022, \mn@doi [MNRAS] {10.1093/mnras/stac2139}, 515,
  5416

\bibitem[\protect\citeauthoryear{Yung et~al.,}{Yung et~al.}{2023}]{Yung2023}
Yung L. Y.~A.,  et~al., 2023, \mn@doi [MNRAS] {10.1093/mnras/stac3595}, 519,
  1578

\bibitem[\protect\citeauthoryear{Yung, Somerville, Finkelstein, Wilkins  \&
  Gardner}{Yung et~al.}{2024}]{Yung2023a}
Yung L. Y.~A.,  Somerville R.~S.,  Finkelstein S.~L.,  Wilkins S.~M.,   Gardner
  J.~P.,  2024, \mn@doi [MNRAS] {10.1093/mnras/stad3484}, 527, 5929

\bibitem[\protect\citeauthoryear{Zhang, Behroozi, Volonteri, Silk, Fan,
  Hopkins, Yang  \& Aird}{Zhang et~al.}{2023}]{Zhang2023}
Zhang H.,  Behroozi P.,  Volonteri M.,  Silk J.,  Fan X.,  Hopkins P.~F.,  Yang
  J.,   Aird J.,  2023, \mn@doi [MNRAS] {10.1093/mnras/stac2633}, 518, 2123

\bibitem[\protect\citeauthoryear{Zhao, Jing, Mo  \& Brner}{Zhao
  et~al.}{2003}]{Zhao2003}
Zhao D.~H.,  Jing Y.~P.,  Mo H.~J.,   Brner G.,  2003, \mn@doi [ApJ]
  {10.1086/379734}, 597, L9

\bibitem[\protect\citeauthoryear{Zhao, Jing, Mo  \& B{\"{o}}rner}{Zhao
  et~al.}{2009}]{Zhao2009}
Zhao D.~H.,  Jing Y.~P.,  Mo H.~J.,   B{\"{o}}rner G.,  2009, \mn@doi [ApJ]
  {10.1088/0004-637X/707/1/354}, 707, 354

\bibitem[\protect\citeauthoryear{van~der Walt, Colbert  \& Varoquaux}{van~der
  Walt et~al.}{2011}]{VanderWalt2011}
van~der Walt S.,  Colbert S.~C.,   Varoquaux G.,  2011, \mn@doi [Comput. Sci.
  Eng.] {10.1109/MCSE.2011.37}, 13, 22

\makeatother
\end{thebibliography}




\appendix

\section{Fitting parameters to high-redshift halo mass function}
\label{sec:AppA}

We provide best-fit parameters for simulated ultra- high $z$ HMFs following the same parameterization from \citet{Tinker2008} and \citet{Rodriguez-Puebla2016}. The comoving number density of halos of mass between $M_\text{vir}$ + $dM_\text{vir}$ is given by

\begin{equation}
	\frac{dn_\text{h}}{dM_\text{vir}} = f(\sigma) \frac{\bar{\rho}_\text{m}}{M_\text{vir}^2} \left| \frac{d\ln \sigma^{-1}}{d\ln M_\text{vir}} \right| \text{,}
\end{equation}
where $\bar{\rho}_\text{m}$ is the critical matter density in the Universe, $\sigma$ is the amplitude of the perturbations, and $f(\sigma)$ is called the halo multiplicity function, which takes the form of

\begin{equation}
	f(\sigma) = A \left[ \left(\frac{\sigma}{b}\right)^{-a} + 1\right] e^{-c/\sigma^2} \text{,}
\end{equation}
where $\chi_{i} = A$, $a$, $b$, and $c$ are free parameters, given by $\chi_{i} = \chi_{0,i} + \chi_{1,i}z + \chi_{2,i}z^2$ with the best-fit parameters fitted to \gureft+MultiDark HMFs between $z = 6$ to 19.  
Here, $\sigma \equiv \sigma(M_\text{vir})D(z)$, where $D(z)$ is the linear growth-rate factor.
We have also updated the approximation for $\sigma(M_\text{vir})$ using the same functional form given by \citet{Rodriguez-Puebla2016} to fit to a wider halo mass range of $6 < \log(M_\text{vir}/\text{M}_\odot h^{-1}) < 13$:
\begin{equation}
    \sigma(M_\text{vir}) = \frac{22.87988436y^{0.4069515}}{1 + 5.65110544y^{0.23076429} + 4.02321628y^{0.36760932}} \text{,}
\end{equation}
where $ y \equiv 1/M_\text{vir,12}$ and $M_\text{vir,12} \equiv M_\text{vir}/(10^{12}$ \Msun\ $h^{-1}$).
We note that the HMF fitting parameters provided by previous studies were fitted to HMFs in conventional simulation units (e.g. masses in M$_\odot\,$h$^{-1}$ and distances in Mpc\,h$^{-1}$). However, we find it more helpful to also show fits for HMFs in physical units (e.g. masses in M$_\odot$ and distances in Mpc). 
For $\sigma(M_\text{vir})$ fitting function with no $h$ scaling: 
\begin{equation}
    \sigma(M_\text{vir}) = \frac{26.80004233y^{0.40695158}}{1 + 6.18130098y^{0.23076433} + 4.64104008y^{0.36760939}} \text{,}
\end{equation}
where $ y \equiv 1/M_\text{vir,12}$ and $M_\text{vir,12} \equiv M_\text{vir}/(10^{12}$ \Msun).
Here we present fitted parameters for both h scaled simulation units and physical units in Table~\ref{tab:hmf_fit_parm} and Table~\ref{tab:hmf_fit_parm_physical}, respectively.

\begin{figure}
    \includegraphics[width=\columnwidth]{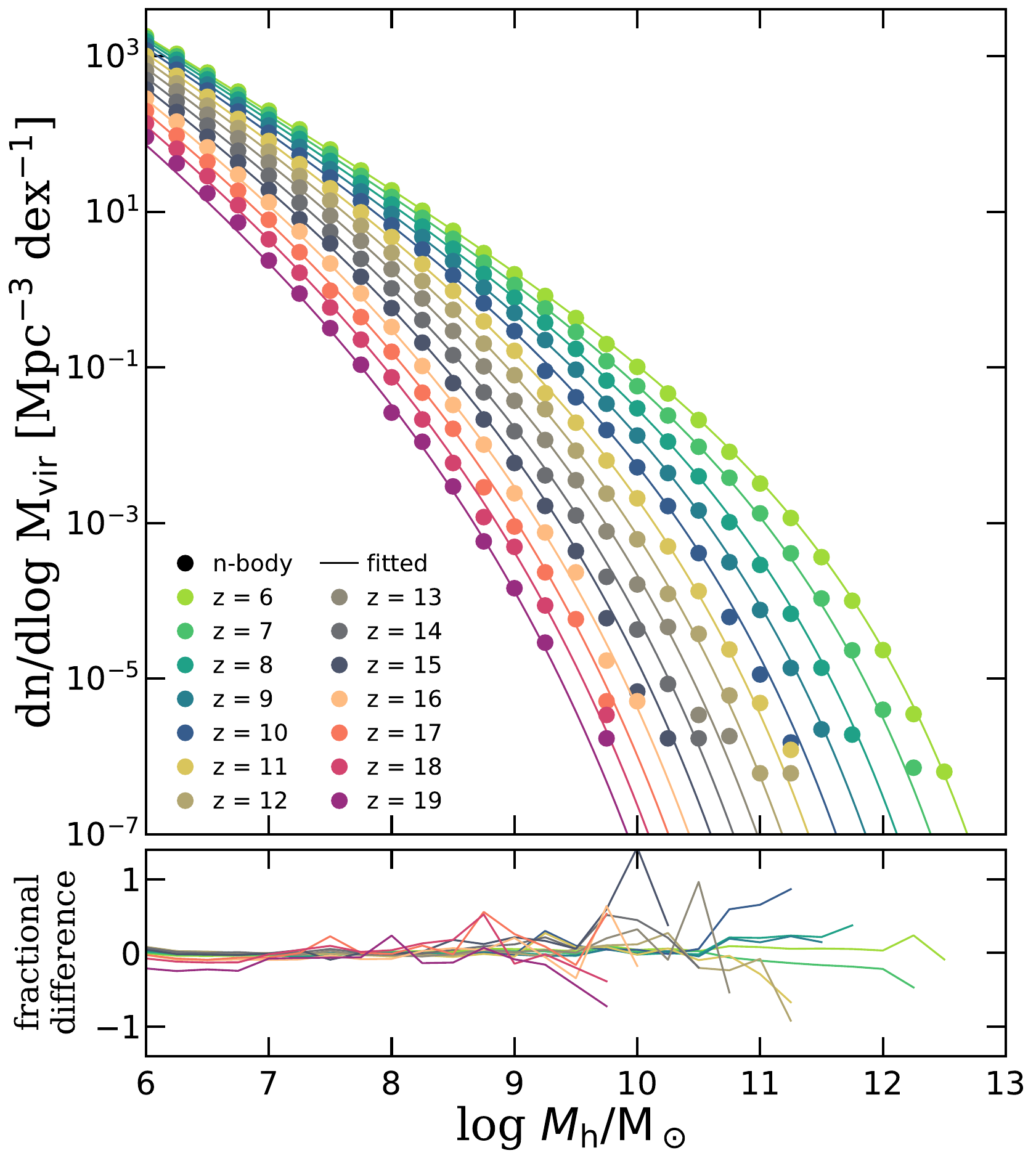}
    \caption{
        \textit{Top panel:} Halo mass function from $n$-body simulations (data points; \gureft + MultiDark, see Fig.~\ref{fig:hmf} and associated text) and fitting functions (solid lines) between $z = 6$ and 19. The lines and data points are colour-coded by redshift. \textit{Bottom panel:} The fractional differences between the $n$-body HMFs and fitting functions ($\phi_\text{fitted} - \phi_\text{n-body})/\phi_\text{n-body}$.
    }
    \label{fig:hmf_fitted}
\end{figure}

\begin{table}
    \centering
    \caption{
        Fitting parameters for the combined \gureft+MultiDark HMFs fitted in conventional simulation units of [M$_\odot\, h^{-1}$] and [Mpc$^{-3}\,h^3\,$dex$^{-1}$].
    }
    \label{tab:hmf_fit_parm}
    \begin{tabular}{cccc}
        \hline
        $\chi_{i}$ & $\chi_{0,i}$ & $\chi_{1,i}$ & $\chi_{2,i}$  \\
        \hline
        A & $0.11416632$ & $-0.01486746$ & $ 0.00137191$   \\
        a & $1.05274399$ & $ 0.02803087$ & $-0.00306126$   \\
        b & $8.62813020$ & $ 0.00384969$ & $-0.02349983$   \\
        c & $1.13138924$ & $ 0.01713172$ & $-0.00113630$   \\
        \hline
    \end{tabular}
\end{table}

\begin{table}
    \centering
    \caption{
        Fitting parameters for the combined \gureft+MultiDark HMFs fitted in physical units of [$M_\odot$] and [Mpc$^{-3}$ dex$^{-1}$].
    }
    \label{tab:hmf_fit_parm_physical}
    \begin{tabular}{cccc}
        \hline
        $\chi_{i}$ & $\chi_{0,i}$ & $\chi_{1,i}$ & $\chi_{2,i}$  \\
        \hline
        A & $0.13765772$ & $-0.01003821$ & $ 0.00102964$   \\
        a & $1.06641384$ & $ 0.02475576$ & $-0.00283342$   \\
        b & $4.86693806$ & $ 0.09212356$ & $-0.01426283$   \\
        c & $1.19837952$ & $-0.00142967$ & $-0.00033074$   \\
        \hline
    \end{tabular}
\end{table}

\section{Effective mass range of \gureft\ boxes for scaling relations }
\label{sec:AppB}

\begin{figure*}
    \centering
    \begin{subfigure}[b]{0.33\textwidth}
        \centering
        \includegraphics[width=\textwidth]{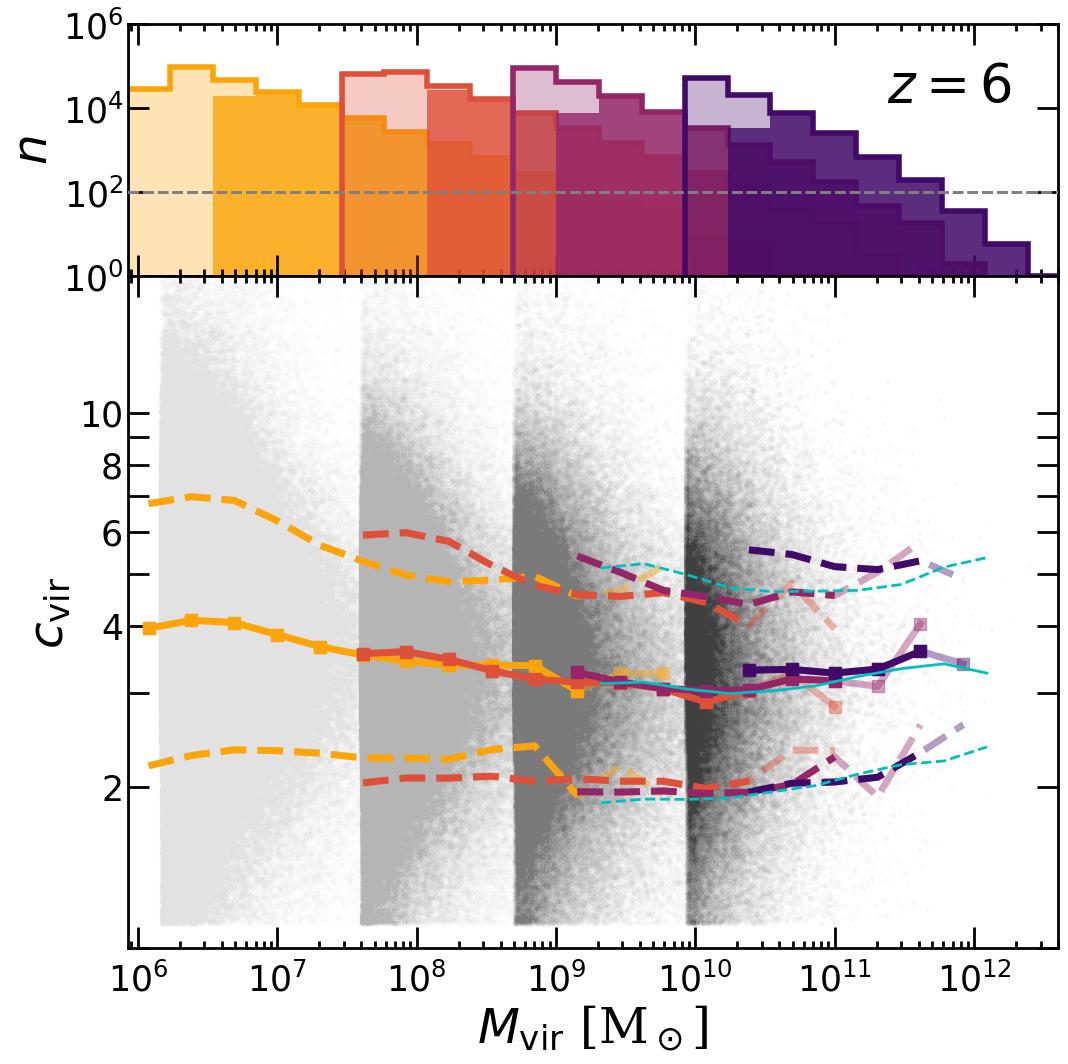}
    \end{subfigure}
    \begin{subfigure}[b]{0.33\textwidth}
        \centering
        \includegraphics[width=\textwidth]{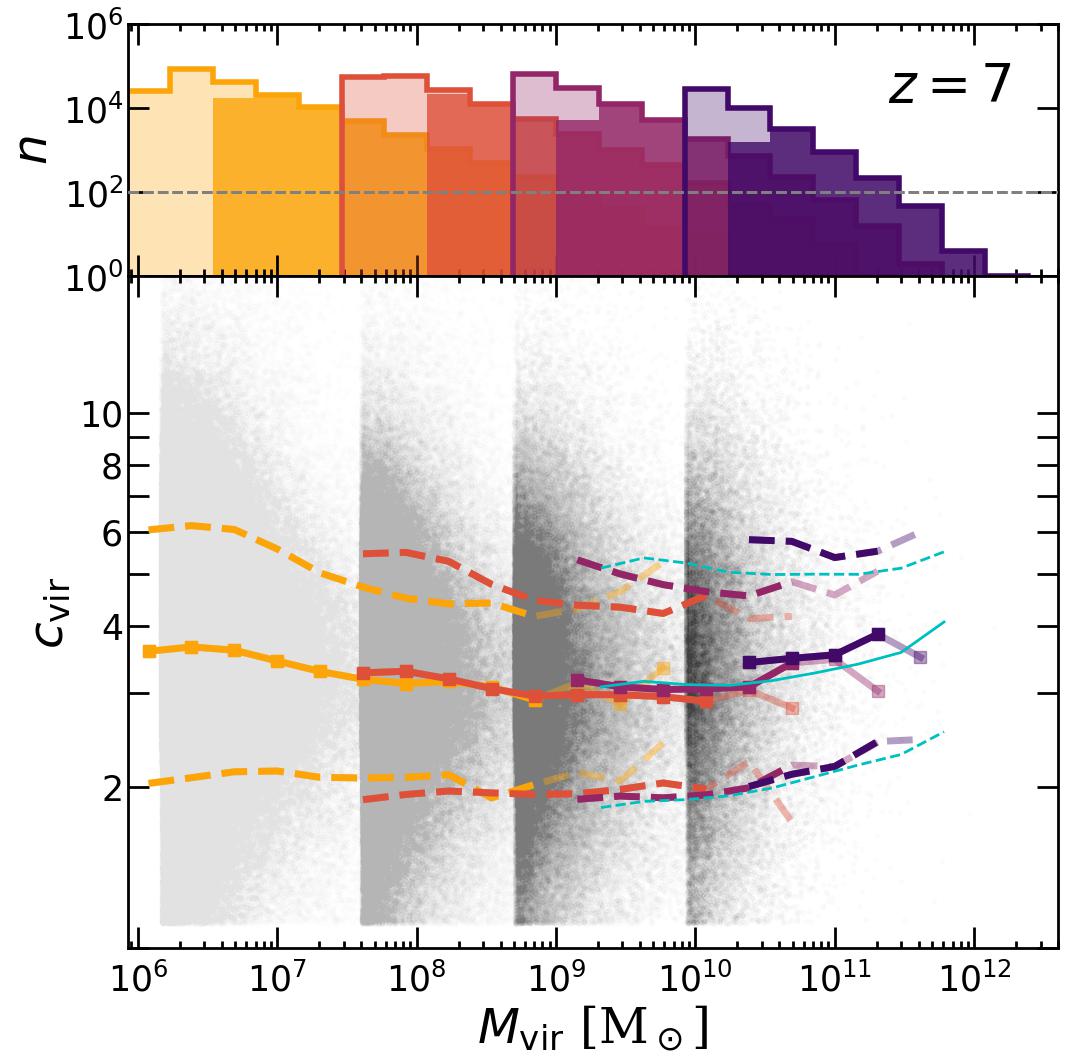}
    \end{subfigure}
    \begin{subfigure}[b]{0.33\textwidth}
        \centering
        \includegraphics[width=\textwidth]{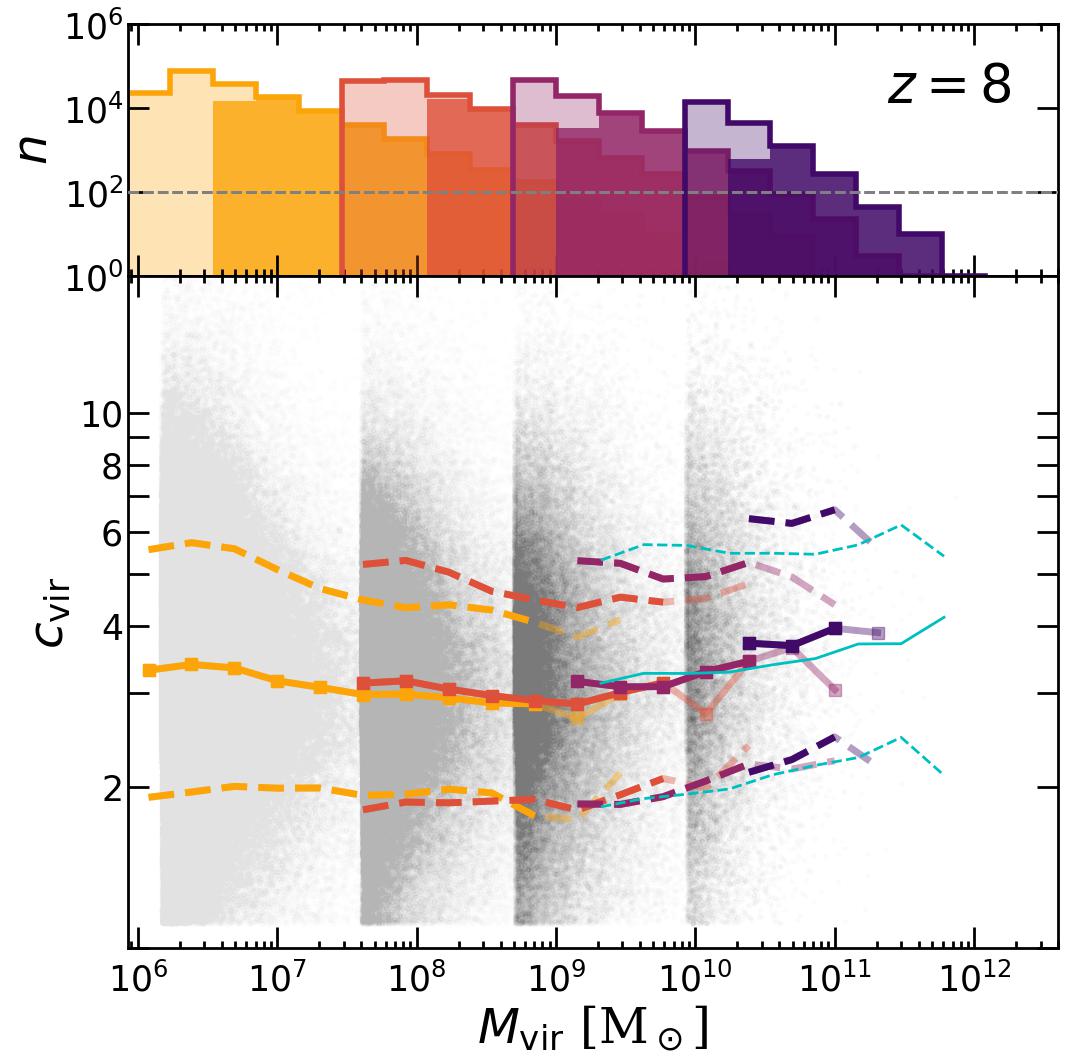}
    \end{subfigure}
    \begin{subfigure}[b]{0.33\textwidth}
        \centering
        \includegraphics[width=\textwidth]{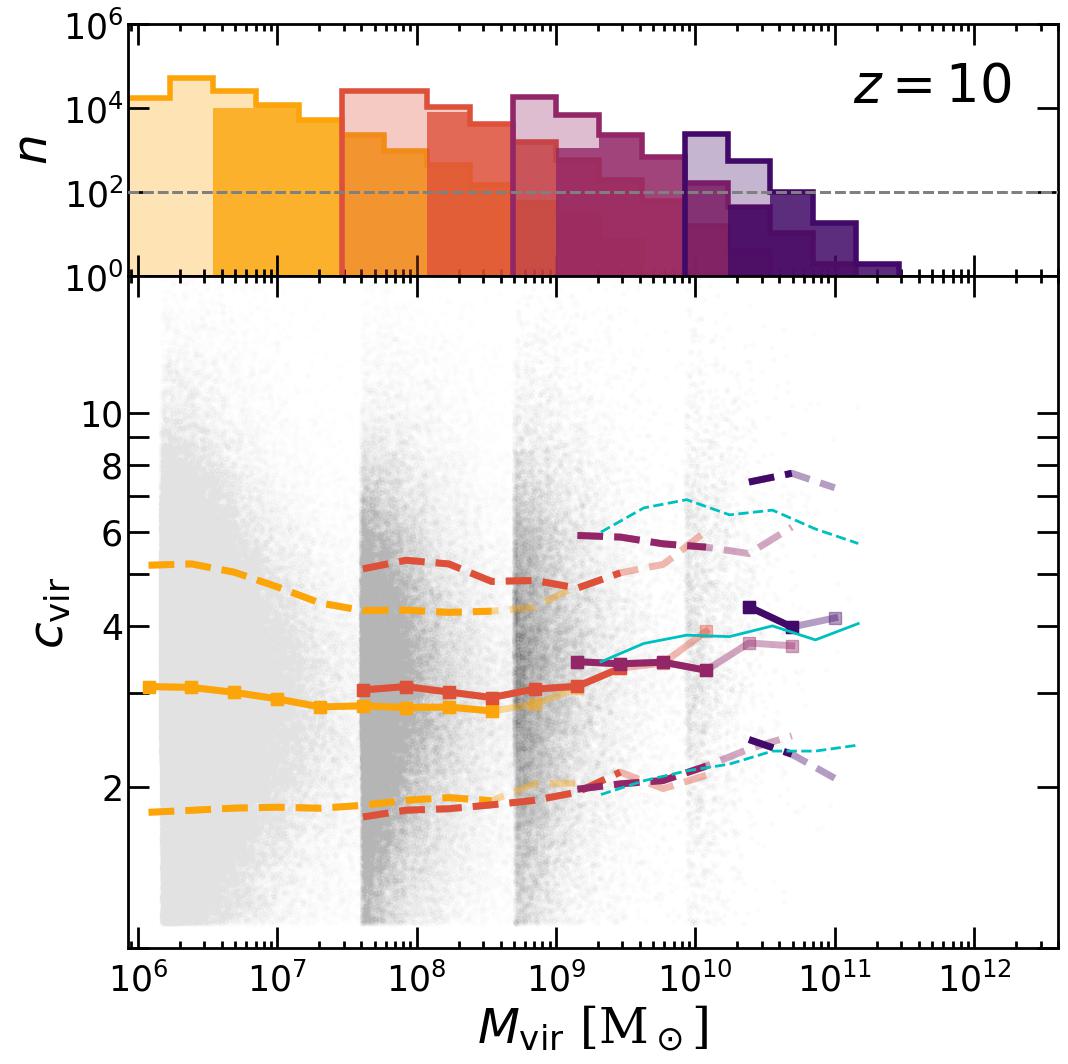}
    \end{subfigure}
    \begin{subfigure}[b]{0.33\textwidth}
        \centering
        \includegraphics[width=\textwidth]{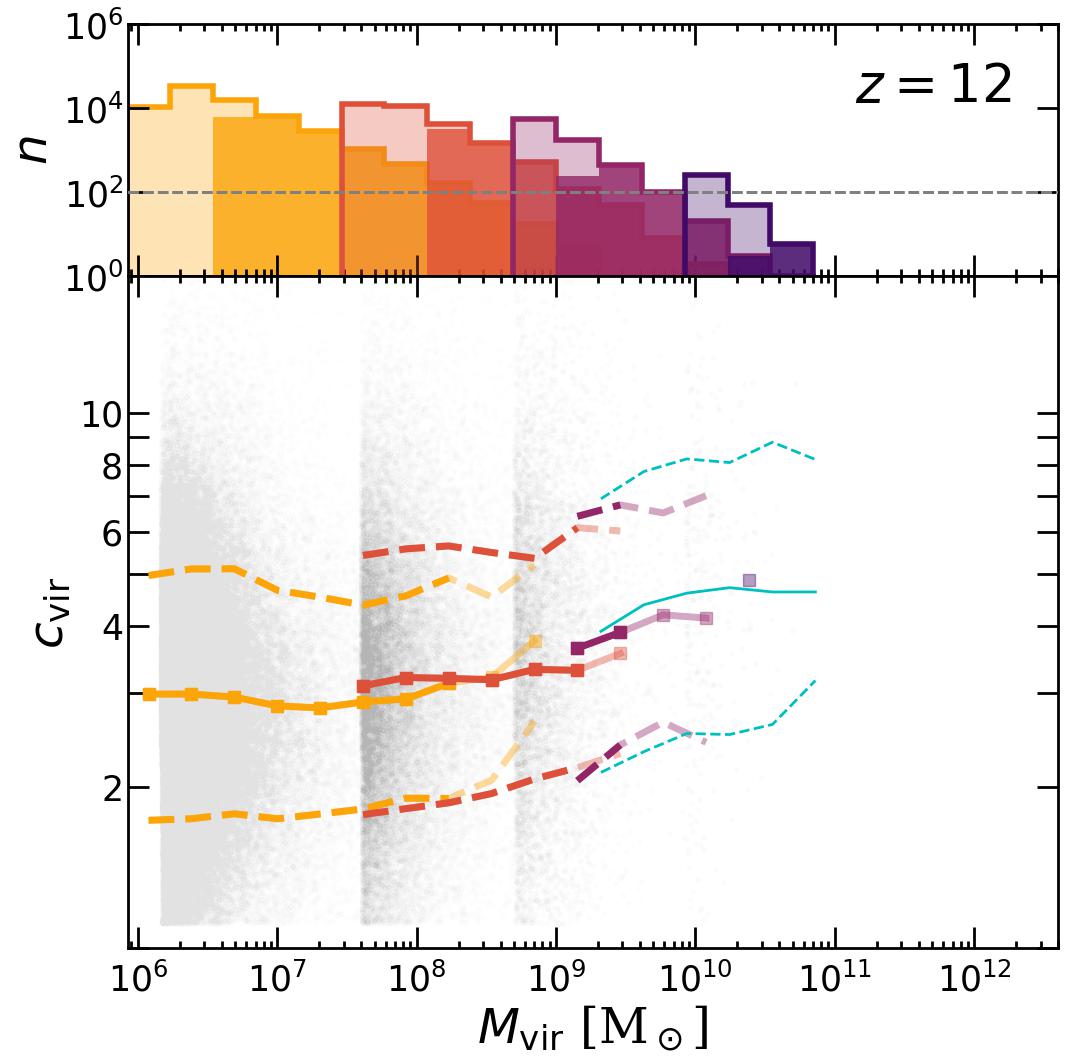}
    \end{subfigure}
    \begin{subfigure}[b]{0.33\textwidth}
        \centering
        \includegraphics[width=\textwidth]{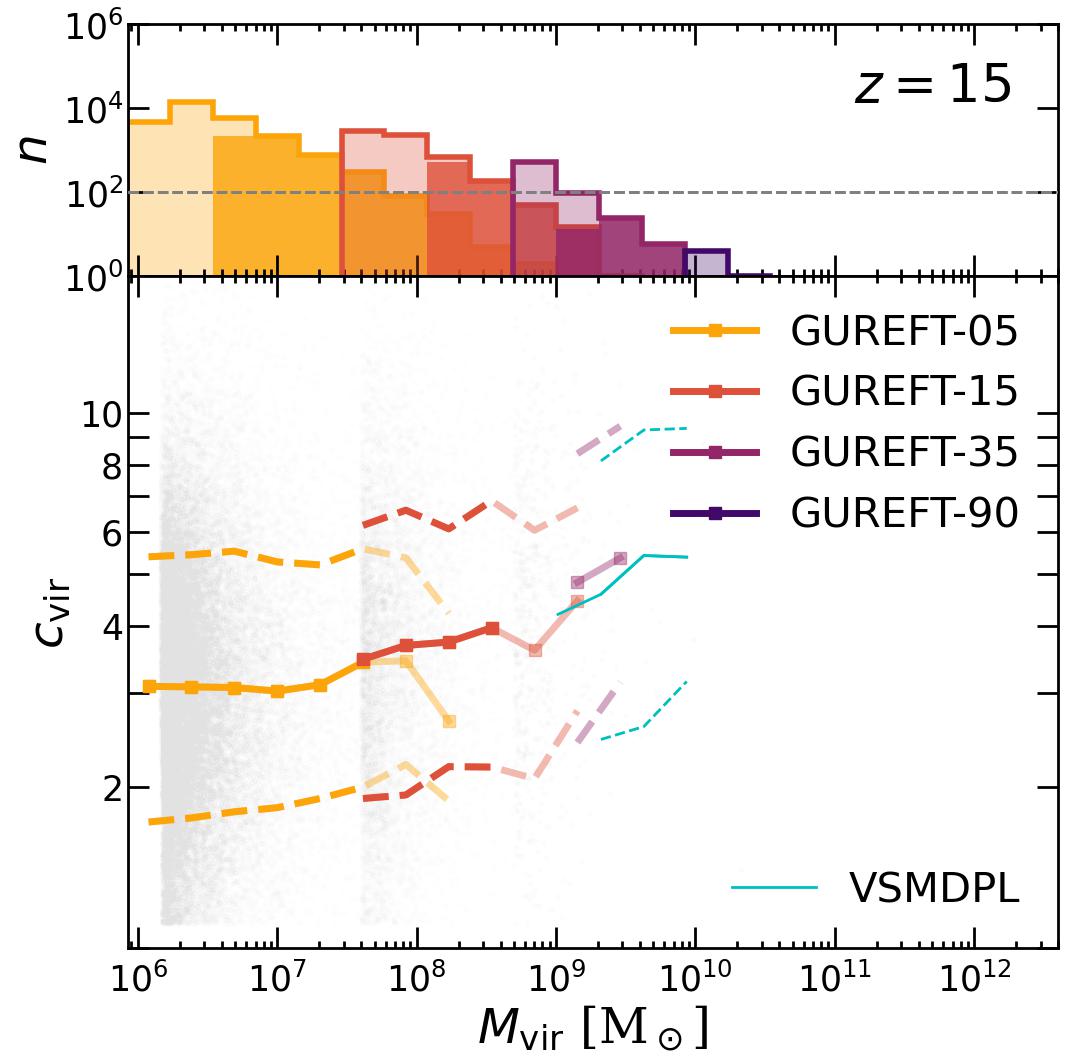}
    \end{subfigure}
    \caption{
        This set of plots shows the convergence diagnostics for simulated halo populations found in the suite of \gureft\ boxes at $z = 6, 7, 8, 10, 12, 15$. The \textit{top panels} of these plots show histograms simply show counts of halos found in \gureft\ (not normalised to binsize or simulated volume). These histograms include halos identified with at least 100 dark matter particles (shown in semi-transparent colours) and at least 500 particles (shown in solid colours). The horizontal grey dashed line marks where the halo count equals 100. The \textit{bottom panels} show the predicted $M_\text{vir}$--$c_\text{vir}$ relations for halos from \gureft, shown in matching colour as the histograms above. The median and 16th and 84th percentiles are computed in bins matching the histograms and are marked by the solid and dashed lines, respective. These lines turn semi-transparent for bins containing few than 100 halos. The underlying distribution of data points from \gureft\ are shown in distinct shades of grey.
    }
    \label{fig:gureft_stitch}
\end{figure*}

In this work, we presented selected scaling relations for key physical properties for halos from the suite of \gureft\ simulations. With different simulated volume and mass resolution, these boxes are expected to cover different mass ranges, with the low-mass end limited by the mass of DM particles and the number of particles required to resolve a halo (a 100-particle threshold is adopted in this work) and the massive end limited by the simulated volume.
In this appendix, we provide in-depth diagnostics of the behaviour of the halo populations in the \gureft\ boxes and explain the decisions made in combining the scaling relations predictions as presented in \ref{sec:properties}.

We experiment with the threshold on DM particle number used to consider a halo resolved to assess its impact on the predicted physical quantities. In Fig.~\ref{fig:gureft_stitch}, we show the median scatter of the $M_\text{vir}$--$c_\text{vir}$ at various redshifts for all \gureft\ volumes. The halos included in this plot contain $\geq 100$ dark matter particles. The histogram in the top of each panel can be used to indicate the 500-particle threshold, which occurs when the histograms transition from solid to semi-transparent. For \gureft-15, -35, and -90, the mass range in between 100 and 500 particles can be cross-checked with the massive-end of \gureft-05, -15, and -35, respectively, between $z = 6$ to 10, where there are sufficient halos to predict the $M_\text{vir}$--$c_\text{vir}$ relation. Given that there are no significant resolution effects observed in the $M_\text{vir}$--$c_\text{vir}$ relation below the 500-particle threshold, we can safely conclude that halos with 100 particles are sufficient for \textsc{rockstar} to reliably measure $c_\text{vir}$. In an internal test, we find that the $c_\text{vir}$ measured by \textsc{rockstar} demonstrates a sharp jump for halos identified with $\lesssim 85$ particles, where these halos have systematically higher $R_\text{s}$. This indicates that $c_\text{vir}$ is cannot be accurately measured for halos made up of fewer DM particles.

In this experiment, we also find that the sample size has a significant impact on the scatter in the $M_\text{vir}$--$c_\text{vir}$ relation. It is clear that when the low-mass end of a larger box overlaps with the massive end of a smaller box, even though both halo populations are considered well-resolved, scatter in the predicted $M_\text{vir}$--$c_\text{vir}$ relation seems to be affected by the sample size available, given that the massive end of a smaller box would contain significantly fewer halos than the low-mass end of a larger box. This discrepancy systematically persists across all boxes and redshifts. We also note that this is a consequence that follows the halo populations that naturally occur in cosmological simulations, where there are fewer massive halos than low-mass halos because of hierarchical structure formation. Since we also find that the low-mass end of a larger box can more reliably predict scaling relations than the massive-end of a smaller box, we prioritise results from the smaller box over the mass range where two boxes overlap.

\section{Fitting the distribution of spin}
\label{sec:AppC}

The distribution of the halo spin parameter at $z\sim0$ has been shown the be well-fitted with a modified log-normal distribution \citep{Bullock2001a} and a Schechter-like function \citetalias{Rodriguez-Puebla2016}. 
Here, we provide fitting functions and parameters for the high- to ultra-high-redshift spin distributions with a non-linear least-squares method.
Following the approach of \citet{Bullock2001a}, we fit the distribution with a log-normal distribution function 
\begin{equation}
    P(\log\lambda) = \frac{1}{\lambda\sqrt{2\pi\sigma^2}} \exp\left( -\frac{\log^2(\lambda/\lambda_0))}{2\sigma^2}\right)\text{,}
\end{equation}
where $\log$ denotes natural log.
We note that this is different from the (modified) log-normal distribution adopted by \citetalias{Rodriguez-Puebla2016} 
\begin{equation}
    P(\log\lambda) = \frac{1}{\sqrt{2\pi\sigma^2}} \exp\left( -\frac{\log_{10}^2(\lambda/\lambda_0))}{2\sigma^2}\right)\text{,}
\end{equation}
which differs by a factor of $1/\lambda$ and $\log_{10}$ is used instead of natural log. Thus, the fitting parameters provided in Table \ref{tab:spin_hist_fit_logN} are not directly comparable to the ones presented in table 7 in \citetalias{Rodriguez-Puebla2016}. We also adopted a Schechter-like fit similar to the one presented by \citetalias{Rodriguez-Puebla2016}
\begin{equation}
    P(\log\lambda) = \frac{f(\lambda)}{\int_{-\infty}^\infty \! f(\lambda) \, \mathrm{d}\lambda}\text{,}
\end{equation}
where
\begin{equation}
    f(\lambda) = \left( \frac{\lambda}{\lambda_0} \right)^\alpha \exp \left[ -\left(\frac{\lambda}{\lambda_0}\right)^{\beta} \right] \text{.}
\end{equation}
Here, we modified the sign of the exponent from $-\alpha$ to $\alpha$. We find that integrating $f(\lambda)$ in its original form as presented in \citetalias{Rodriguez-Puebla2016} with $\alpha > 0$ does not converge.
We note that this modification is also needed in order to reproduce the results presented in fig.~21 in \citetalias{Rodriguez-Puebla2016}. The best-fit parameters using this Schechter-like fitting function are presented in Table \ref{tab:spin_hist_fit_schechter}.
We also note that the colour-label of fig.~21 in \citetalias{Rodriguez-Puebla2016} should be blue (black) for $\lambda_\text{B}$ ($\lambda_\text{P}$), opposite to the caption indicated.

In this Appendix, we explore fitting the distribution of $\lambda$ at higher redshifts. Similar to the findings reported by \citetalias{Rodriguez-Puebla2016}, the log-normal distribution consistently under fits the distribution of low $\lambda$. On the other hand, the log-normal fit seems to reproduce the distribution of high $\lambda$ for both $\lambda_\text{P}$ and $\lambda_\text{B}$. Overall, the Schechter-like fit is able to well-represent the overall distribution of $\lambda_\text{B}$.

\begin{figure*}
    \centering
    \begin{subfigure}[b]{0.49\textwidth}
        \centering
        \includegraphics[width=\textwidth]{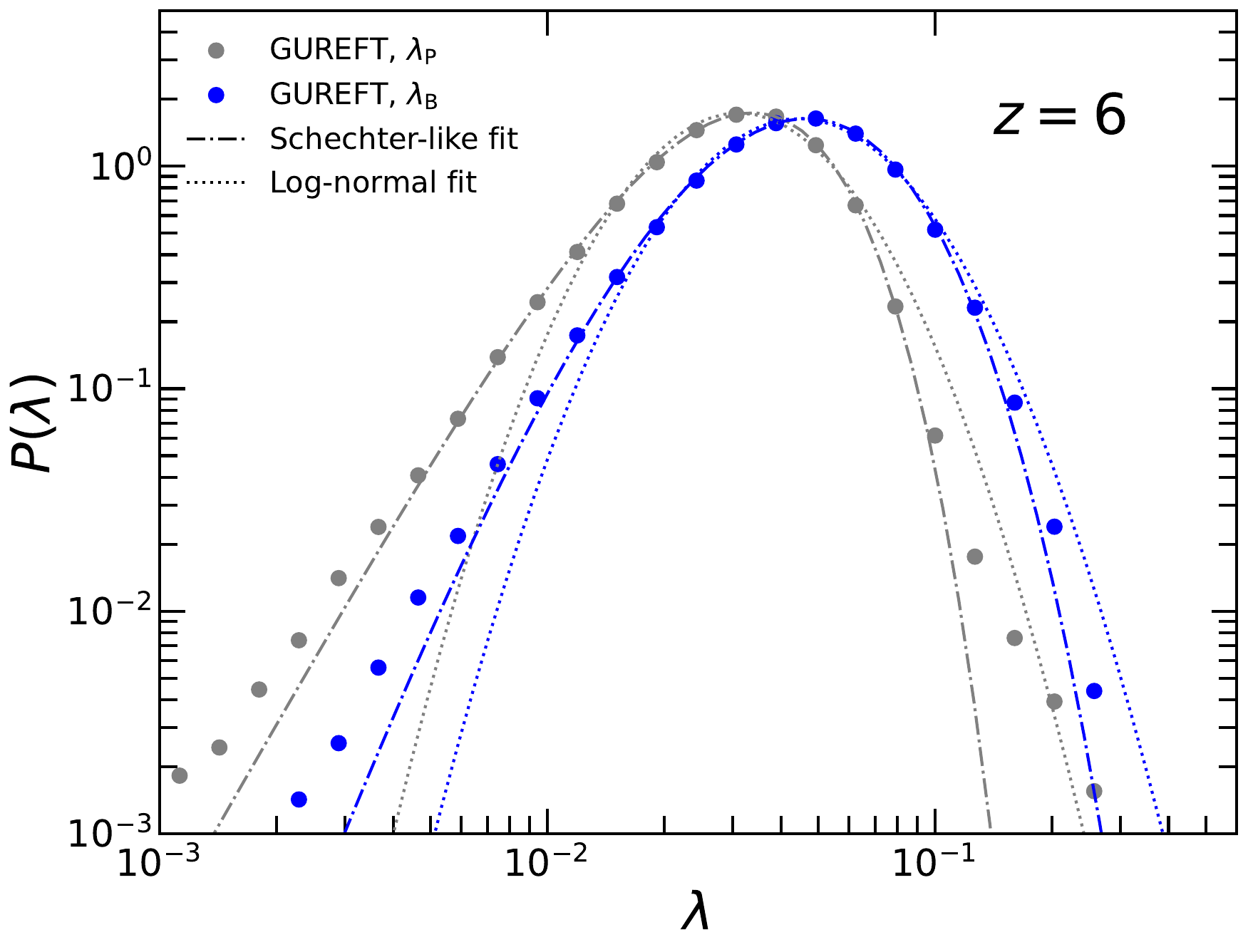}
    \end{subfigure}
    \hspace{0.01in}
    \begin{subfigure}[b]{0.48\textwidth}
        \centering
        \includegraphics[width=\textwidth]{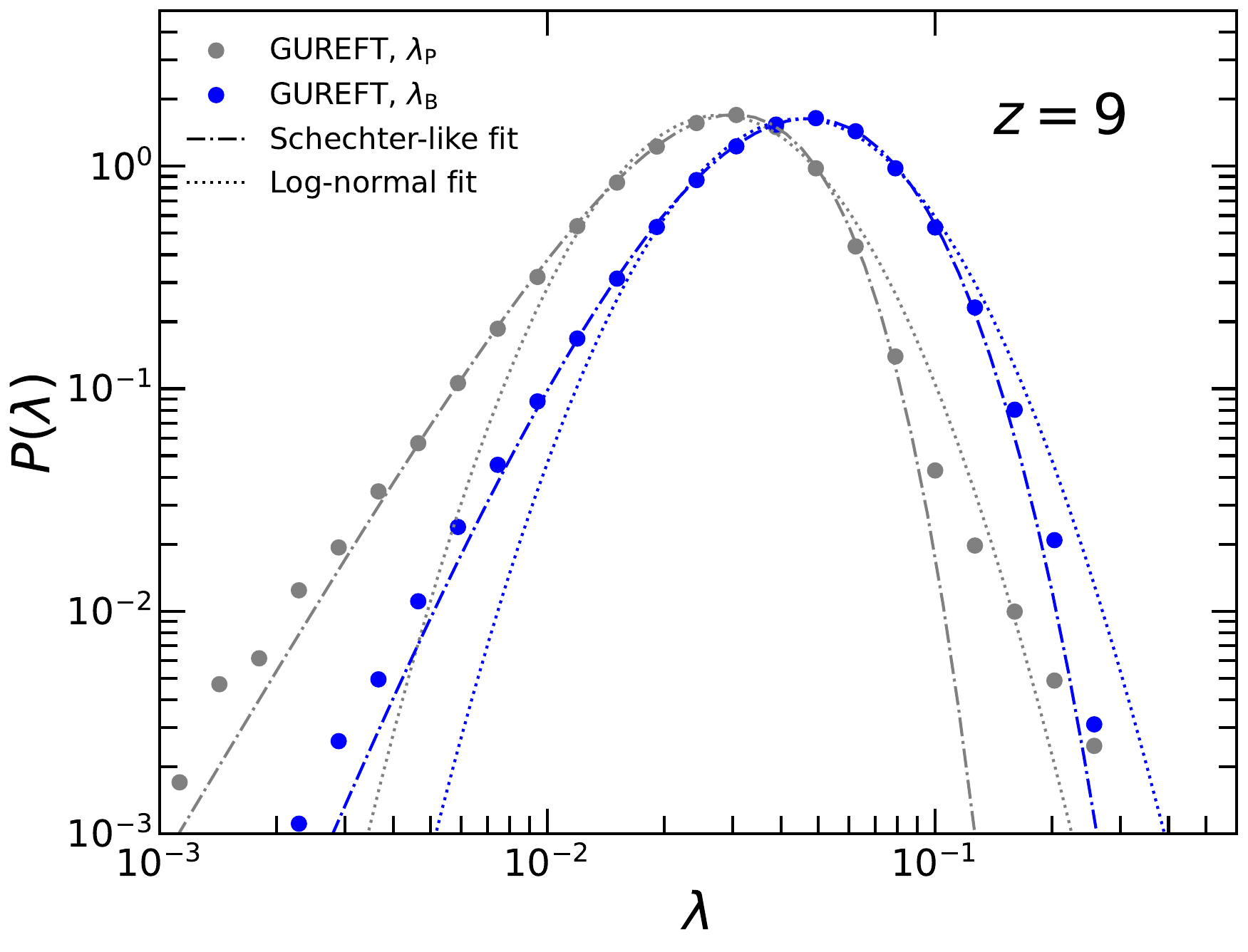}
    \end{subfigure}
    \hspace{0.01in}
    \begin{subfigure}[b]{0.48\textwidth}
        \centering
        \includegraphics[width=\textwidth]{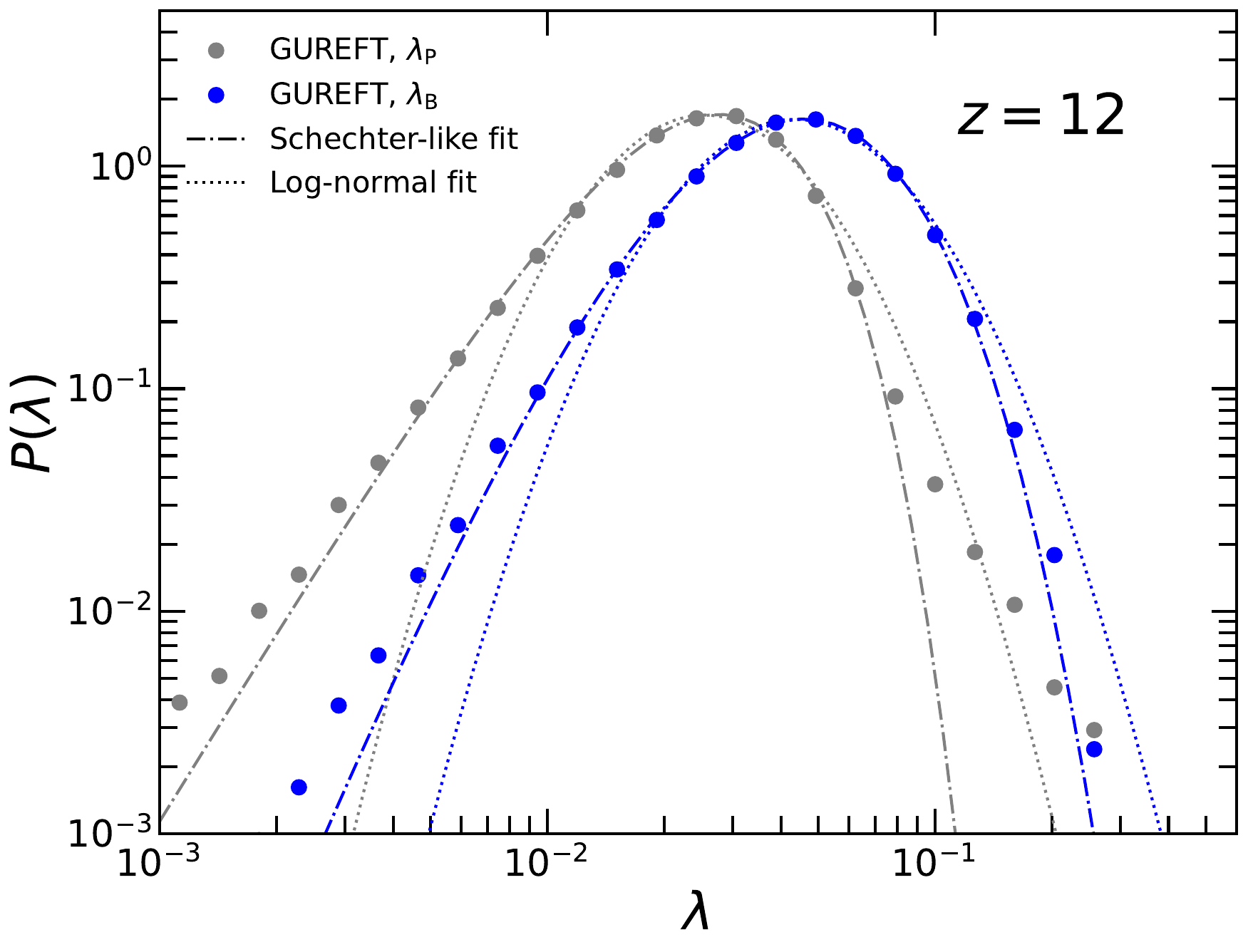}
    \end{subfigure}
    \hspace{0.01in}
    \begin{subfigure}[b]{0.48\textwidth}
        \centering
        \includegraphics[width=\textwidth]{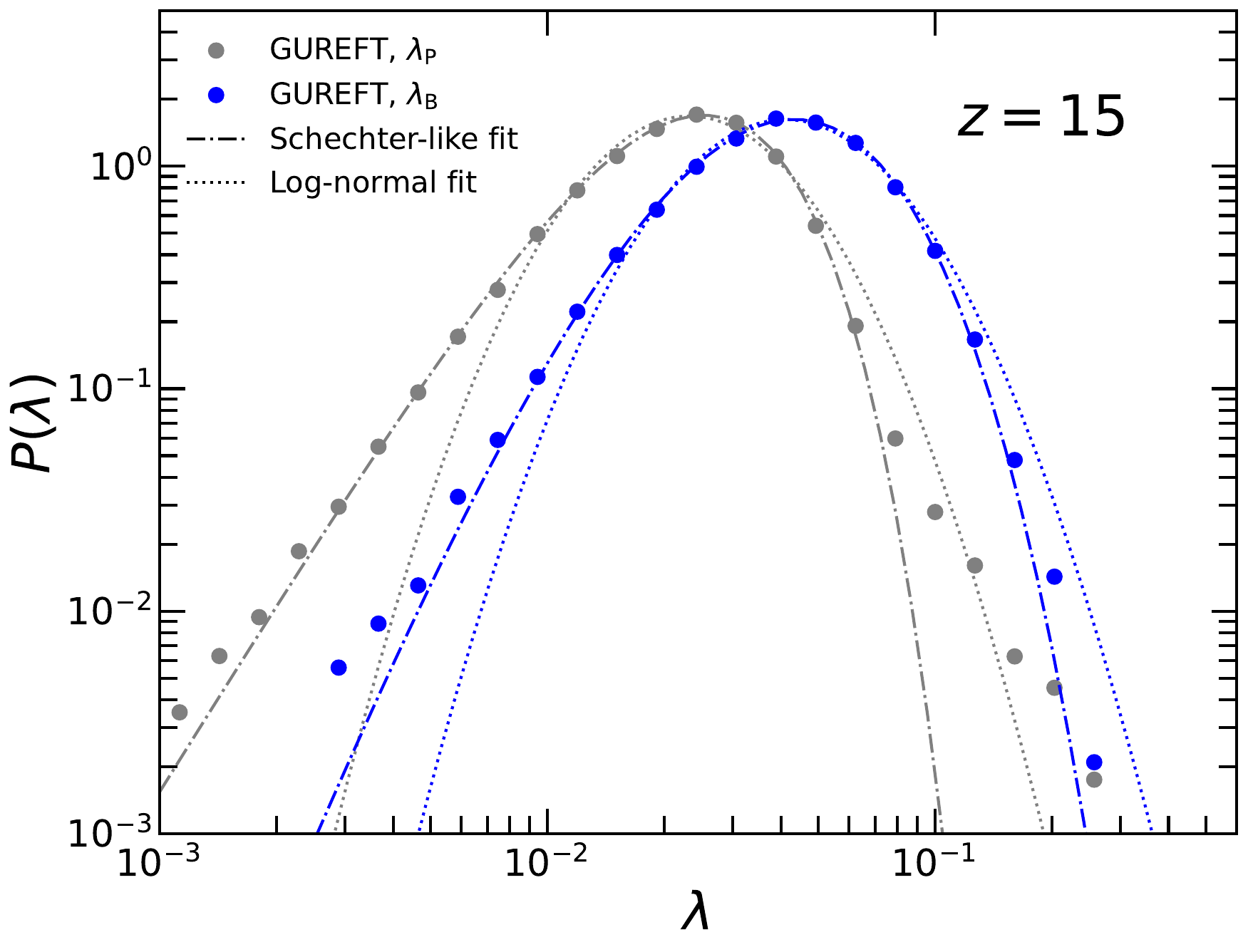}
    \end{subfigure}
    \caption{Halo spin distribution for halos in the \gureft\ simulations at $z = 6$, 9, 12, and 15, where $\lambda_\text{B}$ and $\lambda_\text{P}$ are denoted by blue and grey, respectively. The data points shows the probability distribution function from the simulated halos, and fitted distribution with a log-normal and Schechter-like function are shown by dotted and dot-dashed lines, respectively.
    }
    \label{fig:gureft_Spin}
\end{figure*}

\begin{table}
    \centering
    \caption{Best-fitting parameters for the distribution of $\lambda_\text{P}$ and $\lambda_\text{B}$ using the log-normal distribution function between $z = 6$ to 19.}
    \label{tab:spin_hist_fit_logN}
    \begin{tabular}{ccccc}
        \hline
        & \multicolumn{2}{c}{fitting for $P(\lambda_\text{P})$} & \multicolumn{2}{c}{fitting for $P(\lambda_\text{B})$}\\
        $z$ & $\sigma_\text{P}$ & $\log_{10}\lambda_\text{0,P}$ &  $\sigma_\text{B}$ & $\log_{10}\lambda_\text{0,B}$\\
        \hline
        6 & 0.5310 & -1.5067 & 0.5622 & -1.3517\\
7 & 0.5355 & -1.5231 & 0.5619 & -1.3475\\
8 & 0.5388 & -1.5395 & 0.5621 & -1.3450\\
9 & 0.5420 & -1.5552 & 0.5624 & -1.3483\\
10 & 0.5390 & -1.5677 & 0.5631 & -1.3517\\
11 & 0.5427 & -1.5808 & 0.5657 & -1.3562\\
12 & 0.5411 & -1.5944 & 0.5656 & -1.3617\\
13 & 0.5399 & -1.6063 & 0.5642 & -1.3682\\
14 & 0.5419 & -1.6221 & 0.5710 & -1.3773\\
15 & 0.5466 & -1.6345 & 0.5669 & -1.3862\\
16 & 0.5408 & -1.6462 & 0.5705 & -1.3926\\
17 & 0.5399 & -1.6582 & 0.5674 & -1.4001\\
18 & 0.5374 & -1.6634 & 0.5642 & -1.4078\\
        \hline
    \end{tabular}
\end{table}

\begin{table}
    \centering
    \caption{Best-fitting parameters for the distribution of $\lambda_\text{P}$ and $\lambda_\text{B}$ using the Schechter-like distribution function between $z = 6$ to 18.}
    \label{tab:spin_hist_fit_schechter}
    \begin{tabular}{ccccccc}
        \hline
        & \multicolumn{3}{c}{fitting for $P(\lambda_\text{P})$} & \multicolumn{3}{c}{fitting for $P(\lambda_\text{B})$}\\
        $z$ & $\alpha_\text{P}$ & $\beta_\text{P}$ & $\log_{10}\lambda_\text{0,P}$ &  $\alpha_\text{B}$ &  $\beta_\text{B}$ & $\log_{10}\lambda_\text{0,B}$\\
        \hline
        6 & 2.1144 & 1.2226 & -1.8039 & 4.1486 & 0.6279 & -2.7865\\
7 & 2.0305 & 1.2399 & -1.8002 & 3.9323 & 0.6577 & -2.6564\\
8 & 2.0705 & 1.2052 & -1.8414 & 3.9215 & 0.6582 & -2.6509\\
9 & 1.9712 & 1.2387 & -1.8252 & 3.6211 & 0.7037 & -2.4868\\
10 & 2.0256 & 1.2248 & -1.8529 & 3.8948 & 0.6596 & -2.6497\\
11 & 1.9453 & 1.2478 & -1.8432 & 4.1280 & 0.6223 & -2.8074\\
12 & 1.8490 & 1.3056 & -1.8161 & 3.6061 & 0.6971 & -2.5149\\
13 & 1.9517 & 1.2585 & -1.8642 & 3.5591 & 0.7095 & -2.4836\\
14 & 1.9173 & 1.2647 & -1.8725 & 3.3529 & 0.7261 & -2.4237\\
15 & 1.8367 & 1.2879 & -1.8624 & 3.6823 & 0.6820 & -2.5905\\
16 & 2.0303 & 1.2158 & -1.9373 & 3.8166 & 0.6533 & -2.6986\\
17 & 2.2248 & 1.1333 & -2.0254 & 5.1768 & 0.5088 & -3.5139\\
18 & 1.9551 & 1.2694 & -1.9170 & 3.9704 & 0.6471 & -2.7550\\
        \hline
    \end{tabular}
\end{table}

\section{\textsc{gureft} Pilot Simulations}
\label{sec:AppD}

Prior to carrying out the full suite of \gureft\ simulations, we ran a set of pilot simulations to understand the behaviour of individual simulation tools. In this appendix, we document the pilot simulations and the tests we conducted with them, which played a significant role in informing the design and final execution of the simulations.

\begin{figure}
    \includegraphics[width=\columnwidth]{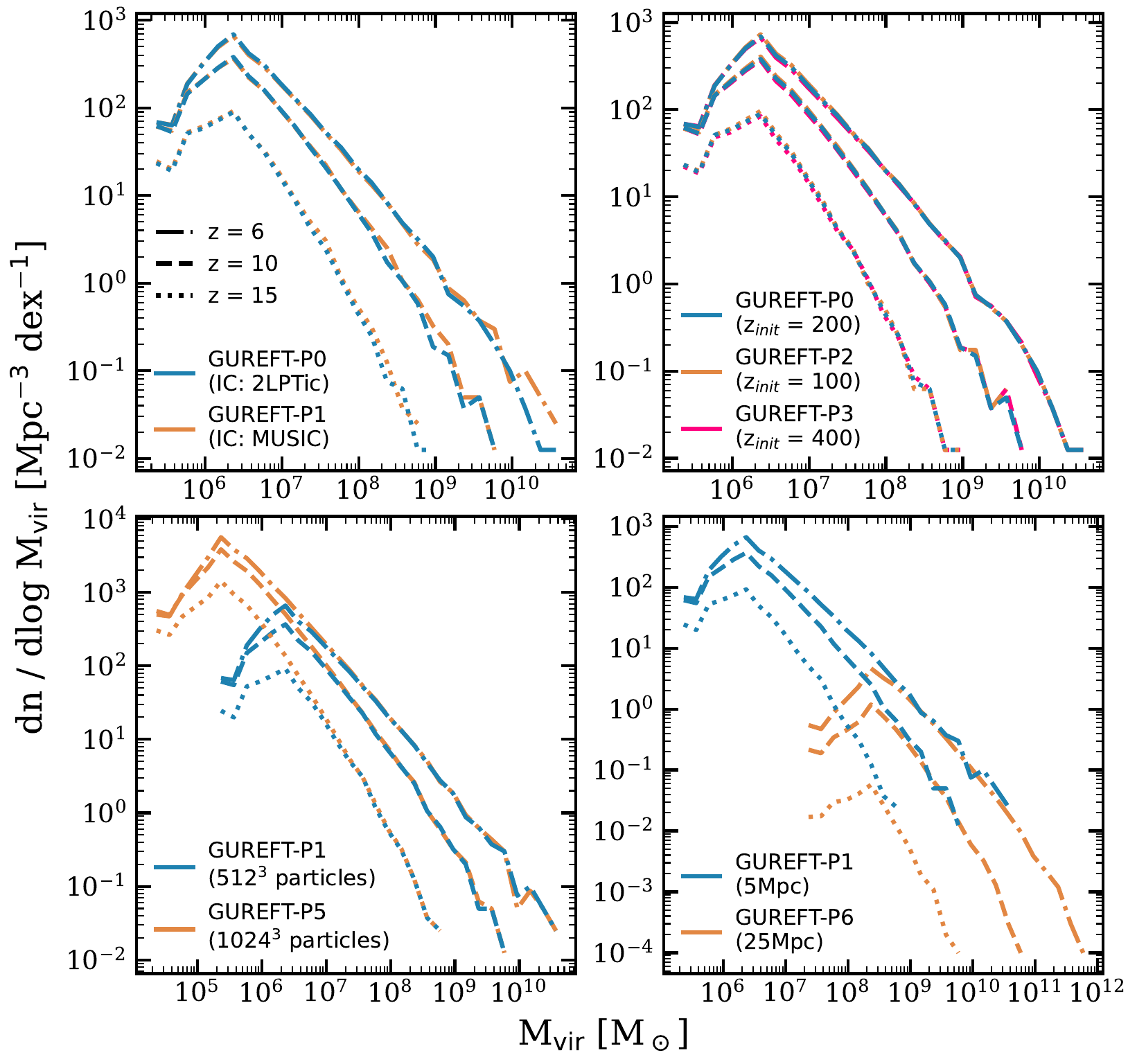}
    \caption{
        This figure shows a comparison of HMFs among the \gureft Pilot simulations. In these pilot simulations, we compared two different initial condition generation tools (\textit{upper-left panel}), three different initial redshifts (\textit{upper-right panel}), simulations of the same volume but different numbers of particles (\textit{lower-left panel}), and simulations with the same number of particles but different simulated volumes (\textit{lower-right panel}). The resultant HMFs demonstrate the high level of convergence that is needed to carry out the \gureft\ simulations.
    }
    \label{fig:GUREFT_Pilot}
\end{figure}

With \gureft\ pilot, we explored two initial condition generation tools: 2LPT\textsc{ic}\footnote{\url{https://cosmo.nyu.edu/roman/2LPT/}}, a second-order Lagrangian Perturbation Theory (2LPT) initial condition generation tool based on n-genic \citep{Springel2015}, and the Multi-Scale Initial Conditions \citep[MUSIC;][]{Hahn2011} code. We also explored the effect of various initial redshift choices $z_\text{init}$, as well as the convergence of simulations with different numbers of particles and box sizes.
The configurations of these simulations are summarised in Table \ref{tab:gureft_pilot_specs}, and we show the HMFs of these simulations in Fig.~\ref{fig:GUREFT_Pilot}.

As shown in Fig.~\ref{fig:GUREFT_Pilot}, there are no significant differences between 2LPT\textsc{ic} and MUSIC. However, we were unable to generate initial conditions with $1024^3$ particles with 2LPT\textsc{ic} (\gureft-P4) and thus opted for MUSIC for the production runs. We also see no significant differences between the different choices $z_\text{init}$ = 100, 200, and 400. 
The resultant HMFs demonstrate the high level of convergence that is needed to carry out the \gureft\ simulations.

\begin{table}
    \centering
    \caption{Configurations of the suite of \gureft\ Pilot (\gureft-P) simulations.
    }
    \label{tab:gureft_pilot_specs}
    \begin{tabular}{lcccccc}
        \hline
                        & Box size         &  N       & $z_{\rm init}$ & IC Generator     \\
                        & [Mpc\,$h^{-1}$]  &          &                                   \\
        \hline
        \gureft-P0      & 5                & $512^3$  & 200            & 2LPT\textsc{ic}  \\
        \gureft-P1      & 5                & $512^3$  & 200            & MUSIC            \\
        \gureft-P2      & 5                & $512^3$  & 100            & 2LPT\textsc{ic}  \\
        \gureft-P3      & 5                & $512^3$  & 400            & 2LPT\textsc{ic}  \\
        \gureft-P4      & 5                & $1024^3$ & 200            & 2LPT\textsc{ic}  \\
        \gureft-P5      & 5                & $1024^3$ & 200            & MUSIC            \\
        \gureft-P6      & 25               & $512^3$  & 200            & MUSIC            \\
        \hline
    \end{tabular}
\end{table}

\bsp	
\label{lastpage}
\end{document}